\documentclass[here,a4paper,11pt]{article}
\renewcommand{\vec}[1]{\mbox{\boldmath $#1$}}
\usepackage{here,psfig,times}

\input epsf
\epsfverbosetrue

\def\eg{{e.g.}}                                    
\def\ie{{ i.e.}\ }                                   
\def\cf{{cf.}\ }

\def\gsim{\lower.4ex\hbox{$\;\buildrel >\over{\scriptstyle\sim}\;$}} 
\def\lsim{\lower.4ex\hbox{$\;\buildrel <\over{\scriptstyle\sim}\;$}} 
 
\def\~  {$\sim$}

\def\rot{\mathop{\rm rot}\nolimits}
\def\div{\mathop{\rm div}\nolimits}
\def\alf{$\alpha$}

\def\Om{$\Omega$}

\def\L{$\Lambda$}

\def\R{R\"udiger}
\def\B{Brandenburg}

\def\R{R\"udiger}

\def\K{Kitchatinov}
\def\qq{\qquad\qquad}                      
\def\q{\qquad}
\def\F{Ferri\`{e}re}

\def\start{\begin{itemize}}
\def\stop{\end{itemize}}
\def\beg{\begin{equation}}
\def\ende{\end{equation}}

\newcommand{\ea}{{\it et al.\ }}

\begin{document}

{\huge \bf Physics of the solar cycle}\\

\ \ \ \\

{\Large G\"unther R\"udiger and Rainer Arlt}\\

Astrophysikalisches Institut Potsdam,\\
An der Sternwarte 16, D-14482 Potsdam, Germany


\bigskip
\bigskip

\begin{scriptsize}
The theory of the solar/stellar activity cycles is presented,
based on the mean-field concept in magnetohydrodynamics. A new
approach to the formulation of the electromotive force as well
as the theory of differential rotation and meridional  circulation
 is described for use in dynamo theory. Activity cycles of  dynamos in the overshoot layer (BL-dynamo)  and distributed dynamos are compared, 
with the latter including the influence of meridional flow. The 
overshoot layer dynamo is able to reproduce the solar cycle periods 
and the butterfly diagram only if $\alpha=0$ in the convection zone.
The problems of too many magnetic belts and too short cycle times 
emerge if the overshoot  layer is too thin. The distributed dynamo 
including meridional flows with a magnetic Reynolds number 
${\rm Rm}\gsim 20$ (low magnetic Prandtl number) reproduces
the observed butterfly diagram even with a positive dynamo-$\alpha$
in the bulk of the convection zone.

The nonlinear feedback of strong magnetic fields on  differential
rotation in the mean-field conservation law of angular momentum leads 
to grand minima in the cyclic activity  similar to those observed. 
The 2D model described here contains the large-scale interactions as 
well as the small-scale feedback of magnetic fields on differential 
rotation and induction in terms of a mean-field formulation 
($\Lambda$-quenching, $\alpha$-quenching). Grand minima may also 
occur if a  dynamo occasionally falls below its critical eigenvalue. 
We expressed this idea by an on-off $\alpha$ function which is 
non-zero only in a certain range of magnetic fields near the 
equipartition value. We never found any indication that the dynamo 
collapses by this effect after it had once been excited.

The full quenching of turbulence by strong magnetic fields in
terms of reduced induction ($\alpha$) and reduced turbulent 
diffusivity ($\eta_{\rm T}$) is studied with a 1D model. 
The full quenching results in a stronger dependence of cycle period
on dynamo number compared with the model with $\alpha$-quenching alone
giving a very weak cycle period dependence.

Also the temporal  fluctuations of $\alpha$ and $\eta_{\rm T}$ from
a random-vortex simulation were applied to a dynamo model. Then the  
low `quality' of the solar cycle can be explained with  a
relatively small number of giant cells acting as the dynamo-active turbulence. 
The simulation contains  the transition from almost regular magnetic 
oscillations (many vortices) to a more or less chaotic time series (very 
few vortices).

\end{scriptsize}

\noindent
\section{Introduction}

Explaining the characteristic period of the quasicyclic activity 
oscillations of stars with the Sun included is one of the challenges 
for stellar physics. The main period is an essential property
of the dynamo mechanism. 
Solar dynamo theory is reviewed here in  the
special context of the cycle-time problem. 
The parameters of the convection zone turbulence do not
easily provide us with the 22-year time scale for the solar dynamo.
Even for the boundary layer dynamos, it is  only possible
if a `dilution factor' in the turbulent electromotive
force smaller than unity  is  introduced which parameterizes
the intermittent character of the MHD-turbulence.    

\begin{figure}
\psfig{figure=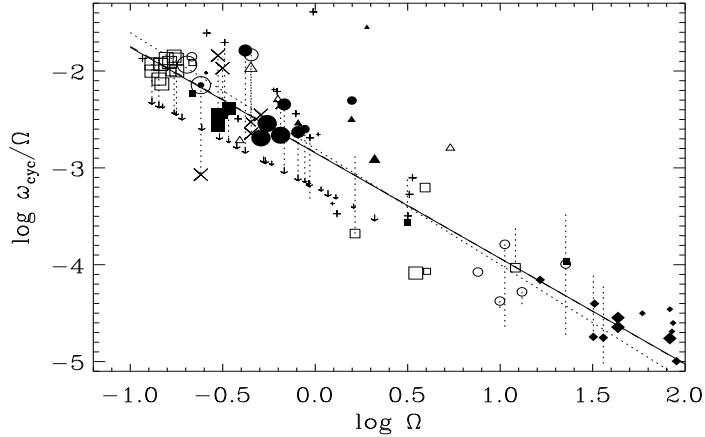,width=12.5cm,height=6.5cm,bbllx=-50pt,bblly=360pt,bburx=558pt,bbury=720pt}
\caption{Stellar cycles: the ratio of the cycle time to the
rotation period after Saar and Brandenburg (1999). The solid
line fits the observations for all stars and leads to the weak proportionality
$\omega_{\rm cyc} \propto \Omega^{-0.1}$ while a previous fit
(dashed line) provided $\omega_{\rm cyc} \propto \Omega^{-0.2}$
(Brandenburg {\it et al.}, 1998).} 
\label{cycles}
\end{figure}

A notable number of interesting phenomena have been investigated
in the search for the solution of this problem: flux-tube dynamics, 
magnetic quenching, parity breaking, and chaos. Nevertheless, even the 
simplest observation -- the solar cycle period of 22 years -- is hard 
to explain (cf. DeLuca and Gilman, 1991;
Stix, 1991; Gilman, 1992; Levy, 1992; Schmitt, 1993;
Brandenburg, 1994a; Weiss, 1994).
How can we understand the existence of the
large ratio of the mean cycle period and the correlation time of the
turbulence? Three  main observations are basic in this respect:
\vspace*{-0.3truecm}
\begin{itemize}
\parskip0pt
\parsep0pt
\itemsep0pt
\item There is a factor of about 300 between the solar cycle time 
and the Sun's rotation period. 
\item This finding is
confirmed by stellar observations (Fig.~\ref{cycles}).
\item The convective turnover
time near the base of the convection zone is very similar to the 
solar rotation period.
\end{itemize}
\vspace*{-0.3truecm}
The problem of 
the large {\it observed\/} ratio of cycle and correlation times,
\begin{equation}
\frac{\tau_{\rm cyc}}{\tau_{\rm corr}} \gsim 10^2,
\label{cyc}
\end{equation}
constitutes the primary concern of dynamo models. In a thick
convection shell this number reflects (the square of) the
ratio of the stellar radius to the correlation length and
numbers of the order 100 in (\ref{cyc}) are possible. For the thin boundary layer
dynamo, however, the problem becomes more dramatic and is in need
of an extra hypothesis. 

\begin{figure}
\mbox{\psfig{figure=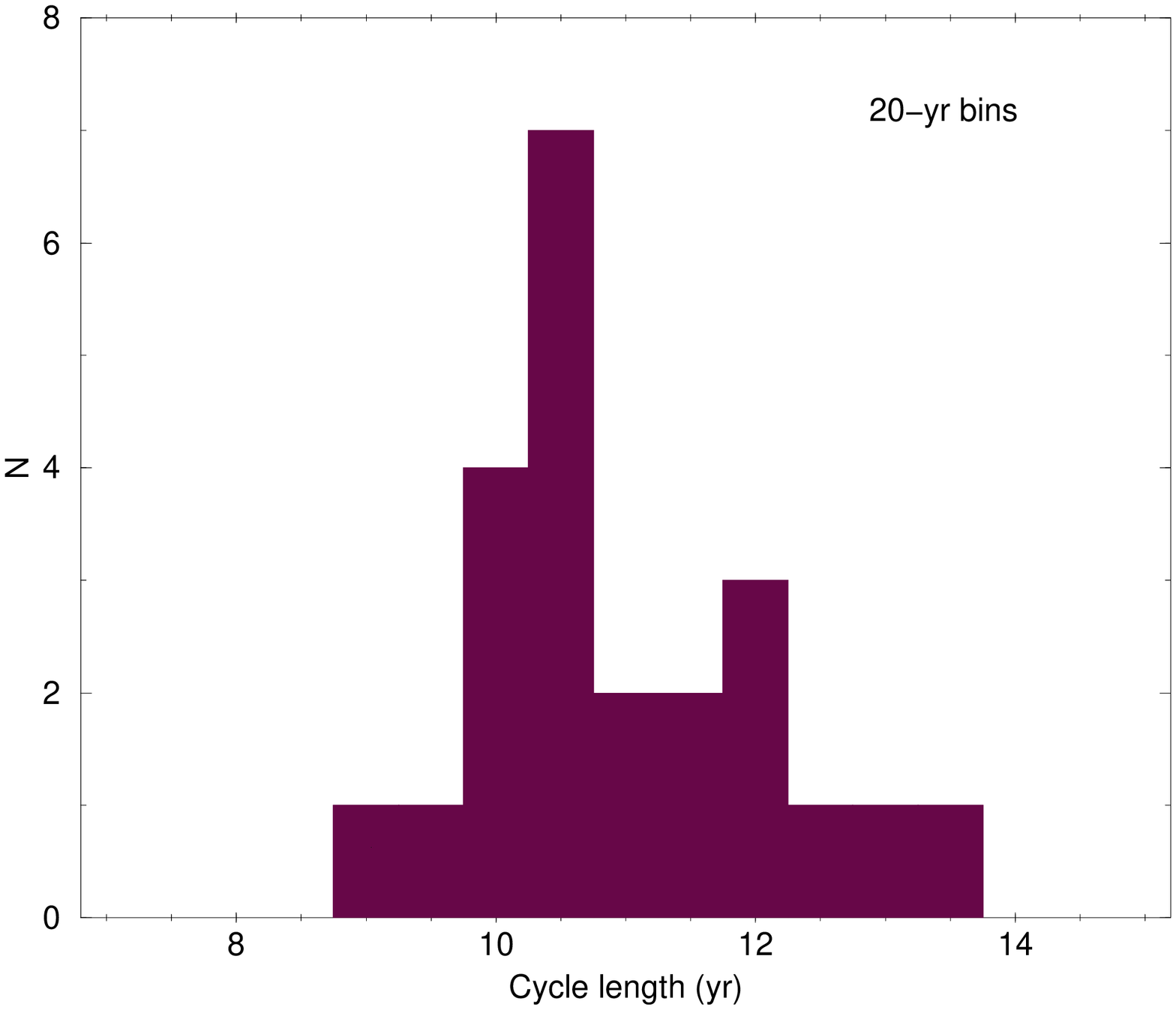,angle=270,height=6truecm,width=7.5truecm,bbllx=12pt,bblly=62pt,bburx=576pt,bbury=757pt}
\hfill
\psfig{figure=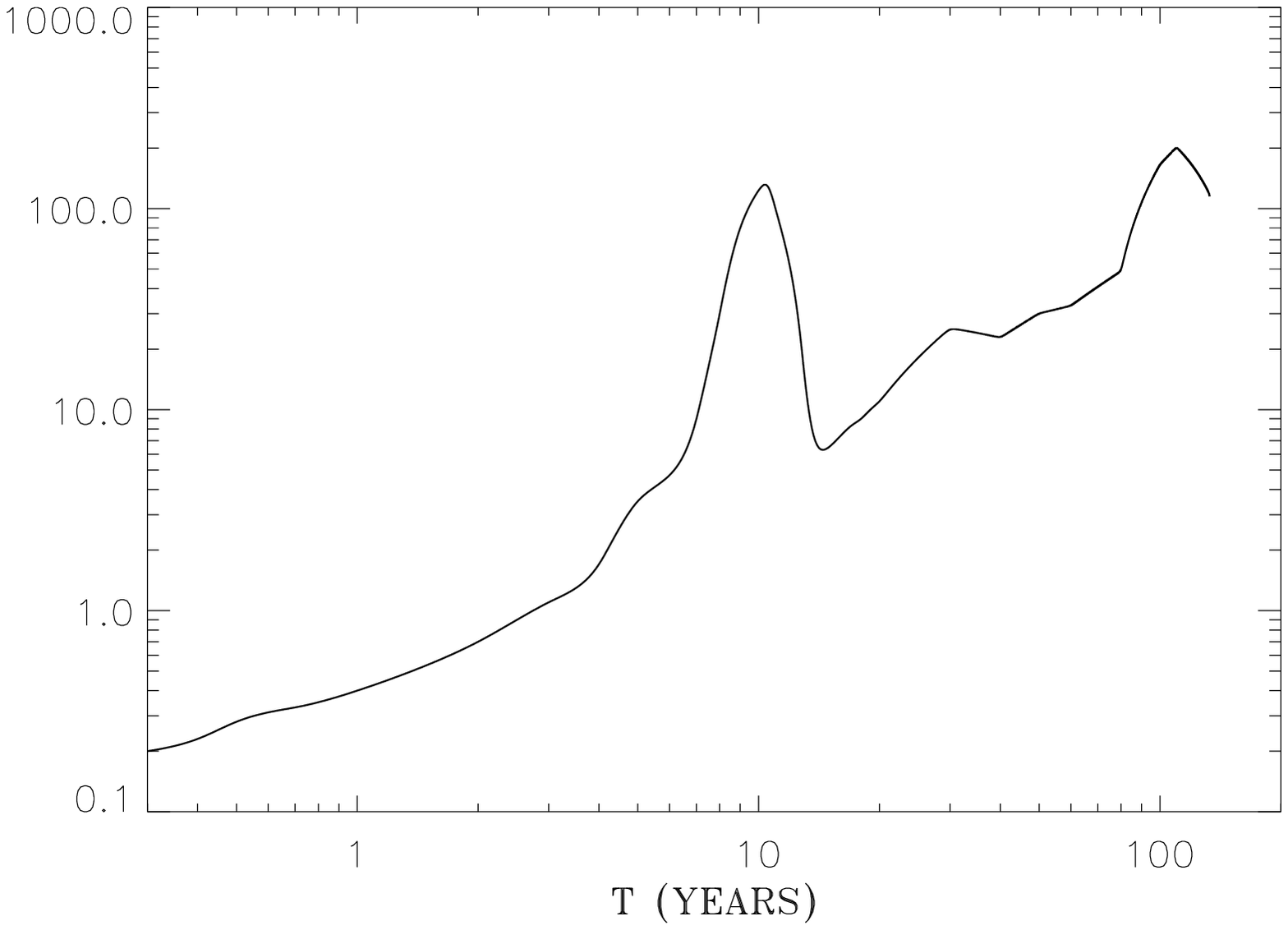,height=5.5truecm,width=7.5truecm,bbllx=54pt,bblly=380pt,bburx=558pt,bbury=727pt}}
\caption{LEFT: The distribution of the solar cycle length does
not approach a Dirac function, the `quality' of the cycle only
gives values of about 5. RIGHT: The wavelet spectrum of the
sunspot-number time series shows two  peaks for both 10~yr 
and 100~yr (Frick {\it et al.}, 1997a).}  
\label{quality}
\end{figure}

The activity period of the Sun varies strikingly about its average
from one cycle to another. Only a nonlinear theory will be able to 
explain the non-sinusoidal (chaotic or not) character of the activity
cycle (Fig.~\ref{quality}). A linear theory is only concerned with
the {\it mean} value of the oscillation frequency.

\section{Basic Theory}
Kinematic dynamo theory utilizes only one equation to advance
the mean magnetic field in time, i.e.
\begin{equation}
\frac{\partial \langle{\bf B}\rangle}{\partial t} =  \rot\left(
\langle{\bf u}\rangle \times \langle{\bf B}\rangle  + \vec{\cal E}\right).
\label{1}
\end{equation}
Here only a non-uniform rotation will be imposed on the mean flow, 
$\langle{\bf u}\rangle$; any meridional flow shall be
introduced later.
The turbulent electromotive force (EMF), $\vec{\cal E}  =
\langle{\bf u'}\times {\bf B'}\rangle$,  
contains induction $\alpha_{ij}$ and dissipation 
$\eta_{ijk}$, i.e.
\beg
{\cal E}_i = \alpha_{ij}
\langle B_j\rangle + \eta_{ijk} 
\langle B_j\rangle_{,k} + \dots \ .  
\label{2}
\ende
Both tensors are pseudo-tensors. While for $\eta_{ijk}$ an elementary 
isotropic pseudo-tensor exists (\lq\lq $\varepsilon_{ijk}$"), the same is not true for $\alpha_{ij}$.  
An odd number of $\Omega$'s is, therefore, {\it required\/} for the 
$\alpha$-tensor, which is only possible with an odd number of
another preferred direction, (say) ${\bf g}$. The \alf-effect 
can thus only exist in stratified, rotating turbulence. The first
formula reflecting this situation,
\beg
\alpha = c_\alpha \frac{l^2_{\rm corr}~ \Omega}{H_\rho}
\cos{\theta},
\label{a3}
\ende
was given by Krause (1967) with
$\Omega$ being the angular velocity of the basic rotation, $\theta$
the colatitude, and $H_\rho$ the density scale height.
Evidently, \alf \ is a complicated effect, where the effective $\alpha$
might really be very small; the unknown factor $c_\alpha$ in (\ref{a3})  
may be much smaller than unity. The strength of this effect was
computed in recent analytical and  numerical simulations for both 
convectively unstable as well as stable stratifications. 
While Brandenburg \ea (1990) worked with a box heated from
below, Brandenburg and Schmitt (1998) considered magnetic
buoyancy in the transition region between the radiative solar core
and the convection zone, Brandenburg (2000) probed the
Balbus-Hawley instability for dynamo-\alf\ production.
\F\ (1993), Kaisig \ea (1993), and Ziegler \ea (1996) used 
random supernova explosions to drive the galactic turbulence.  
The magnitudes of the \alf-effects do not reach the given estimate 
in these cases: the dimensionless factor $c_\alpha$ seems
really to be much smaller than unity. 

\subsection{Simple Dynamos}

\subsubsection{Dynamo waves}
An illuminating example of kinematic dynamo theory for the
cycle period is the dynamo wave solution. In  plane
Cartesian geometry there is a mean magnetic field subject
to a (strong) shear flow and an \alf-effect --
all quantities vary only in the $z$-direction with a given
wave number (Parker, 1975). The amplitude equations can then be written as
\beg
\dot{A} + A = C_\alpha B, \ \ \ \dot{B} + B = iC_\Omega A,
\label{4}
\ende
with $A$ representing the poloidal magnetic field component and $B$ the
toroidal component. $C_\alpha$ is the normalized \alf \ and 
$C_\Omega$ is the normalized shear. 
The eigenfrequency of the equation system is the complex number
\beg
\omega_{\rm cyc} = \sqrt{{{\cal D} \over 2}} + i \left(1 -
\sqrt{{{\cal D}\over 2}}\right).
\label{5}
\ende
A marginal solution of the magnetic field is found for a
`dynamo number' ${\cal D} \equiv C_\alpha C_\Omega = 2$. In that case the 
field is not steady but oscillates with the (dimensionless)
frequency $\omega_{\rm cyc} = 1$.   
 
The (re-normalized) cycle period is thus given simply by
combining the eddy diffusivity and the 
wave number, i.e. $\omega_{\rm cyc} \simeq \eta_{\rm T} k^2$ (which is
simply the 
skin-effect relation).  With a  mixing-length expression for  
the eddy diffusivity, $\eta_{\rm T} = c_\eta l^2_{\rm corr}/\tau_{\rm
corr}$, one 
finds the basic relation between the cycle time and the
correlation time as
\beg
\frac{\tau_{\rm cyc}}{\tau_{\rm corr}} =
\frac{1}{2\pi c_\eta}\left(\frac{R}{l_{\rm corr}}\right)^2.
\label{6}
\ende
Note that $c_\eta \le 0.3$. The ratio between the global scale, $R$, and
the correlation length of the turbulence, $l_{\rm corr}$,
determines the cycle time.  As this ratio has a minimum value
of 10, it is thus no problem to reach a factor of 100 between the cycle period
and the correlation time.\footnote{On the other hand, in order
to reproduce a factor of order $10^2$ the correlation length
must not be too small.}  
The rotation period does not enter --  it only influences the
cycle period in the nonlinear regime (for dynamo numbers
exceeding 2).  For such numbers the frequency increases -- and
the cycle time becomes shorter.

\subsubsection{Shell dynamo}
The simplest assumptions about the \alf-effect and the differential rotation
are used in the shell dynamo, i.e.
\beg
\alpha = \alpha_0 \cos{\theta},  \qq \Omega = {\rm const.} +
\Omega' x
\label{7}
\ende
for $x_i < x < 1$ (Roberts, 1972; Roberts and
Stix, 1972).  
Positive $\Omega'$ means radial super-rotation and negative $\Omega'$
means sub-rotation. The solution with the lowest eigenvalue for the 
sub-rotation case in a thick shell is a solution with dipolar symmetry, which is, 
in the vicinity of the bifurcation point, the only stable one
(Krause and Meinel, 1988).  
 
The cycle time grows with the linear thickness $D$ of the shell.
Compared with the dynamo wave, we find a reduction of (\ref{6}) by the
normalized thickness, $d \le 0.5$. As long as convection zones of
main-sequence stars are considered, that is not too dramatic,
but what about rather thin layers like the solar overshoot region? 
As the cycle period is found to vary linearly with $D$ in 
similar shell models,
\beg
\tau_{\rm cyc} \simeq 0.26\ \frac{ R D}{\eta_{\rm T}},
\label{8}
\ende
the cycle time must become dramatically short in very thin boundary
layers.

\subsubsection{The dynamo dilemma}
The sign of \alf\ has been presumed as positive in 
the northern hemisphere and the differential rotation has been
considered unknown (Steenbeck and Krause, 1969) from the early 
years of dynamo theory.
Only sub-rotation -- with angular velocity increasing inwards -- 
could then produce a suitable butterfly diagram from the
latitudinal migration 
of the toroidal field. In the other case, e.g. where the surface
rotation law is applied in the entire convection zone, one cannot 
reproduce the migration of the toroidal field towards the equator
(K\"ohler, 1973).  

The correct field migration according to the sunspot butterfly
diagram is generally produced by a negative dynamo number only,
which may be either due to positive \alf\ in the northern hemisphere 
and sub-rotation, or negative \alf\ and super-rotation.  
For all dynamos with super-rotation, however, the phase
relation between the radial field component and the toroidal field
component is $\langle B_r\rangle \langle B_\phi\rangle > 0$, again disagreeing
with the observations (Stix, 1976). The prediction of dynamo theory for the solar 
differential rotation was thus 
a clear sub-rotation, $\partial \Omega / \partial r < 0$.
This prediction contrasts with the results of the helioseismology which
revealed sub-rotation only near the poles, whereas near the solar 
equator there is a super-rotation at the base of the convection zone 
(Fig.~\ref{dr}). 

Agreement with the butterfly diagram shall thus be satisfied by a 
negative \alf\ which can be only explained for the overshoot
region.  But  the phase 
relation of the field components does then not agree with the known one (Yoshimura,
1976; Parker, 1987; Schlichenmaier and Stix, 1995). With the
incorrect phase relation it is 
even problematic to  derive  the observed properties of the
torsional oscillations (Howard and LaBonte, 1980), which are
suggested to arise as a
result of the backreaction of the mean-field Lorentz force
(Sch\"ussler, 1981; Yoshimura, 1981; R\"udiger {\it et al.},
1986; K\"uker {\it et al.}, 1996).

In the light of recent developments for dynamos with meridional
flow included, there could easily exist a solution of this
dynamo dilemma in quite an unexpected way (see Section 4).
\subsection{ Differential Rotation}
The theory of the `maintenance' of differential rotation in convective
stellar envelopes might be an instructive detour from dynamo theory.
Differential rotation is also turbu\-lence-induced but without the
complications due to the magnetic fields.  
It is certainly unrealistic to expect a solution of the complicated problem of the solar dynamo
without an understanding of mean-field hydrodynamics. There
is even no hope for the {\it stellar} dynamo concept if the internal
stellar rotation law cannot be predicted.

The main observational features of the solar differential rotation are
\vspace*{-0.3truecm}
\begin{itemize}
\parskip0pt
\parsep0pt
\itemsep0pt
\item surface equatorial acceleration of about 30\%,
\item strong polar sub-rotation and weak equatorial super-rotation,
\item reduced equator-pole difference in $\Omega$ at the
lower convection-zone boundary. 
\end{itemize}
\vspace*{-0.3truecm}
The characteristic Taylor-Proudman structure in the equatorial region
and the characteristic disk-like structure in the polar region
are comprised by the results.
In the search for {\it stellar} surface differential rotation,
chromospheric activity  has been monitored for more than a decade.
Surprisingly enough, there is not yet a very clear picture. For
example, the rotation pattern of the solar-type star 
HD 114710 might easily be reversed compared with that of the Sun 
(Donahue and Baliunas, 1992).\footnote{if the same butterfly
diagram is applied.}   

In close correspondence to dynamo theory we develop the theory of 
differential rotation in a mean-field formulation starting from 
the conservation of angular momentum,
\begin{equation}
\frac{\partial}{\partial t}\left(\rho r^2
\sin^2{\theta \Omega}\right) + \frac{\partial}{\partial x_i} \left(\rho r
\sin{\theta} Q_{i \phi}\right) = 0,
\label{9}
\end{equation}
where the Reynolds stress $Q_{i\phi}$ derived from the correlation tensor
\beg
Q_{ij} = \langle u'_i({\bf x},t)u_j'({\bf x},t)\rangle 
\ende
corresponds to the EMF in mean-field electrodynamics. 

\begin{figure}
\mbox{\psfig{figure=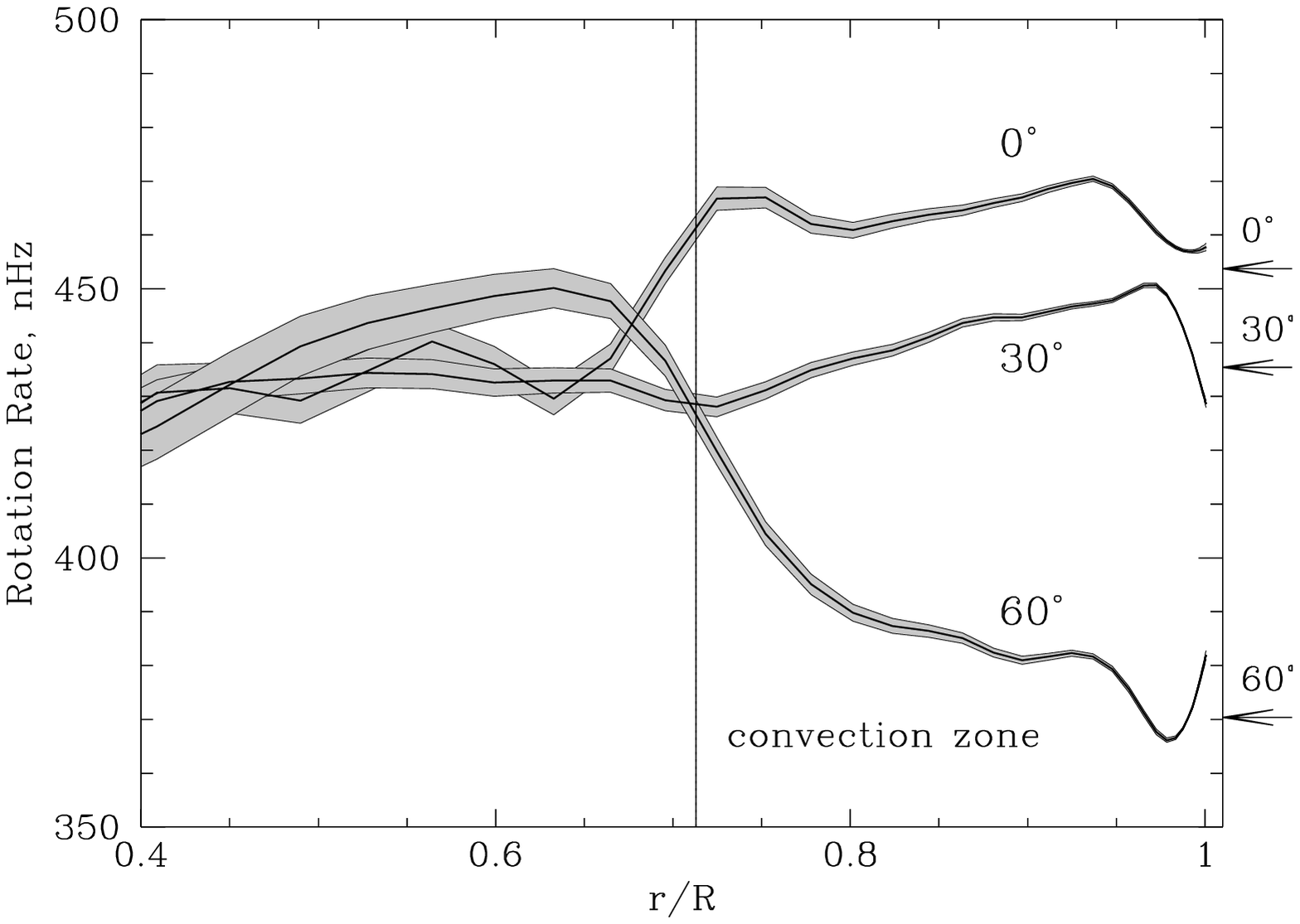,width=8.0cm,height=3.0cm} \hfill
     \psfig{figure=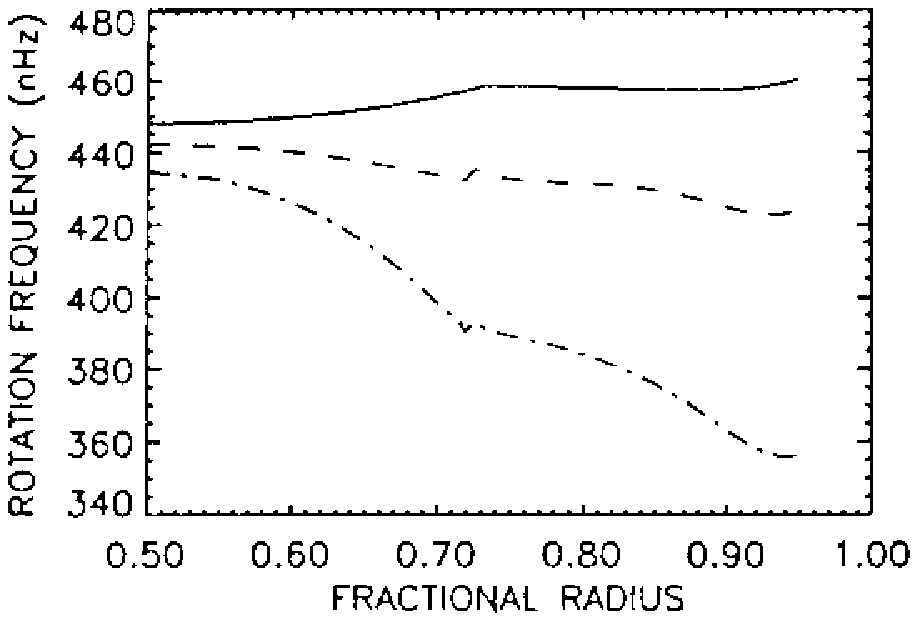,width=7cm,height=4.0cm,bbllx=166pt,bblly=342pt,bburx=440pt,bbury=514pt}}
\caption{LEFT: The internal rotation of the Sun after the
inversion of SOHO data  (Kosovichev {\it et al.}, 1997). RIGHT:
Theory of the solar internal rotation by Kitchatinov and
R\"udiger (1995). The rotation frequency is given for the
equator (solid), mid-latitudes (dashed) and poles (dashed-dotted).} 
\label{dr}
\end{figure}

The correlation tensor involves both dissipation (`eddy viscosity') 
as well as `induction' ($\Lambda$-effect):
\beg
Q_{ij} = \Lambda_{ijk} \Omega_k - {\cal N}_{ijkl}\Omega_{k,l} .
\label{10}
\ende
 Both effects are represented by tensors and  must be 
computed carefully. For anisotropic and rotating
turbulence the zonal fluxes of angular momentum can be written as
\begin{eqnarray}
Q_{r\phi} &=&  - \nu_\perp r \sin{\theta} \frac{\partial
\Omega}{\partial r} - (\nu_\parallel - \nu_\perp) \sin{\theta} 
\cos{\theta} \left(r \cos{\theta} \frac{\partial
\Omega}{\partial r} - \sin{\theta} \frac{\partial
\Omega}{\partial \theta}\right) +\nonumber\\
& +& \nu_{\rm T} \left(V^{(0)} +
\sin^2{\theta V^{(1)}}\right) \Omega \sin{\theta},
\label{Q}
\end{eqnarray}
\begin{equation}
Q_{\theta \phi} =  - \nu_\perp  \sin{\theta} \frac{\partial
\Omega}{\partial \theta} - (\nu_\parallel - \nu_\perp) \sin^2{\theta}
\left(\sin{\theta} \frac{\partial \Omega}{\partial \theta} - r
\cos{\theta} \frac{\partial \Omega}{\partial r}\right)
 + \nu_{\rm T}
H^{(1)} \Omega \sin^2{\theta} \cos{\theta}
\label{Q1}
\end{equation} 
(see \K, 1986; Durney, 1989; \R, 1989).  All coefficients are found to be
strongly dependent on the Coriolis number
\beg 
\Omega^* = 2\tau_{\rm corr}~ \Omega
\label{omst}
\ende
with $\tau_{\rm corr}$ as the convective
turnover time (Gilman, 1992; \K\ and \R, 1993). Moreover, the most important terms of the 
$\Lambda$-effect $(V^{(0)}, H^{(0)})$ correspond to  higher orders of the Coriolis number.
The Coriolis number exceeds unity almost everywhere in the
convection zone except the surface layers.
That is true for all stars -- in this sense all main-sequence stars are
rapid rotators. {\em Theories linear in $\Omega$ are not appropriate 
for stellar activity physics.}
The  \alf-effect will thus never run with $\cos{\theta}$ in its
first power.
\begin{figure}
\mbox{\psfig{figure=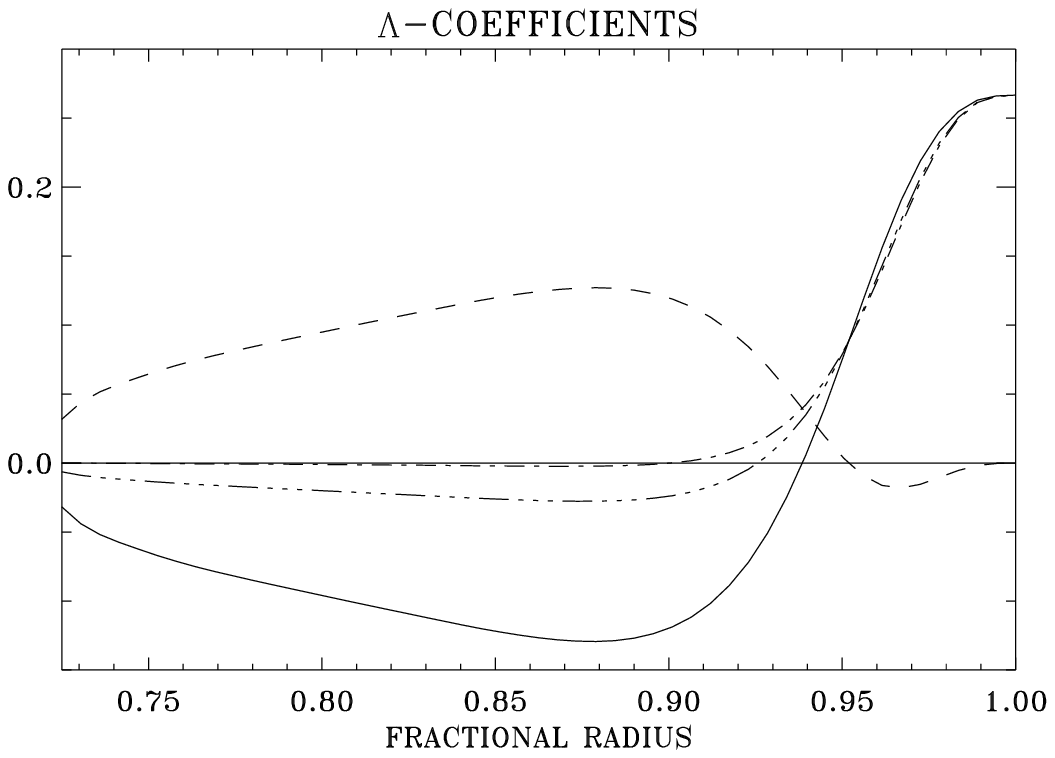,width=7.0cm,height=4.0cm} \hfill
     \psfig{figure=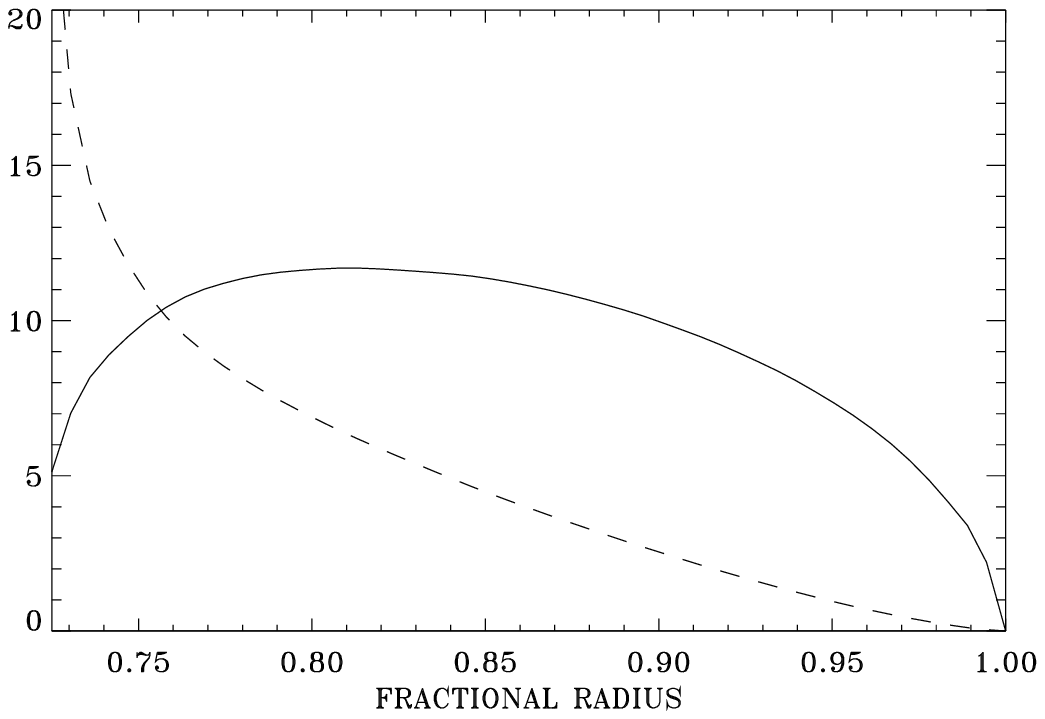,width=7cm,height=4.0cm}}
\caption{LEFT: The turbulence-originated coefficients  in the
non-diffusive zonal fluxes of angular momentum in the solar
convection zone. $V^{(0)}$ (solid), $V^{(1)}$ (dashed),
$V^{(0)} + V^{(1)}$ (dot-dashed).
RIGHT: Eddy viscosity  in units of $10^{12}$ cm$^2$/s (solid) and Coriolis number $\Omega^*$ in the solar convection zone.
} 
\label{Ro}
\end{figure}

The Coriolis number 
$\Omega^*$ is smaller than unity at the top of the convection 
zone and larger than unity at its bottom. At that depth we find
minimal eddy viscosities (`$\Omega$-quenching') and maximal
$V^{(1)} = H^{(1)}$ (Fig.~\ref{Ro}). Since  the latter are
known as responsible for 
pole-equator differences in $\Omega$, we can state that the
differential rotation  
is produced in the deeper layers of the convection zone where
the rotation must be considered as rapid ($\Omega^* \lsim 10$).

\begin{figure}
\mbox{\psfig{figure=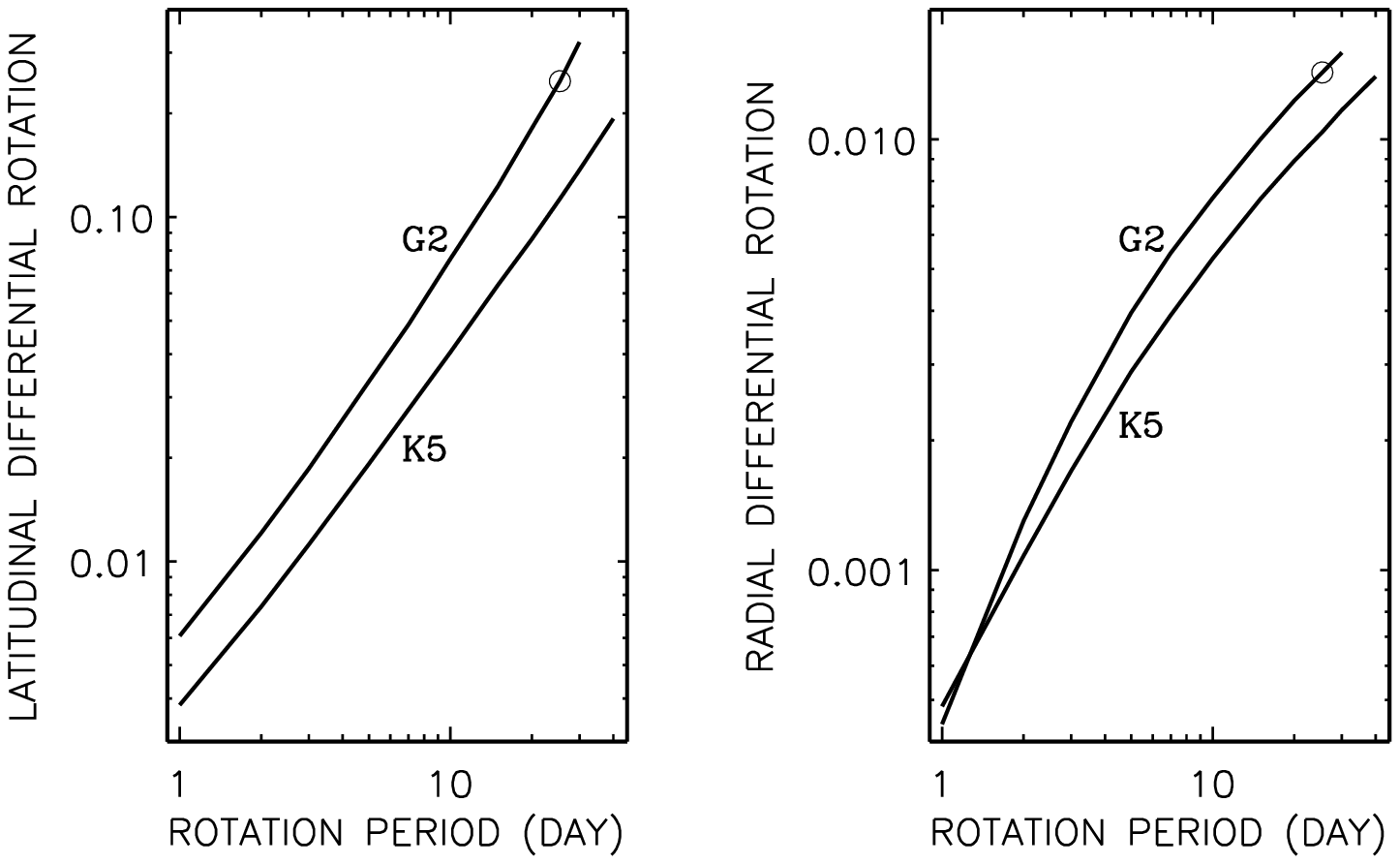,width=9.0cm,height=5.0cm} \hfill
     \psfig{figure=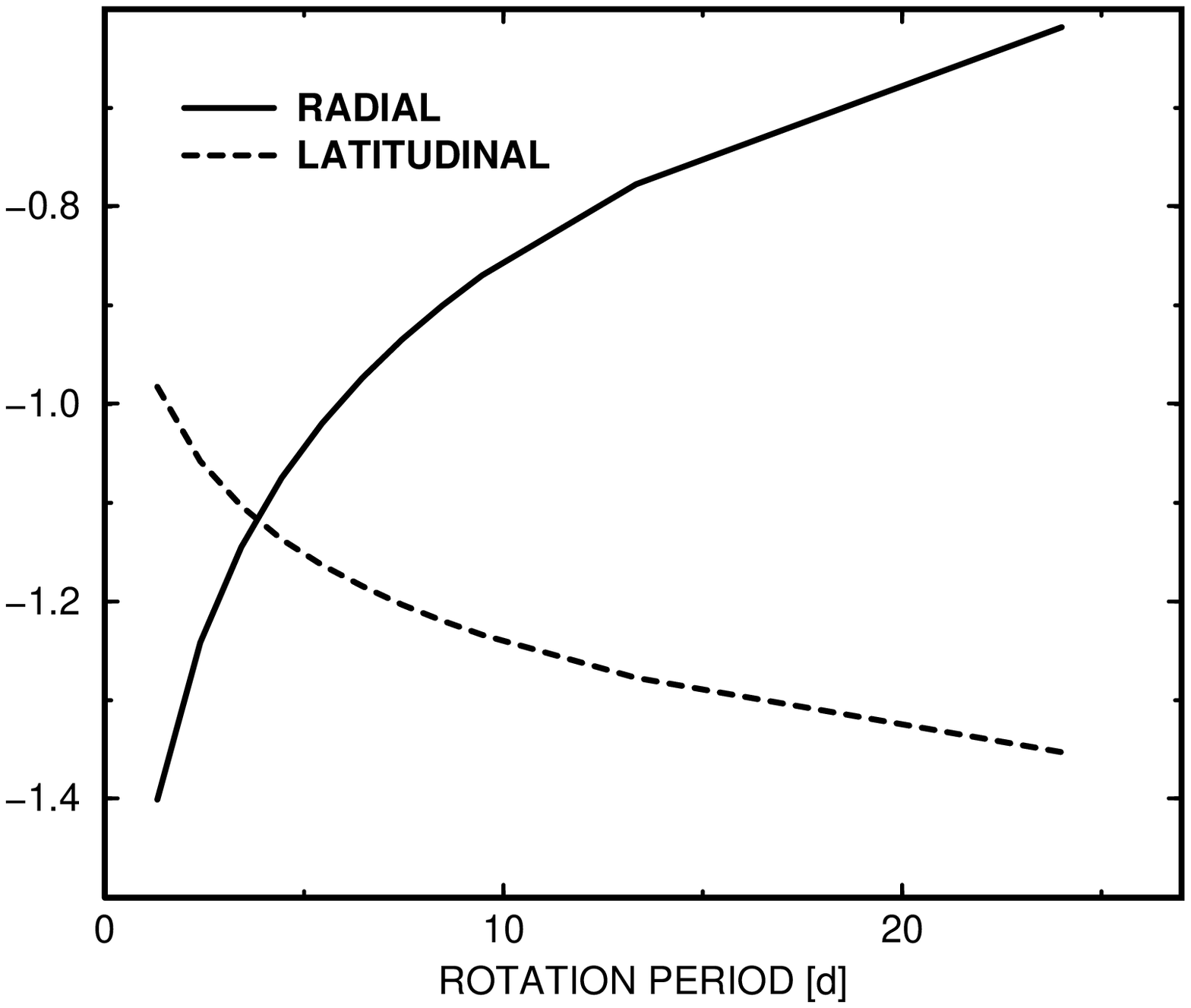,width=6.1cm,height=5.4cm,angle=0,bbllx=-25pt,bblly=29pt,bburx=576pt,bbury=468pt}}
\caption{The rotation differences $\delta \Omega / \Omega$ for both a solar model and a K5 main-sequence star 
vs. basic rotation after \K\ and \R\ (1999). LEFT:  Normalized equator-pole difference of surface 
rotation, CENTER: The normalized radial differential rotation, RIGHT: The scaling  exponent
$\kappa$ in (\ref{15.1}) for the  solar model. In both cases 
$\kappa \simeq -1$ is approximated. } 
\label{fdr1}
\end{figure}

The solution of (\ref{9}) with the turbulence quantities in
Fig.~\ref{Ro} is given in K\"uker \ea (1993) and  
\K\ and \R\ (1995) using a mixing-length model by Stix
and Skaley (1990). 
With a mixing-length  ratio  ($\alpha_{\rm MLT} = 5/3$)
we find the correct equatorial acceleration of about 30\,\%.
There is a clear radial sub-rotation ($\partial\Omega/\partial r < 0$)
below the poles while in mid-latitudes and below the equator the 
rotation is basically rigid (right panel in Fig.~\ref{dr}). In this way 
the bottom value of the pole-equator difference is reduced and the 
resulting profiles  of the internal angular
velocity are close to the observed ones. Fig.~\ref{fdr1} presents the 
results of an extension of the
theory to a sample of main-sequence stellar models given in \K\
and \R\ (1999). Simple scalings like
\beg
{\delta \Omega \over \Omega} \propto \Omega^\kappa
\label{15.1}
\ende
may be introduced for both the latitudinal as well as the radial
differences of the angular velocity. The $\kappa$-exponents prove 
to be negative and of order $-1$
(right panel of Fig. \ref{fdr1}). So we find that for one and
the same spectral type the approximation
\beg
\delta \Omega \simeq {\rm const.} 
\label{15.2}
\ende
should be not too rough. Recent observations of AB~Dor (Donati and
Cameron, 1997) and PZ~Tel (Barnes {\it et al.}, 1999) seem to
confirm this surprising and unexpected result where in all 3
cases the constant value approaches 0.06 day$^{-1}$. 

\subsection{The EMF for Turbulence}
We now step forward to the turbulent EMF using the same turbulence
model as in Section~2.2. We are working with
the same quasilinear approximation and the same convection
zone  model. The additional equation for the magnetic
fluctuations is
\beg
{\partial {\bf B}'\over \partial t} - \eta_{\rm t} \Delta {\bf B}' =
\rot ({\bf u}' \times \langle{\bf B}\rangle),
\label{16}
\ende
where  $\eta_{\rm t} \simeq \nu_{\rm t}$ is the small-scale diffusivity. 
In this configuration we deal with an \alf-tensor being highly 
anisotropic, even for the most simple case of slow rotation ($\Omega^* \ll 1$),
\beg
\alpha_{ij}  =  \  \gamma \varepsilon_{ijk} G_k
- \alpha_1({\bf G\Omega}) \delta_{ij} - \alpha_2 (G_i\Omega_j +G_j\Omega_i)
 +\  \hat{\gamma} \varepsilon_{ijk} U_k - \hat{\alpha}_1({\bf
U\Omega}) \delta_{ij} -  \hat{\alpha}_2 (U_i\Omega_j
+U_j\Omega_i)
\label{17.1}
\ende
(Moffatt, 1978; Krause and R\"adler, 1980; W\"alder {\it et al.}, 1980) with
the stratification vectors ${\bf G} = \nabla \log \rho$ and 
 ${\bf U} = \nabla \log u_{\rm T}$ with the rms velocity $u_{\rm
T}=\sqrt{\langle {\bf u}'^2 \rangle}$. 
The $\gamma$-terms  describe advection effects, such as turbulent diamagnetism 
(R\"adler, 1968; \K\ and \R, 1992) and buoyancy (\K\ and Pipin, 1993) 
and cover the off-diagonal elements of the $\alpha$-tensor. 

The diagonal elements are essential for the induction. The 
mixing-length approximation used in \R\ and \K\ (1993) leads to
\beg
\alpha_{rr} = \hat{\alpha} \left({\bf U} +
\frac{\bf G}{4}\right) {\bf \Omega}, \qq \alpha_{\phi \phi} =
\alpha_{\theta \theta} = - 
\hat \alpha \left({\bf U} + \frac{3
{\bf G}}{2}\right) {\bf \Omega}.
\label{18}
\ende
While the most important component $\alpha_{\phi \phi}$ becomes 
positive (if density stratification dominates), the component $\alpha_{rr}$ 
becomes negative in the northern hemisphere. In contrast to the
standard formulation, the $\alpha_2$-components in (\ref{17.1}) 
are dominant as confirmed by numerical simulations by Brandenburg 
\ea (1990). Our  quasilinear theory of the \alf-effect provides 
$\hat \alpha = 8/15\ \tau_{\rm corr}^2 u_{\rm T}^2$.

The general case of an arbitrary rotation rate is not easy to
present. The `overshoot' region, however, is of particular 
interest and is characterized by $|{\bf U}|> |{\bf G}|$ and
$\Omega^* \gg 1$. In cylindrical coordinates ($s, \phi, z$) and
for very fast rotation one gets
\beg
\alpha =  c_\alpha\ \ \left(
   \matrix{
-{3 \pi\over 8} \cos\theta & {3\pi\over 8 \Omega^*} \cos\theta     & 0\cr
-{3\pi\over 8\Omega^*} \cos \theta  & -{3 \pi\over 8}  \cos\theta & 0\cr     
0   &  -  {3\pi\over 8\Omega^*} \sin\theta    &0 \cr
}\right) \ \  
 \frac{d}{dr}
\left(\tau_{\rm corr} u_{\rm T}^2\right)
\label{19}
\ende
(\R\  and \K, 1993). Note that
\vspace*{-0.3truecm}
\begin{itemize}
\parskip0pt
\parsep0pt
\itemsep0pt
\item all components with index $z$ disappear,
\item the remaining diagonal terms (the \alf-effect) do not
vanish for very rapid rotation,
\item the \alf-terms are negative (in the northern hemisphere)
for outward increase of the 
turbulence intensity (like in the overshoot region),
\item the advection terms tend to vanish for rapid rotation;
their relation to the \alf-effect is given by the Coriolis
number $\Omega^*$. 
\end{itemize}
\vspace*{-0.3truecm}
\begin{figure}
\psfig{figure=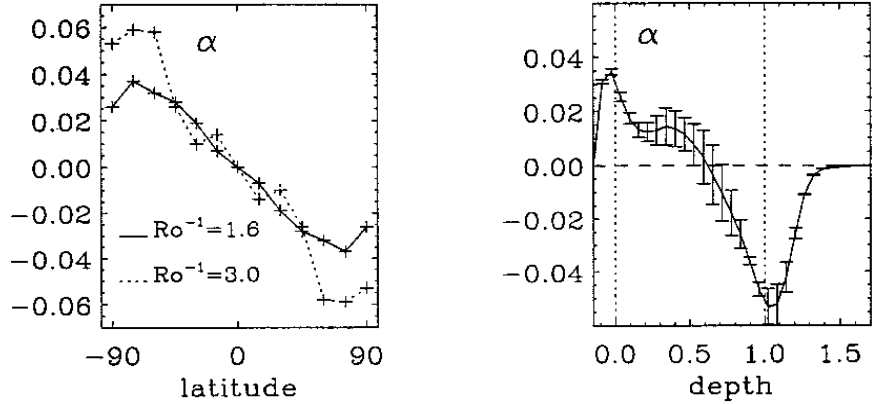,width=13cm,height=5.3cm,bbllx=30pt,bblly=331pt,bburx=479pt,bbury=529pt}
\caption{Simulation of the \alf-effect after Brandenburg (1994b)
versus latitude and depth. Note the negative values at the
bottom of the convection zone (RIGHT) and the maximum for high
latitudes (LEFT).}  
\label{alf}
\end{figure}
It might seem that the \alf-effect is now in our hands.
There are, however, shortcomings beyond the use of the
quasilinear approximation. The main point is the absence of
terms higher than  first order in ${\bf G}$ or
${\bf U}$, respectively. In particular,  all terms with
(${\bf G \Omega})^3 \sim \cos^3{\theta}$ are missing so that we
have no information  about the \alf-effect at
the poles where for slow rotation
$\alpha_{\phi \phi}$ is largest. However, this need not to be
true for `rapid' rotation. Simulations of $\alpha_{\phi \phi}$ with
its latitudinal profile by Brandenburg (1994b) indeed showed
that the maximum occurs at mid-latitudes (Fig.~\ref{alf}). 
Schmitt (1985, 1987) derived a similar profile for a different type 
of instability.  

A new discussion of the $\alpha$-effect in shear flows
recently emerged (cf.\ Brandenburg and Schmitt, 1998). A quasilinear computation of the influence
of the differential rotation on the $\alpha$-effect leads to 
\beg
\alpha \simeq -l_{\rm corr}^2   {\partial \Omega \over \partial
\theta}  {d \log(\rho u_{\rm T})\over dr} \sin\theta 
\label{15.3}
\ende
for a sphere or 
\beg
\alpha \simeq -l_{\rm corr}^2\ s {d\Omega \over ds} {d\log\rho
\over dz}
\label{16.1}
\ende
for a disk  (R\"udiger and Pipin, 2000). The latter  relation  
for accretion disks with  $ \partial \Omega/\partial s < 0$
yields negative \alf-values in the northern hemisphere and positive
values in the southern hemisphere (see Brandenburg and Donner, 1997).
According to (\ref{19}) or (\ref{15.3}), negative values for the northern 
$\alpha_{\phi\phi}$ in the solar overshoot region are only achievable with a very  
steep decrease of the turbulence intensity with depth.

There are very interesting observational issues concerning the current helicity 
\beg
{\cal H}_{\rm curr}= \langle \vec{j}'\cdot \vec{B}'\rangle
\label{al1}
\ende
with  the electric current $\vec{j}=\rot \vec{B}/\mu_0$. The current helicity  
has the same kind of equatorial \mbox{(anti-)} symmetry as the
dynamo-\alf. For  
{\em homogeneous} global magnetic fields, the dynamo-\alf\ 
 is related to the turbulent EMF 
according to 
$ \alpha_{ij} \bar B_i \bar B_j = {\vec{\cal E}}\cdot \bar{\vec{B}}$.
After R\"adler and Seehafer (1990) we  read this equation as
$\alpha_{\phi\phi} = {\vec{\cal E}} \cdot\bar{\vec{B}} / \bar
B_\phi^2$,
where $\alpha_{\phi\phi}$ is the dominant component of the
$\alpha$-tensor. 
We are, in particular,  interested in checking
their and Keinigs' (1983) antiphase relation,
\beg
{\alpha_{\phi\phi} \bar B^2 \over \mu_0 {\cal H}_{\rm curr}} = - \eta <0,
\label{al2}
\ende
between $\alpha$-effect and current helicity. In addition, we
have recently shown  that indeed the negativity of the
left-hand-side of (\ref{al2}) is preserved for magnetic-driven
turbulence fields which do not fulfill the conditions for which
the Keinigs relation (\ref{al2}) originally  has been derived
(R\"udiger {\it et al.}, 2000).

An increasing number of   
papers presents  
observations of the current helicity at the solar surface, all showing that 
it is {\em negative} in
the northern hemisphere and  positive in the southern
hemisphere (Hale, 1927; Seehafer, 1990;
Pevtsov {\it et al.}, 1995; Abramenko {\it et al.}, 1996; Bao and Zhang,
1998; see Low (1996) for a review). There 
is thus a strong empirical evidence that the
\alf-effect is {\em positive} in the bulk of the solar convection zone
in the northern hemisphere.

For a given field of {\em magnetic} fluctuations the dynamo-\alf\, as well 
as the kinetic and current helicities,  have been computed by 
\R\ \ea\ (2000) assuming  
that the turbulence is subject to  magnetic buoyancy and global rotation.
In particular, the role of magnetic buoyancy  
appears quite important for the generation of $\alpha$-effect. So
far, only the role of density stratification has been
discussed for both $\alpha$-effect and $\Lambda$-effect, and
their relation to kinetic helicity and turbulence
anisotropy. If density fluctuations shall  be included then  it 
makes no sense here to adopt the anelastic approximation 
$\div \rho \vec{u}' =0$ but one has to work with the  mass 
conservation law in the form 
\beg
{\partial \rho'\over \partial t}  + \bar \rho \div \vec{u}' = 0.
\label{2.1}
\ende
For the turbulent energy equation one can  simply adopt the
polytropic relation $
p' = c_{\rm ac}^2  \rho',
$
where $c_{\rm ac}$ is the isothermal speed of sound. The turbulence 
is thus assumed to be driven by the Lorentz force and it is subject to 
a global rotation. The result is an \alf-effect of the form  
\beg
\alpha_{\phi\phi} = - {1\over 5} {\tau_{\rm corr}^2\over c_{\rm ac}^2}
 {\langle {B^{(0)}}^2\rangle \over \mu_0
\rho}  (\vec{g} \cdot \vec{\Omega}).
\label{11.1}
\ende
Here $\vec{g}$ is the gravitional acceleration, and the energy of 
the magnetic fluctuations forcing the turbulence is also given. 
For rigid rotation  the \alf-effect 
proves thus to be {\em positive} in the northern hemisphere
and negative in the southern hemisphere.  Again we do not find 
any possibility to present a negative \alf-effect in the northern 
hemisphere.

The kinetic helicity 
\beg
{\cal H}_{\rm kin} = \langle \vec{u}' \cdot \rot \vec{u}'\rangle
\label{hel}
\ende
has just the same latitudinal distribution as the \alf-effect,
but the magnetic model does not provide  the  minus sign 
between the dynamo-\alf\ and the
kinetic helicity which is characteristic within the conventional 
framework, $\alpha \propto - \tau_{\rm corr} \ {\cal H}_{\rm kin}$. If a rising 
eddy can expand in density-reduced surroundings, then a {\em
negative} value of the kinetic helicity is expected. The
magnetic-buoy\-an\-cy model, however, leads to another result. 

If the real convection zone turbulence is formed by a mixture 
of both dynamic-driven and magnetic driven turbulence, the 
kinetic helicity should also be a mixture of positive parts 
and negative parts so that it should be a rather small quantity. 
The opposite is true for the current helicity which also changes its 
sign at the equator. Again it turns out to be negative in the 
northern hemisphere and positive in the southern hemisphere. 
The  current helicity (due to fluctuations) and the \alf-effect are  
thus always out of phase. The current helicity at the solar 
surface might thus be considered a much more robust 
observational feature than the kinetic helicity (\ref{hel}). 

Strong magnetic fields are known to suppress turbulence and 
the $\alpha$-effect is also expected to decrease. The field strength
at which the $\alpha$-effect reduces significantly is assumed, in 
energy, to be comparable with the kinetic energy of the velocity 
fluctuations, and is called the equipartition field strength 
$B_{\rm eq}=\left(\mu_0\rho u_{\rm T}^2\right)^{1\!/\!2}$. The magnetic feedback
on the turbulence was studied by R\"udiger \& Kitchatinov (1993)
using a $\delta$-function spectral distribution of velocity
fluctuations. The $\alpha_{\phi\phi}$-component of the $\alpha$-tensor
acting in an $\alpha\Omega$-dynamo is found to be suppressed
by the magnetic field with the $\alpha$-quenching function
\beg\Psi=\frac{15}{32\beta^4}\left(1-\frac{4\beta^2}{3(1+\beta^2)^2}-
  \frac{1-\beta^2}{\beta}\arctan\beta\right)
  \label{alfquench}
\ende
with
\beg
  \beta=|\langle B\rangle|/B_{\rm eq}.
\ende
The quenching function decreases as $\beta^{-3}$ for strong magnetic
fields. The function is similar to the heuristic approach
$(1+\beta^2)^{-1}$ which is artificial but nevertheless frequently used in $\alpha$-effect
dynamos. For small fields, $\Psi$ runs like $1-\frac{12}{7}\beta^2$,
the mentioned heuristic approximation runs like $1-\beta^2$.

We can also compute the eddy diffusivity using the same turbulence
theory as in the previous Section. The simplest expression  is
\beg
\eta_{\rm T} = \ c_\eta \tau_{\rm corr} u_{\rm T}^2
\label{20}	
\ende
with $c_\eta \lsim 0.3$. This expression, however, fails to include
rotation {\it and} stratification.  Even a
non-uniform but weak magnetic field induces a turbulent EMF, without 
rotation and stratification in accordance to 
$\eta_{ijk} = \eta_{\rm T} \varepsilon_{ijk}$. A much stronger `magnetic
quenching' of such an eddy diffusivity has been discussed by
Vainshtein and Cattaneo (1992) and Brandenburg (1994a)
and is still a matter of debate. A more conventional theory of
the $\eta_{\rm T}$-quenching and its consequences for the stellar activity
cycle theory is given in Section 6 below.

\section{A Boundary-Layer Dynamo for the Sun}
The spatial location of the dynamo action is unknown unless
helioseismology will reveal the exact position of the
magnetic toroidal belts beneath the solar surface (cf.
Dziembowski and Goode, 1991). There are,
however, one or two arguments in favour of locating it deep
within or below  the convection zone:
\vspace*{-0.3truecm}
\begin{itemize}
\parskip0pt
\parsep0pt
\itemsep0pt
\item Hale's law of sunspot parities can only be fulfilled if
the toroidal magnetic field belts are very strong (10$^5$
Gauss, see Moreno-Insertis, 1983; Choudhuri, 1989; Fan {\it et al.}, 1993; Caligari
{\it et al.}, 1995). 
\item The dynamo field strength is approximately the equipartition value
$B_{\rm eq} = 
\left(\mu_0\rho u_{\rm T}^2\right)^{1\!/\!2}$. 
Using mixing-length theory arguments $\rho u_{\rm T}^3 
\simeq\ $const., hence $B_{\rm eq}$ is  increasing inwards with
$\rho^{1/6}$ (one  order of magnitude, $\lsim 10^4$ Gauss).  
\item  The radial gradient of \Om\ is maximal below the convection zone.
\end{itemize}
\vspace*{-0.27truecm}
As a consequence of these arguments high field
amplitudes might be generated only  in the    layer
 between the convection zone and the radiative
interior (Boundary layer (BL), `tachocline', see van
Ballegooijen, 1982). On the other hand, if there is some form
of   `turbulence'  in this layer, it will be  hard to understand
the present-day finite value of lithium in the solar convection
zone. The lithium burning only starts 40\,000 km below the
bottom of the convection zone. Any 
turbulence  in this domain  would lead to a rapid and complete
depletion of the lithium in the convection zone, which is not
observed. Moreover, as shown by \R\ and \K\ (1997), it is also 
not easy to generate very strong toroidal magnetic fields in
the solar  tachocline as the Maxwell stress tends to deform the
rotation profile to more and more  smooth functions. In our
computation with molecular diffusion coefficient values the
toroidal field amplitude  never did exceed 1 kGauss, 
independent of the radial magnetic field applied. 

\begin{figure}
\psfig{figure=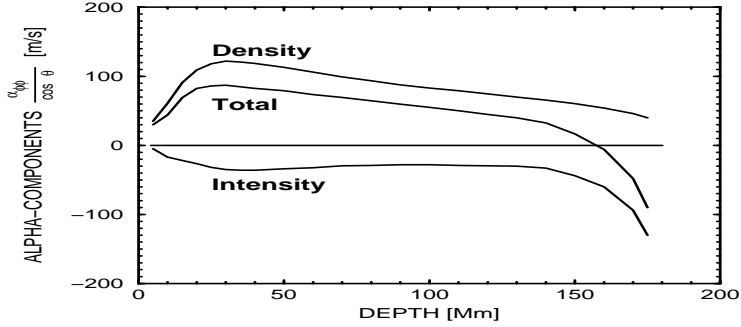,width=13.0cm,height=4.0cm,angle=270,bbllx=100pt,bblly=-100pt,bburx=526pt,bbury=697pt}
\caption{The depth-dependence of the \alf-effect with $C_\alpha
= 1$. Note the different
contributions by density gradient and turbulence intensity distribution.
The \alf-effect becomes negative only at the bottom of the
convection zone (Krivodubskij and Schultz, 1993).} 
\label{alfaphi}
\end{figure}

Krivodubskij and Schultz (1993) (with the inclusion of the
depth-dependence of the Coriolis number $\Omega^*$ and using a
mixing-length model)
derived a  profile of the \alf-effect (Fig.~\ref{alfaphi}) with
a  magnitude of 100 m/s, positive in the convection zone and
negative in the overshoot layer. The negativity of  \alf\ there can
only be relevant for the dynamo  if the bulk of the  convection
zone is free from \alf. It has been argued that the short
rise-times of the flux tubes in the convection zone  prevent
the formation of the \alf-effect (Spiegel and Weiss, 1980;
Sch\"ussler, 1987; Stix, 1991). However, the rise-times may be
much longer  in    the overshoot region (Ferriz-Mas and
Sch\"ussler, 1993, 1995; van Ballegooijen, 1998). Anyway, as
$C_\alpha = 1$ is used in Fig.~\ref{alfaphi}, the given values can
only be considered as maximal values. 

\begin{figure*}[htbp]
\psfig{figure=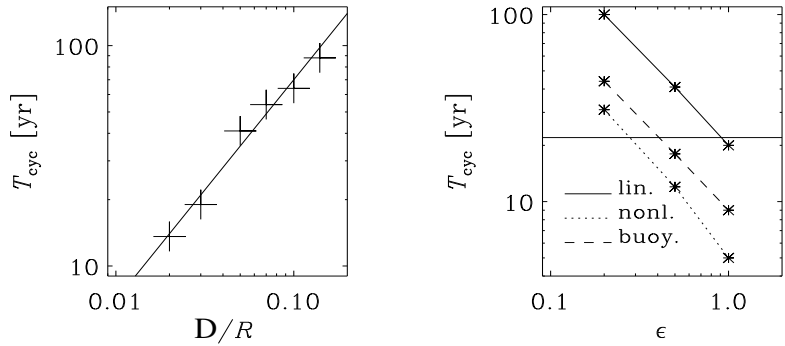,width=13.0cm,height=5.0cm,bbllx=0pt,bblly=380pt,bburx=576pt,bbury=620pt}
\caption[]{
Cycle period $\tau_{\rm cyc}$ as a function of the BL thickness $D/R$
and and the dilution factor $\epsilon$ for $\alpha_{\rm u}=0$.
LEFT: $\epsilon=0.5$, the results are well
represented by $\tau_{\rm cyc}= D\,$[yr/Mm].
RIGHT: $D=35$ Mm. The horizontal line
indicates the solar cycle period of 22 years. 
}\label{Fperiod}
\end{figure*}

\begin{figure*}[htbp]
\psfig{figure=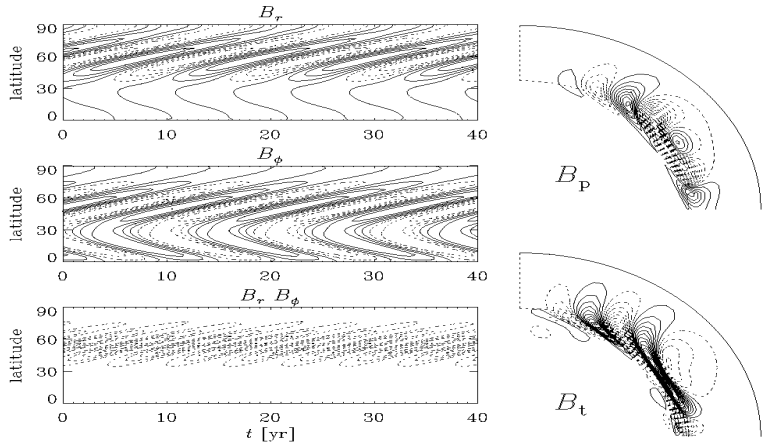,width=13.0cm,height=5.7cm,bbllx=-10pt,bblly=260pt,bburx=595pt,bbury=610pt}
\caption[]{
The magnetic field geometry for the nonlinear ($\alpha$-quenched) solution
without buoyancy and  $\epsilon=0.5$, $D=35$ Mm, $\alpha_{\rm u}=1$.
An obvious problem of the BL dynamo is the large number of toroidal
field belts.
}\label{Fbut3}
\end{figure*}

There is a serious  shortcoming of the BL
concept.  The characteristic scales of the magnetic fields are
no longer much larger than the  scales of the turbulence. The
validity of the local formulations of the mean-field
electrodynamics is not ensured for such thin layers.
Nevertheless, there is an increasing number of corresponding
quantitative models (Choudhuri, 1990; Belvedere {\it et al.}, 1991;
Prautzsch, 1993; Markiel and Thomas, 1999).  
Even in this case a couple of new  questions  will arise.  
For a demonstration of this puzzling situation a model 
has been established by
\R\ and \B\  (1995) under the following assumptions: 
\vspace*{-0.3truecm}
\begin{itemize}
\parskip0pt
\parsep0pt
\itemsep0pt
\item Intermittency (or the weakness  of the turbulence in the
BL) is introduced  by a dilution factor 
$\epsilon$ in ${\cal
E}_i = \epsilon\  (\alpha_{ij}  	
\langle B_j\rangle + \eta_{ijk} \langle B_j \rangle _{,k})$. The
factor $\epsilon$ controls the weight of the turbulent EMF in relation to the
differential rotation.   
\item The ignorance of the $\alpha$-effect in the polar regions is
parameterized with \ $\alpha_{\rm u}$ \ with \
$\alpha \sim \alpha_{\rm u}\,\cos\theta\sin^2\theta$.
\item  \alf\ exists only in the overshoot region, $\eta_{\rm
T}$ only in
the  convection zone. 
\item The correlation time is taken from $\tau_{\rm corr}
\simeq D/u_{\rm T} \simeq 10^6 $\ s  so that $\Omega^* \simeq$
5 results. 
\item The tensors  \alf\ and $\eta_{\rm T}$ are computed for a
rms velocity profile by Stix (1991), the rotation law is directly
taken from Christensen-Dalsgaard and Schou (1988).  
\end{itemize}
\vspace*{-0.3truecm}
The main results from these models are:
\vspace*{-0.3truecm}
\begin{itemize}
\parskip0pt
\parsep0pt
\itemsep0pt
\item The cycle period has just the same   sensitivity to the BL's
thickness as  the shell dynamo, \ie\  $\tau_{\rm cyc}$[yr]
$\sim D/\epsilon$ for  $D$ in Mm ($D \simeq$
15\dots35 Mm). 
\item Due to the rotational $\eta_{\rm T}$-quenching the linear model 
 yields the correct cycle time.  Generally the
22-year cycle can only exist for a dilution 
of $\epsilon\simeq 0.5$ (\cf Fig.~\ref{Fperiod}). 
\item  For $\alpha_{\rm u} \simeq 0$ the magnetic activity is
concentrated near the poles,  for $\alpha_{\rm u} \simeq 1$ it moves to
the equator.  
\item For too thin BL's there are too many
toroidal magnetic belts in each hemisphere (Fig.~\ref{Fbut3}). 
\end{itemize}
\vspace*{-0.3truecm}

The dependence of the cycle period on $\epsilon$ is shown in
Fig.~\ref{Fperiod}. 
For $\epsilon\approx1$ the linear solutions have the correct 22-year solar
cycle period (right panel). The cycle period increases linearly
with $D$,
\begin{equation} 
\tau_{\rm cyc} \propto D,
\label{period}
\end{equation}
(see left panel of Fig.~\ref{Fperiod}), which is in agreement with the linear
relation (\ref{8}) for spherical shell dynamos.
For too thin boundary layers the cycle time will thus become very short.
Nonlinearity further shortens the cycle period and this effect
has to be compensated for by taking a smaller value of
$\epsilon$, depending on whether or not magnetic buoyancy
is included.

In those cases where $\alpha$-quenching and magnetic buoyancy are included,
the rms-values of the poloidal and toroidal fields are 11 Gauss and 5 kGauss.
Poloidal and toroidal fields as well as their mean and
fluctuating parts are difficult to disentangle observationally.
A toroidal field of 5 kGauss is, however, consistent with the observed
total flux of $10^{24}$ Mx, distributed over 50\ Mm in depth and
400\ Mm in latitude. On the other hand, a poloidal field of 11 Gauss is 
larger than the observed value.

In all cases the magnetic field exceeds the equipartition value
which is here 6 kGauss.
This is partly due to the turbulent diamagnetism leading to an accumulation
of magnetic fields at the bottom of the convection zone, and partly due to
too optimistic estimates for $\alpha$-quenching.

The dilution factor $\epsilon$ is the main free parameter that must be
chosen such that the 22-year magnetic cycle period is obtained.
The thickness of the overshoot layer must be chosen to match
the correct number of toroidal field belts of the Sun.
Our results suggest that the thickness should not be much
smaller than $D \approx35\,$Mm$\,\approx 1/2 H_p$.

Nonlinear effects typically lead to a reduction of the magnetic cycle period.
Unfortunately, there are no consistent quenching expressions that are valid
for rapid rotation.
Magnetic buoyancy also provides a nonlinear feedback
(e.g. Durney and Robinson, 1982; Moss {\it et al.}, 1990), but
computations by Kitchatinov and Pipin (1993) indicate that this
only leads
to a mean field advection of a few m/s, i.e. less than the advection
due to the diamagnetic effect.
However, magnetic buoyancy is important in the upper layers
(where the diamagnetic effect is weak), and can have significant
effects on the field geometry and the cycle period
(Kitchatinov, 1993).

\section{Distributed  Dynamos with Meridional Circulation}

\begin{figure}
\psfig{figure=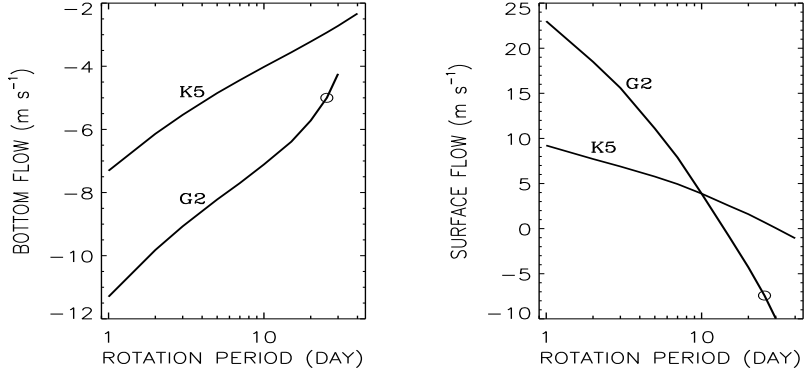,width=12.5cm,height=5.0cm,bbllx=0pt,bblly=368pt,bburx=489pt,bbury=628pt}
\caption{The meridional circulation at mid-latitudes
(45$^\circ$) at the bottom (LEFT) and the top (RIGHT) of the convection zone  for both
a solar model and a K5 main-sequence star as a function of
rotation period (\K\ and \R, 1999). The circle represents the
solar case. The flow at the bottom is equatorwards (as it is
the flow at the surface, there are 2 cells in radius). }   
\label{fdr2}
\end{figure}

Not only differential rotation but also meridional flow  $u^{\rm m}$ will
influence the mean-field dynamo. This influence can be expected
to be only a small modification if and only if its characteristic
time scale $\tau_{\rm drift}$ exceeds the cycle time $\tau_{\rm
cyc}$ of about 11 years. With $\tau_{\rm drift} \simeq R/u^{\rm
m}$ we find $u^{\rm m} 
\simeq$ 2 m/s as a critical value. If the flow is faster (Fig. \ref{fdr2}) the modification can be drastic.
In particular, if the flow counteracts the diffusion wave, i.e.
if the drift at the bottom of the convection zone is polewards,
the dynamo might come into trouble. The behaviour of such a
solar-type dynamo (with superrotation at the bottom of the
convection zone and perfect conduction beneath the convection
zone) is demonstrated in the present Section. 

In the following the dynamo equations are given with the
inclusion of the induction by meridional circulation. For
axisymmetry  the mean flow in spherical coordinates is given by 
${\bf u}  = \left( u_r, u_\theta, r \sin\theta\ \Omega\right)$.

\begin{figure}
\psfig{figure=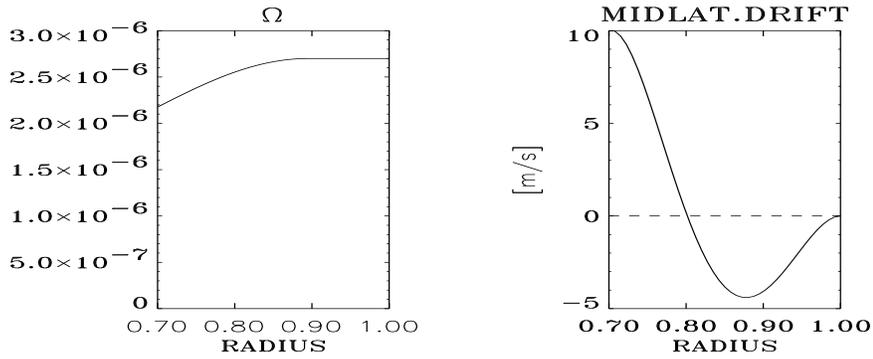,width=13.0cm,height=4.5cm,bbllx=10pt,bblly=370pt,bburx=558pt,bbury=700pt}
\caption[]{Rotation law and  the pattern of a meridional flow 
at 45$^\circ$
for the Steenbeck-Krause dynamo model of Section 4. The meridional circulation is
prescribed (\R, 1989) and {\it not} yet a result of
differential rotation theory. The rotation profile always has a
positive slope.  
}
\label{midlat}
\end{figure}

With a similar notation the  magnetic field is
\begin{equation}
{\bf B}  =  \left( {1\over r^2\sin\theta}{\partial A
	     \over\partial\theta},
-{1\over r \sin\theta}{\partial A\over\partial r}\ 
     ,\ B \right) 
\label{dynb}
\end{equation}
with  $A$ as the poloidal-field potential and $B$ as the toroidal field.
Their evolution is   described by 
\begin{eqnarray}
{\partial A \over \partial t} + ({\bf u}\cdot\nabla)A &=& \alpha s
B + \eta_{\rm T} \left({\partial^2 A \over \partial r^2} +
{\sin\theta \over r^2} {\partial \over \partial \theta}
\left({1\over \sin\theta} {\partial A \over \partial
\theta}\right)\right) ,
\label{dyna}\\
{\partial B \over \partial t} + s \rho ({\bf u}\cdot\nabla ) {B\over {s \rho}}
&=& {1\over r} \left({\partial \Omega \over \partial r}
{\partial A \over \partial \theta} - {\partial \Omega \over
\partial \theta} {\partial A \over \partial r}\right) - {1\over
s} {\partial \over \partial r} \left(\alpha {\partial A \over
\partial r}\right) - {1\over r^3} {\partial \over \partial
\theta} \left({\alpha \over \sin\theta} {\partial A \over
\partial \theta}\right)+ \nonumber\\
&+& {\eta_{\rm T} \over s} \left({\partial^2(sB) \over \partial
r^2} + {\sin\theta \over r} {\partial \over \partial \theta}
\left({1\over s} {\partial (sB) \over \partial \theta}\right)\right)
\end{eqnarray}
for uniform $\eta_{\rm T}$ and with $s=r \sin\theta$ (cf.
Choudhuri {\it et al.}, 1995).  
In order to
produce here a solar-type butterfly diagram with ${\bf u} = 0$
the $\alpha$-effect must be
taken as negative in the northern hemisphere 
(Steenbeck and Krause, 1969; Parker, 1987). Our rotation law and
the radial meridional flow profile at 45$^\circ$ are given in
Fig.~\ref{midlat}. A one-cell 
flow pattern is adopted in each hemisphere. The $u^{\rm m}$ approximates the
latitudinal drift close to
the convection zone 
bottom, the circulation is counterclockwise (in the first quadrant)
for positive $u^{\rm m}$, i.e. towards the equator at the
bottom of the convection zone. 
Note that the cycle period becomes shorter
for clockwise flow (Roberts and Stix, 1972)  and it becomes
longer for the more realistic  counterclockwise flow. The
latter forms a poleward flow at
the surface in agreement with observations and theory
(Fig.~\ref{fdr2}). For too strong
meridional circulation  ($u^{\rm m}\geq 50$ m/s) our dynamo
stops operation. The same happens already   for $u^{\rm m}\approx -$ 
10 m/s if the flow at the bottom is opposite to the magnetic drift wave.

\begin{figure*}
\psfig{figure=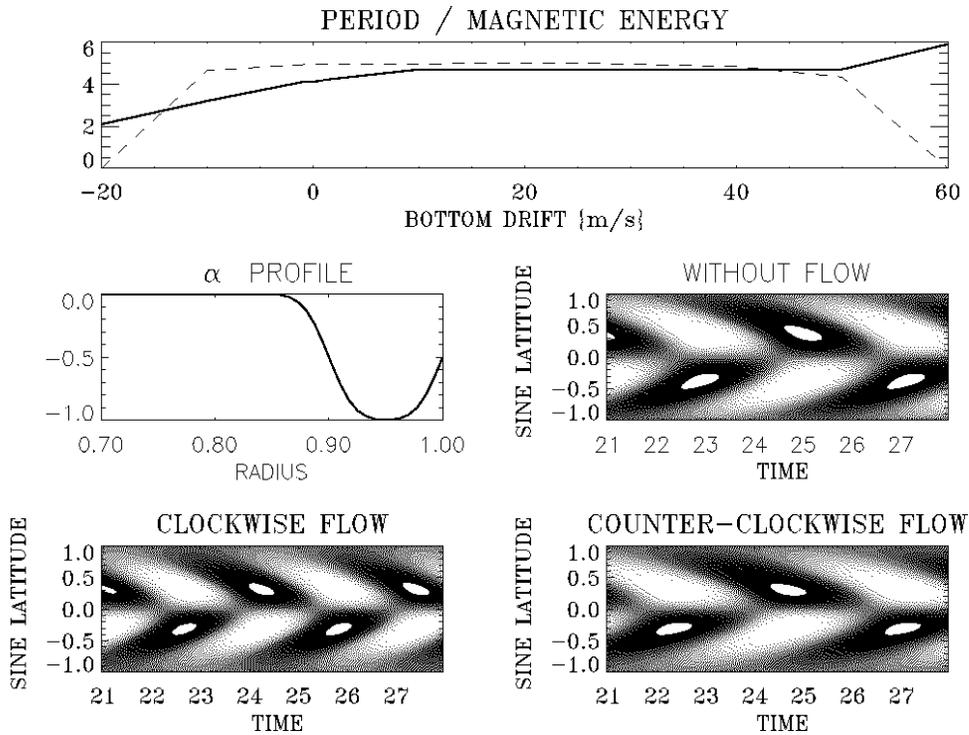,width=14.0cm,height=10.0cm,bbllx=30pt,bblly=250pt,bburx=558pt,bbury=610pt}
\caption[]{
Operation of a Steenbeck-Krause dynamo with {\it negative}
alpha-effect (in northern hemisphere) with and without meridional circulation.
Eddy diffusivity is $5\cdot 10^{12}\ {\rm cm}^2/{\rm s}$.
Positive bottom drift means counterclockwise flow
(equatorwards at the bottom of the convection zone, polewards
at the surface).
TOP: The cycle period (solid, in years) and the magnetic energy
(dashed) vs. the latitudinal drift at the bottom of the
convection zone. MIDDLE: $\alpha$-profile in radius and
butterfly diagram of the circulation-free dynamo. The maximal
toroidal field is plotted vs. the time in years. BOTTOM:
Butterfly diagrams for clockwise and counterclockwise
circulations. 
}\label{drift1}
\end{figure*}

\begin{figure*}[t]
\psfig{figure=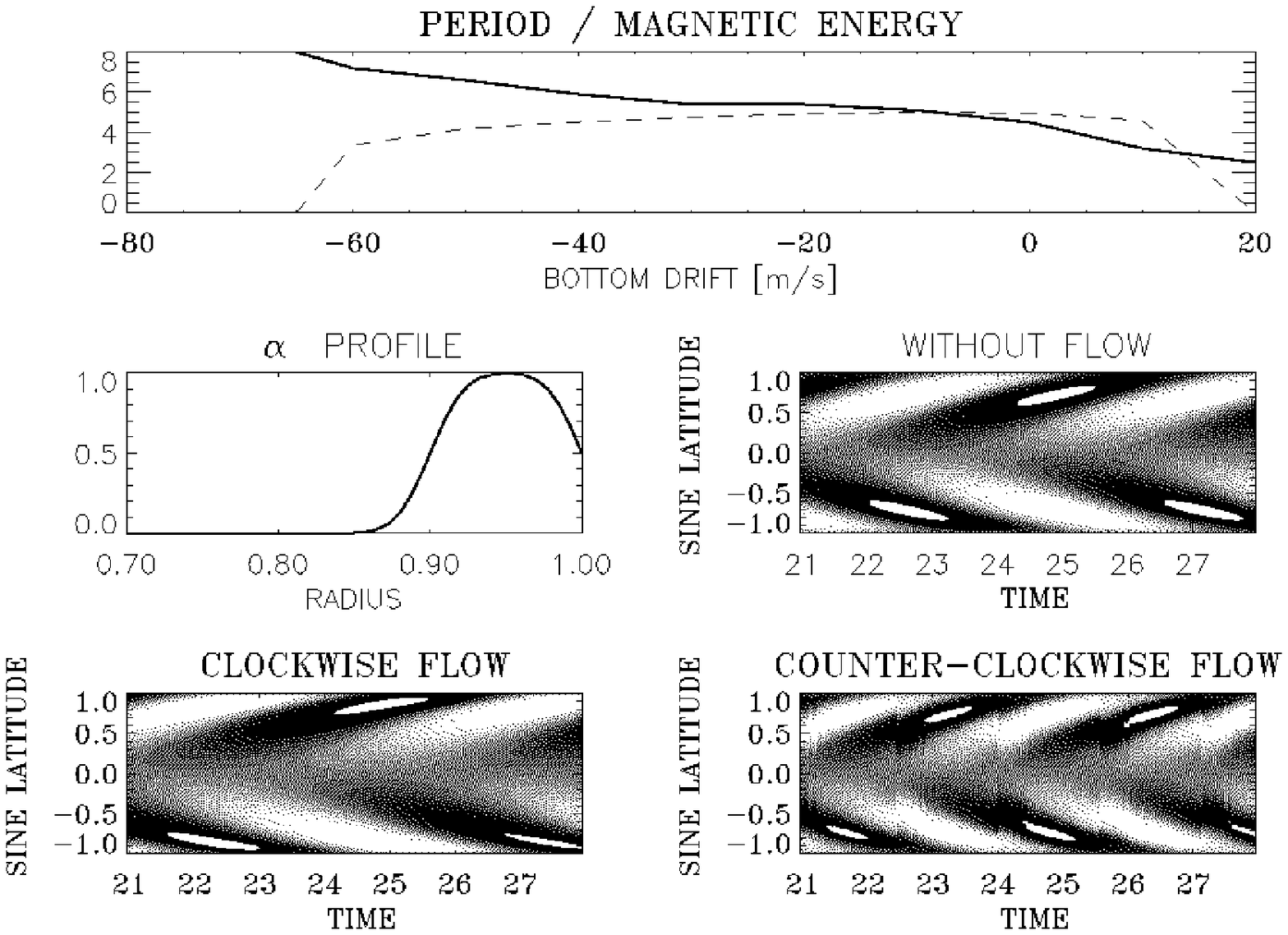,width=14.0cm,height=10.0cm,bbllx=30pt,bblly=250pt,bburx=558pt,bbury=610pt}
\caption[]{
The same as in Fig.~\ref{drift1} but for {\rm  positive} 
$\alpha$-effect (in northern hemisphere). Note the decay of the 
dynamo already for 10\dots20 m/s (equatorwards at the bottom, 
polewards at the top of the convection zone).}
\label{drift2}
\end{figure*}

However, it is hard to explain why the convection zone
itself is producing the {\it negative}
\alf-values used in the model.  More easy to understand are 
the positive values adopted in Fig.~\ref{drift2} (see Durney 1995, 1996). Of course, now
without circulation
the butterfly diagram becomes `wrong', with a poleward migration
of the toroidal magnetic belts. One can also understand that
the counterclockwise flow (equatorwards at the convection zone
bottom) rapidly destroys the dynamo operation, here already for 10 m/s
amplitude. This amplitude is so small that it cannot be a
surprise that such a dynamo might not be realized in the Sun. 
For the eddy diffusivity $5\cdot
10^{12}$ cm$^2$/s is used here, so the magnetic Reynolds
number ${\rm Rm}= u^{\rm m} R/\eta_{\rm T}$ is 
\beg
{\rm Rm} = 14 \ {u^{\rm m} \over 10\ {\rm m/s}},
\label{Rm}
\ende
i.e. ${\rm Rm}=14$ for $u^{\rm m} = 10$ m/s. In their paper Choudhuri \ea (1995) apply Reynolds numbers of order 500 while
Dikpati and Charbonneau (1999) even take ${\rm Rm}= 1400$. Indeed, their
dynamos are far beyond the critical values of
${\rm Rm}\simeq 10$ which are allowed for the weak-circulation
$\alpha\Omega$-dynamo. Their dynamos exhibit the same period
reduction for increasing $u^{\rm m}$ shown in Fig.~\ref{drift2} but
the butterfly diagram is opposite. It seems that quite another
branch of dynamos would exist for ${\rm Rm}\gg 10$. This is
indeed true.

\begin{figure}
\psfig{figure=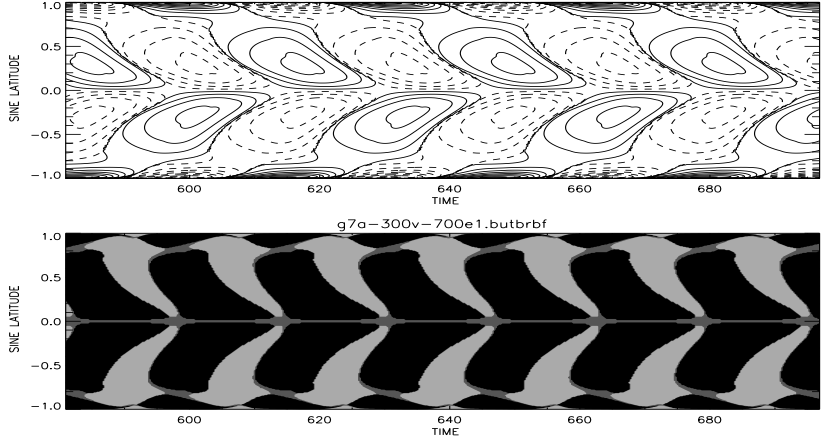,width=13.0cm,height=6.0cm,angle=90,bbllx=54pt,bblly=54pt,bburx=548pt,bbury=838pt}
\caption[]{
TOP: Butterfly diagram of a marginal dynamo with {\it positive}
\alf-effect (3 m/s), with eddy diffusivity of $10^{11}$
cm$^2$/s and a bottom flow of 7 m/s. BOTTOM: $\langle
B_r\rangle \langle B_\phi\rangle$, black colour indicates
negative sign, i.e. the observed phase lay between poloidal and
toroidal field.}
\label{drift3}
\end{figure}

\begin{table}[H]
\caption[]{Cycle periods in years of solar-type-butterfly
dynamos (bold) with {\it positive} \alf-effect (= 3 m/s) and meridional flow
of $u^{\rm m}$  at the bottom of the convection zone, the eddy
diffusivity given in the  headline. Signature `$-$' indicates
polewards migration, `decay'  means decaying fields. Note the
good representation of the $\eta$-dependence of
the cycle time of (\ref{8}) in the first line of the Table.}  
\begin{center}
\begin{tabular}{|r|lll|}\hline
{\rule[-0mm]{0mm}{5mm}
}  $u^{\rm m}$ [m/s] & $1\cdot 10^{11}$ cm$^2$/s & $2\cdot
10^{11}$ cm$^2$/s& $5\cdot 10^{11}$ cm$^2$/s\\[0.5ex] 
\hline
{\rule[-2mm]{0mm}{7mm}
} 1~~~~ &\ \ \ $-$120 &\ \ \ $-$70 & ~\,$-$36 \\
3~~~~ &\ \ \ \phantom{$-$1}{\bf 60} &\ \ \ $-$45 & ~\,$-$24\\
5~~~~ &\ \ \ \phantom{$-$1}{\bf 45} &\ \ \ \phantom{$-$}{\bf 40} & decay\\  
7~~~~ &\ \ \ \phantom{$-$1}{\bf 31} &\ \ \ \phantom{$-$}{\bf 26} & decay\\
11~~~~ &\ \ \ \phantom{$-$1}{\bf 20} &\ \ \ \phantom{$-$}{\bf 20} & decay\\[0.9ex]
\hline
\end{tabular}
\end{center}
\label{periods}
\end{table}
The butterfly diagram of such a dynamo with positive 
\alf-effect (3 m/s) and a magnetic Reynolds number of 450  is
given in  Fig.~\ref{drift3}. It works with low eddy
diffusivity of  $10^{11}\ {\rm cm}^2/{\rm s}$ and the flow
pattern of Fig.~\ref{midlat}. In the lower panel the phase
relation of poloidal and toroidal magnetic fields is shown. In
lower latitudes the observed phase lay between the field
components is obtained. In Table~\ref{periods} the cycle periods 
in years are given for positive-\alf\ dynamos. Bold numbers indicate 
the solar-type butterfly diagram. For 
fast meridional flow and small eddy diffusivity we indeed find 
oscillating dynamos. The eddy diffusivity is the same as that known 
from the sunspot decay, i.e. $10^{11}$ cm$^2$/s (see Brandenburg, 1993). 

\section{The Maunder Minimum}
The period of the solar cycle and its amplitude are far from
constant. The most prominent activity drop was the Maunder
minimum between 1670 and 1715 (Sp\"orer, 1887). The variability 
of the cycle period can be expressed by the `quality', 
$\omega_{\rm cyc}/\Delta\omega_{\rm cyc}$, which is as low as 
5 for the Sun (Hoyng, 1993, cf. Fig.~\ref{quality}).

Measurements of $^{14}$C abundances in sediments and long-lived 
trees provide much longer time series than sunspot counts.
In agreement with Schwarz (1994), Vos \ea (1996, 1997) found
a secular periodicity of 80--90~years as well as a long-duration 
period of about 210~years. The measurements of atmospheric  
$^{14}$C abundances by Hood and Jirikowic (1990) suggested a periodicity of
2400~years which is also associated with a long-term variation of solar
activity. The variety  of frequencies found in solar activity may even
indicate chaotic behaviour as discussed by Rozelot (1995),
Kurths \ea (1993, 1997) and Knobloch and Landsberg (1996).

The activity cycle of the Sun is not exceptional: The
observation of chromospheric Ca-emission of solar-type stars yields
activity periods between 3 and 20~years (Noyes {\it et al.}, 1984;
Baliunas and Vaughan, 1985; Saar and Baliunas, 1992a,b). A few
of these stars do not show any 
significant activity. This suggests that 
even the existence of the grand minima is a typical property of cool main-sequence
stars like the Sun. From ROSAT X-ray data Hempelmann \ea
(1996) find that up to  70\% of the stars with a constant level of activity
exhibit a rather low level of coronal X-ray emission. HD 142373 with
its X-ray luminosity of only $\log F_{\rm X} = 3.8$ is a typical candidate.
We conclude that during a grand minimum not only the magnetic field
in the activity belts is weaker than usual but the total
magnetic field energy is also reduced.

The short-term cycle period appears to decrease at the end of
a grand minimum according to a wavelet analysis of sunspot
data by Frick \ea (1997). There is even empirical evidence
for a very weak but persistent cycle throughout the solar Maunder minimum
as found by Wittmann (1978) and recently Beer \ea (1998).
The latitudinal distribution of the few sunspots observed during 
the Maunder minimum was highly asymmetric (Sp\"orer, 1887;
Ribes and Nesme-Ribes, 1993; Nesme-Ribes {\it et al.}, 1994). Short-term deviations from the 
north-south symmetry in regular solar activity are readily
observable (Verma, 1993), yet a 30-year period of asymmetry in
sunspot positions as seen during 
the Maunder minimum remains a unique property of
grand minima and should be associated with a parity change of the 
internal magnetic fields.

\subsection{Mean-field Magnetohydrodynamics}
The explanation of grand minima in the magnetic activity cycle by a 
dynamo action has been approached by two concepts. The first one 
considers the stochastic character of the turbulence and studies 
its consequences for the variations of the \alf-effect and all 
related phenomena with time (see Section~\ref{acht}). The alternative concept 
includes the magnetic
feedback to the internal solar rotation (Weiss {\it et al.}, 1984;
Jennings and Weiss, 1991). Kitchatinov \ea (1994a) and Tobias
(1996, 1997) even introduced the conservation law of angular
momentum in the convection zone including magnetic
feedback in order to simulate the intermittency of the dynamo cycle.

A theory of differential rotation based on the \L-effect concept
is coupled with the induction equation in a spherical 2D 
mean-field model. 
The mean-field equations for the
convection zone  include the effects of diffusion, \alf-effect,
toroidal field production by differential rotation and the Lorentz
force. They are
\beg
{\partial A\over \partial t}  =  \eta_{\rm T}{\partial^2
A\over\partial r^2}+\eta_{\rm T}{\sin\theta\over
r^2}{\partial\over\partial\theta}
\left({1\over\sin\theta}{\partial A\over\partial\theta}\right) 
+ \alpha r\sin\theta B,
\label{4.4}
\ende

\begin{eqnarray}
{\partial B\over\partial t}  &=&  {1\over r}{\partial\over\partial r}
\left(\eta_{\rm T}{\partial(B r)\over\partial r}\right) +
{\eta_{\rm T}\over r^2}{\partial\over\partial\theta}\bigg({1\over\sin\theta}
{\partial(B\sin\theta)\over\partial\theta}\bigg)
 +\nonumber\\
&  +&  {1\over r} {\partial\Omega\over\partial r}
{\partial A\over\partial\theta} -{1\over r}{\partial\Omega\over\partial\theta}
{\partial A\over\partial r} -
{1\over r \sin\theta}\ {\partial \over \partial r} \left(\alpha
\ {\partial A \over  \partial r}\right) -
 {1\over r^3}\
{\partial \over \partial \theta} \left({\alpha \over
\sin\theta} \ {\partial A \over \partial \theta}\right) ,
\label{3.3}
\end{eqnarray}
\begin{eqnarray}
\lefteqn{\rho r\sin\theta {\partial \Omega\over\partial t}  =
  -{1\over r^3}{\partial\over\partial r}
  \left(r^3\rho\ Q_{r\phi}\right)
 -  {1\over r\sin^2\theta}{\partial\over\partial\theta}
  \left(\sin^2\theta \rho\ Q_{\theta\phi}\right) +}\nonumber\\
&& + {1\over \mu_0 r^2\sin\theta}\bigg({1\over r}{\partial
A\over\partial\theta} {\partial(B r)\over\partial r}
 - {1\over\sin\theta} {\partial A\over\partial r}
{\partial(B\sin\theta)\over\partial\theta}\bigg) .
\label{2.2}
\end{eqnarray}
Since the importance of the large-scale Lorentz force in the momentum 
equation was discussed by Malkus and Proctor (1975), we call this term
`Malkus-Proctor' effect.

The computational domain is a spherical shell covering the outer 
parts of the Sun down to a fractional solar radius, $x=r/R$, of $0.5$. 
The convection zone extends from $x=0.7$ to $x=1$. The 
\alf-effect works only in the lower part 
from $x=0.7$ to $x=0.8$ while turbulent diffusion of the magnetic field,
turbulent viscosity, and the \L-effect are present in the entire
convection zone. Below $x=0.7$ both the magnetic diffusivity
and the viscosity are two orders of magnitude smaller than in the
convection zone.  
The boundary conditions are specified as
$Q_{r\phi} = \partial A/\partial r = B = 0 \q
{\rm at}\ \ r = R$,
and
$Q_{r\phi} = A = B = 0$ at the inner boundary.
The  angular momentum flux is given by 
\beg
Q_{r\phi} = \nu_{\rm T} \ \sin\theta\left(-r \ {\partial
\Omega \over \partial r} + V^{(0)} \Omega\right), \q
Q_{\theta\phi} = - \nu_{\rm T} \ \sin\theta \ {\partial
\Omega \over \partial \theta} .
\label{6.1}
\ende
$V^{(0)}$ determines the radial rotation law without magnetic field.

Five dimensionless numbers define the model, namely the magnetic 
Reynolds numbers of both the differential rotation and the \alf-effect,
$C_\Omega = \Omega_0 R^2/\eta_{\rm T}$, 
$C_\alpha = \alpha_0 R/\eta_{\rm T}$;
the magnetic Prandtl number
${\rm Pm} = \nu_{\rm T}/\eta_{\rm T}$;
the Elsasser number
\beg
{\rm E} = {B_{\rm eq}^2
\over {\mu_0 \rho \eta_{\rm T} \Omega_0}};
\label{13}
\ende
and the  $\Lambda$-effect amplitude, $V^{(0)}$. In the $\alpha$-effect,
\beg
\alpha = \alpha_0 \ \cos\theta \sin^2\theta,
\label{17}
\ende
the factor $\sin^2\theta$ has been introduced to
restrict magnetic activity to low latitudes 
and $\alpha_0 \simeq l_{\rm corr}\ \Omega_0$, so that
$C_\Omega/|C_\alpha| \simeq  R/l_{\rm corr}$,
hence in general $C_\Omega$ exceeds $C_\alpha$ (`\alf\Om\
dynamo'). Our dynamo works with $C_\alpha = -10$ and $C_\Omega
= 10^5$. $V^{(0)}$ is positive in order to produce the
required super-rotation, and its amplitude is 0.37.
With the eddy diffusivity (\ref{20}) the Elsasser number 
reads ${\rm E} = 2/c_\eta \Omega^*$ and is set to unity here. 
(See also K\"uker {\it et al.}, 1999; Pipin, 1999.)
\begin{figure}
\mbox{\psfig{figure=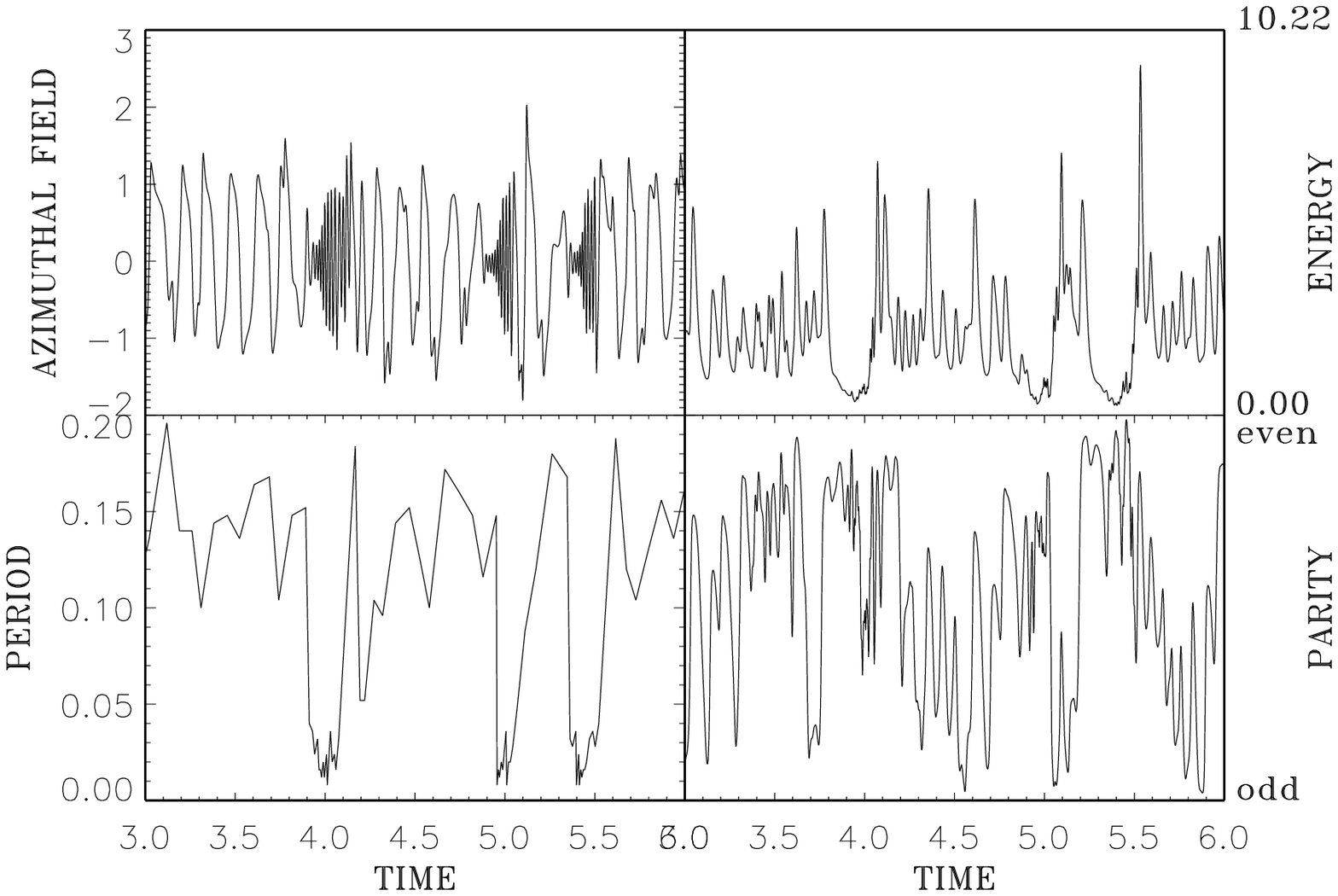,height=6truecm,width=7.2truecm,bbllx=34pt,bblly=360pt,bburx=558pt,bbury=720pt}
\hfill
\psfig{figure=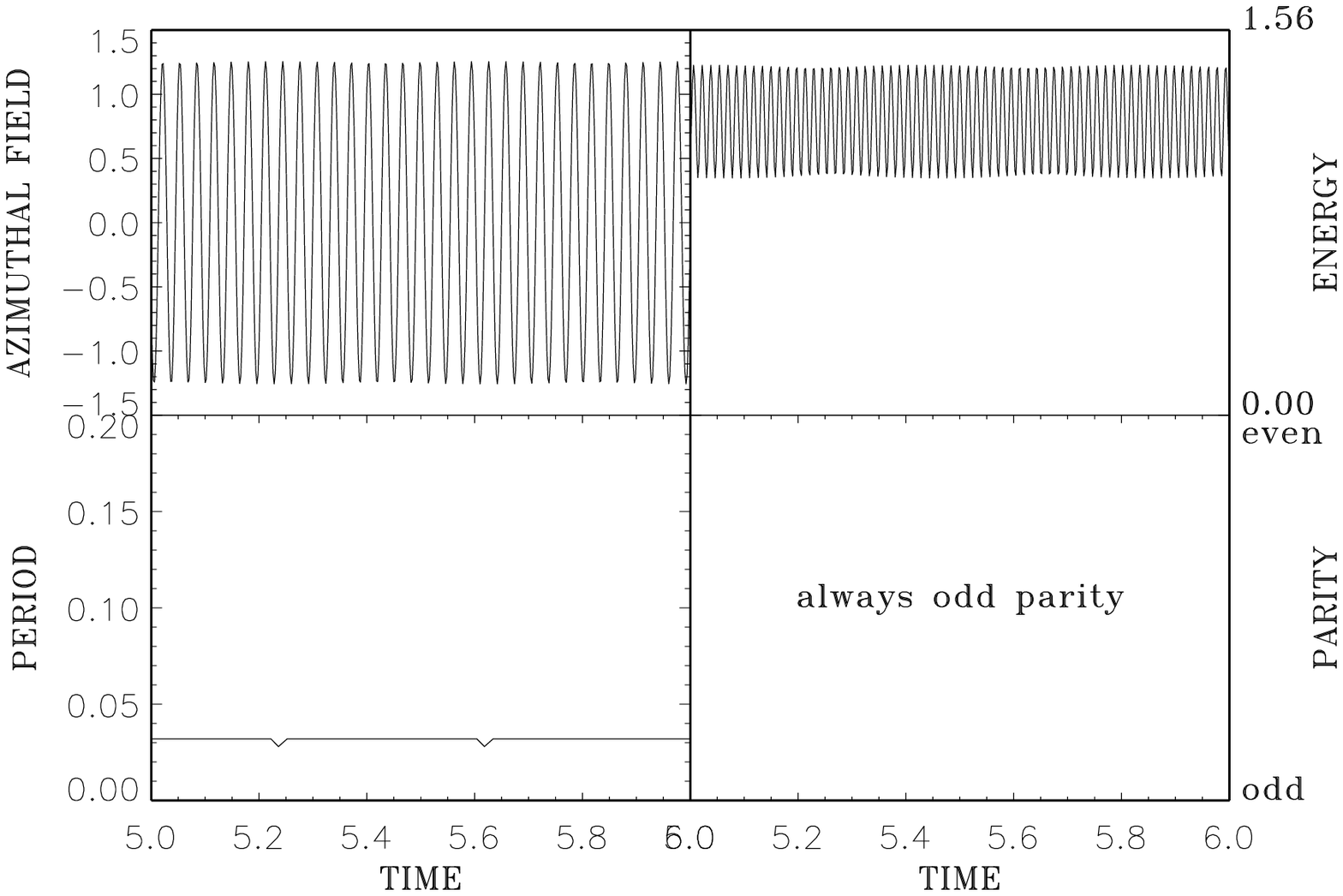,height=6truecm,width=7.4truecm,bbllx=0pt,bblly=360pt,bburx=558pt,bbury=720pt}}
\caption{LEFT: The time dependence of the dynamo for the large-scale
Lorentz-force feedback only (${\rm Pm}=0.1$, $E=1$). Top: Toroidal magnetic
field (left) and magnetic energy (right). Bottom: Cycle time (left)
and magnetic parity (right). RIGHT: The same with $\alpha$-quenching 
but still without $\Lambda$-quenching ($\lambda=0$).} 
\label{f1}
\end{figure}
 \begin{figure}
\psfig{figure=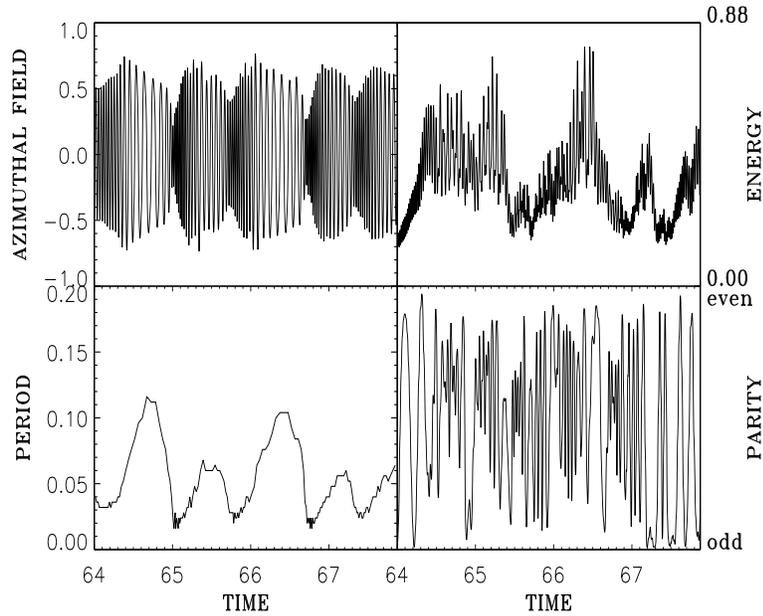,height=8truecm,width=12.5truecm,bbllx=-70pt,bblly=360pt,bburx=558pt,bbury=690pt}
\caption{The same as in Fig.~\ref{f1} but for   strong
$\Lambda$-quenching ($\lambda=25$). 
TOP: Toroidal magnetic field (left) and magnetic energy (right). BOTTOM:
Cycle time (left) and magnetic parity (right).} 
\label{f4}
\end{figure}

\subsection{Results of the Interplay of Dynamo and Angular Velocity}
Figs.~\ref{f1} and \ref{f4} demonstrate the action of
different effects and show the variation of the toroidal magnetic 
field at a fixed point ($x=0.75$, $\theta=30^\circ$), the total
magnetic energy, the variation of the 
cycle period, and the parity 
$	P=(E_{\rm S}-E_{\rm A})/(E_{\rm S}+E_{\rm A}),$ 
derived from the decomposition of the magnetic energy into symmetric
and antisymmetric components (Brandenburg {\it et al.}, 1989). All times
and periods are given in units of a diffusion time $R^2/\eta_{\rm T}$.
Field strengths are measured in units of $B_{\rm eq}$.

If the large-scale Lorentz force (`Malkus-Proctor effect') 
is the only feedback on rotation, 
the time series given in Fig.~\ref{f1} may be compared with 
the results in Tobias (1996) although a number of assumptions 
are different between Tobias' Cartesian approach and our 
spherical model. Fig.~\ref{f1} shows a quasi-periodic behaviour with
activity interruptions like grand minima. This model, however, neglects
the feedback of strong magnetic fields on the $\alpha$-effect
and the differential rotation.

The right-hand side of Fig.~\ref{f1} shows the result of the same
model but with a local $\alpha$-quenching,
\beg
\alpha \propto \frac{1}{1+(B_{\rm tot}/B_{\rm eq})^2},
\ende
where $B_{\rm tot}$ is the absolute value of the magnetic field. 
The variability of the cycles turns into a solution with only one 
period. Similar to the suppression of dynamo action, a quenching of the
$\Lambda$-effect causing the differential rotation is according
to
\beg
V^{(0)} \propto \frac{1}{1+\lambda (B_{\rm tot}/B_{\rm eq})^2}.
\ende
If $\lambda$ is near unity, the maximum field strength and total 
magnetic energy decrease slightly, but the periodic behaviour
remains the same, i.e.\ the effect of the $\Lambda$-quenching is too
small to alter the differential rotation significantly. However,
{\it an increase of $\lambda$ leads to grand minima\/} -- an example 
for $\lambda=25$ is given in Fig.~\ref{f4}. Minima in cycle period 
occur {\it shortly after\/} a grand activity minimum in agreement 
with the analysis of sunspot data by Frick {\it et al.} (1997a,b).
The amplitude of the period fluctuations in Fig.~\ref{f4}
is much lower than in the Malkus-Proctor model but is still 
stronger than that observed. The magnetic Prandtl number used for
the solutions in Figs.~\ref{f1} and \ref{f4} was ${\rm Pm}=0.1$ 
and it is noteworthy that grand minima do {\em not appear} for
${\rm Pm} = 1$. 

The strong variations of the parity between symmetric and 
antisymmetric states appearing in the nonperiodic solutions 
can be explained by the numerous (usually $5\dots6$) magnetic field 
belts migrating towards the equator. Slight shifts of this 
belt-structure against the equator result in strong variations 
in the parity. Averaged over time, dipolar and quadrupolar 
components of the fields have roughly the same strength;
all periodic solutions have strict dipolar structure.

\begin{figure}
\mbox{\psfig{figure=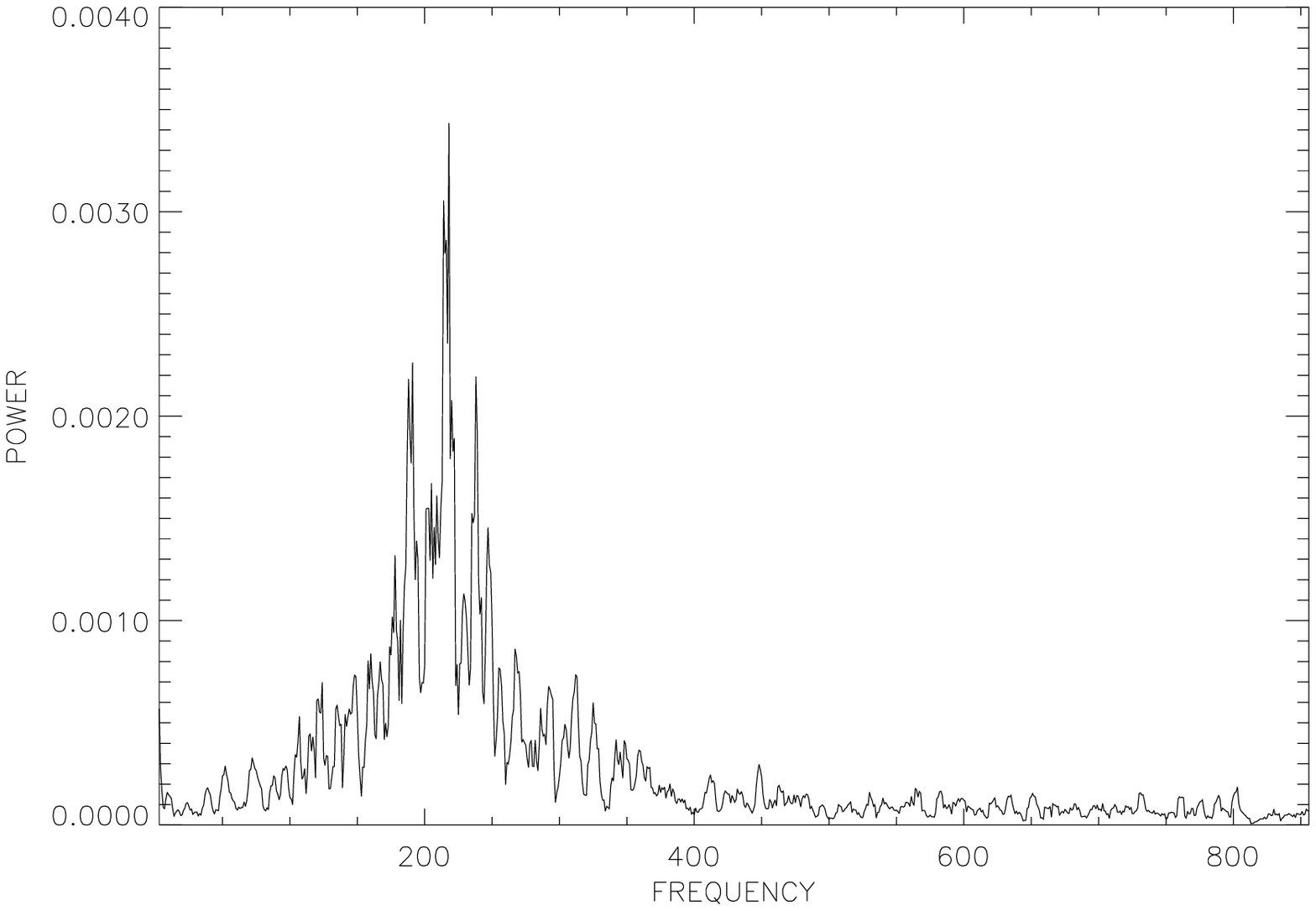,height=6truecm,width=7.2truecm,bbllx=34pt,bblly=360pt,bburx=558pt,bbury=720pt}
\hfill
\psfig{figure=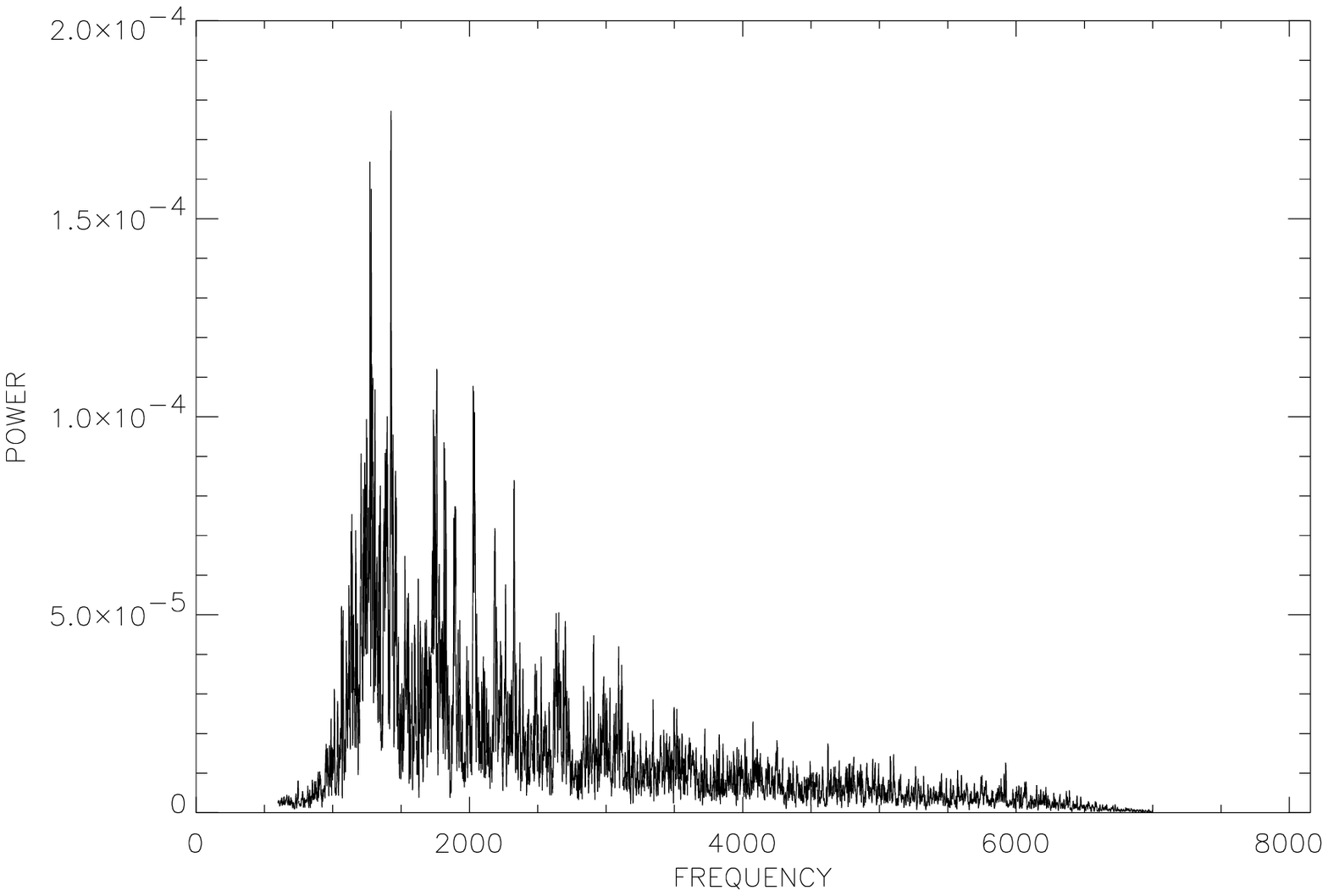,height=6truecm,width=7.4truecm,bbllx=0pt,bblly=360pt,bburx=558pt,bbury=720pt}}
\caption{LEFT: Power spectrum of the magnetic-field amplitude variations
for the Malkus-Proctor model of Fig.~\ref{f1}. The frequency is given
in arbitrary units. RIGHT: The same  but for the model with strong
$\Lambda$-quenching. The highest peaks are the main cycle frequency,
whereas the difference between two peak neighbours indicates the
secular cycle.} 
\label{f5}
\end{figure}

Spectra of long time series of the toroidal magnetic field 
are given in Fig.~\ref{f5} for both the Malkus-Proctor model 
and the strong $\Lambda$-quenching model.
%
%
The long-term variations of the field will be represented by a set of close
frequencies whose difference is the frequency of the grand minima. The
Malkus-Proctor model shows a number of lines close to the main cycle
frequency. The difference between the two highest peaks can be interpreted
as the occurrence rate of grand minima. However, the shape of the 
spectrum indicates that the magnetic field appears rather irregularly.
The spectrum of the model with
all feedback terms and strong \L-quenching is also given in Fig.~\ref{f5} 
and shows a similar behaviour with highest amplitudes near the main cycle
frequency of the magnetic field. The average frequency of the grand minima
is represented by the distance between the two highest peaks.

\begin{figure}
\mbox{\psfig{figure=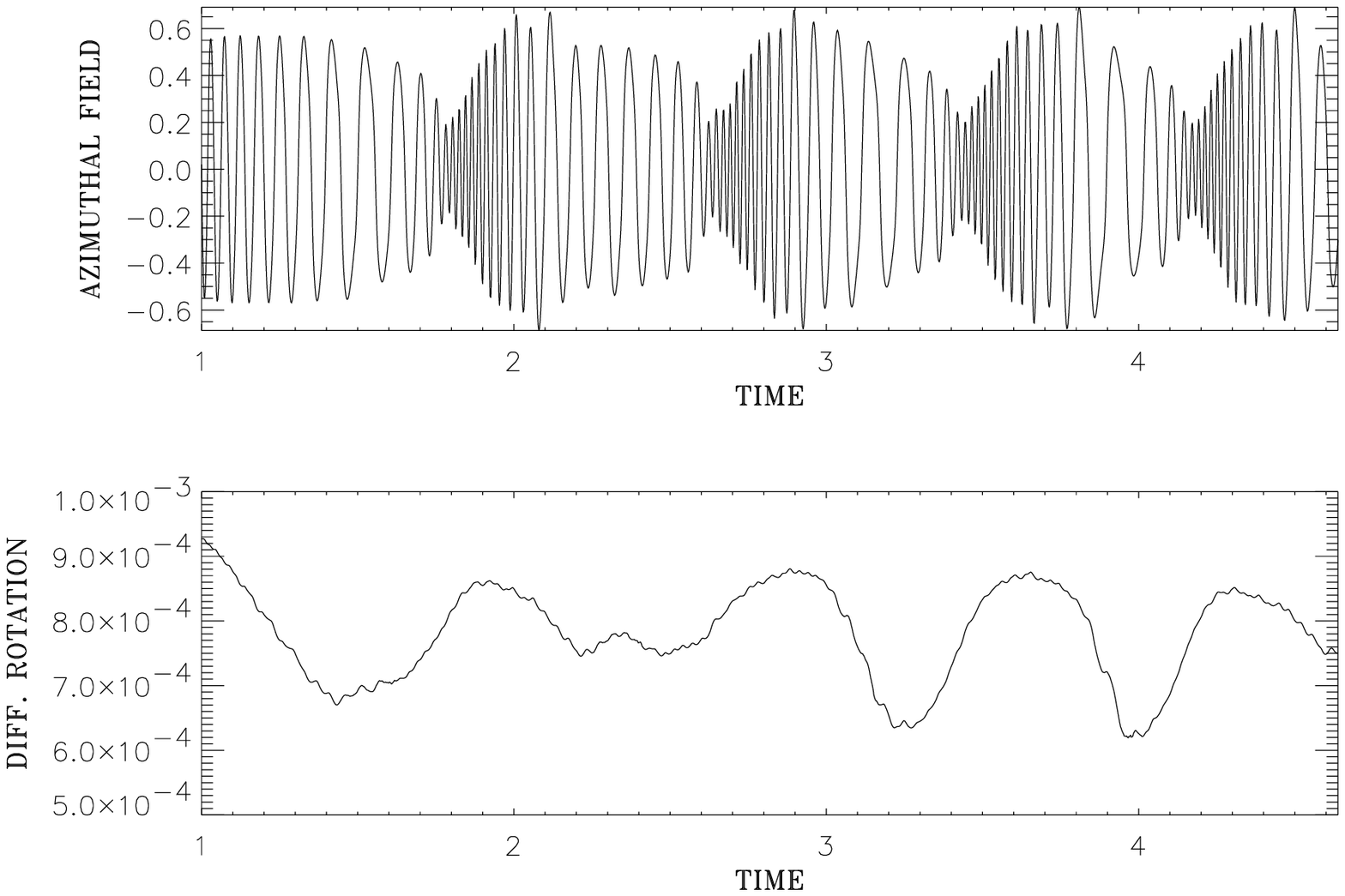,height=6truecm,width=7.2truecm,bbllx=34pt,bblly=360pt,bburx=558pt,bbury=670pt}\hfill
\psfig{figure=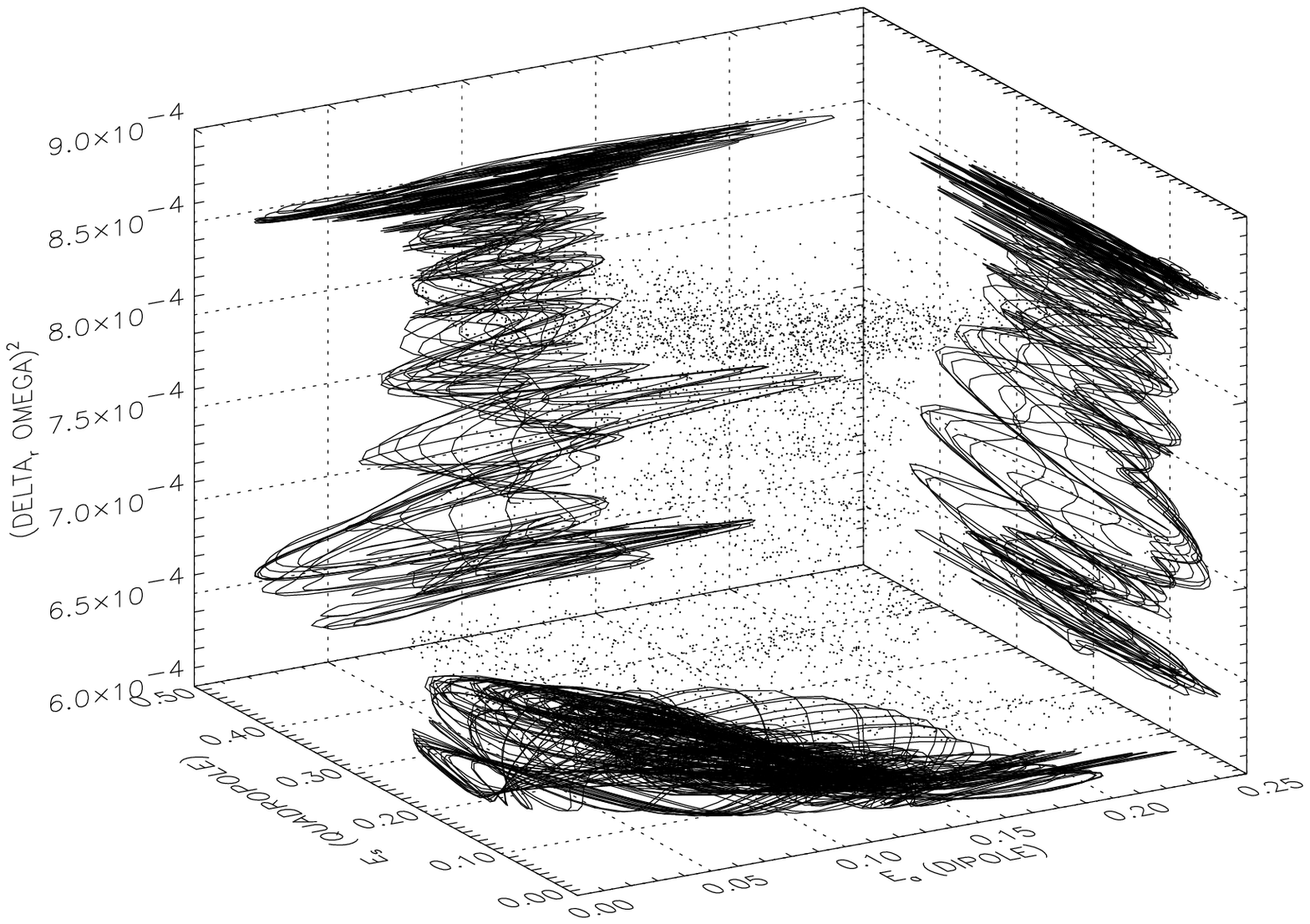,height=6truecm,width=8.0truecm,bbllx=-20pt,bblly=360pt,bburx=558pt,bbury=680pt}}
\caption{LEFT: Correlation between magnetic field oscillations and variations
of the differential rotation measure $(\partial\Omega/\partial r)^2$,
averaged over the latitude $\theta$. RIGHT: Correlations
between magnetic field energies for both parities 
(dipolar and quadrupolar energy component) and the differential rotation.
For most of the time the orbit resides in the state of regular cycles in 
the upper part of the box. Differential rotation is
suppressed during grand minima.}
\label{f7}
\end{figure}
\begin{figure}
\mbox{
\hfill
\psfig{figure=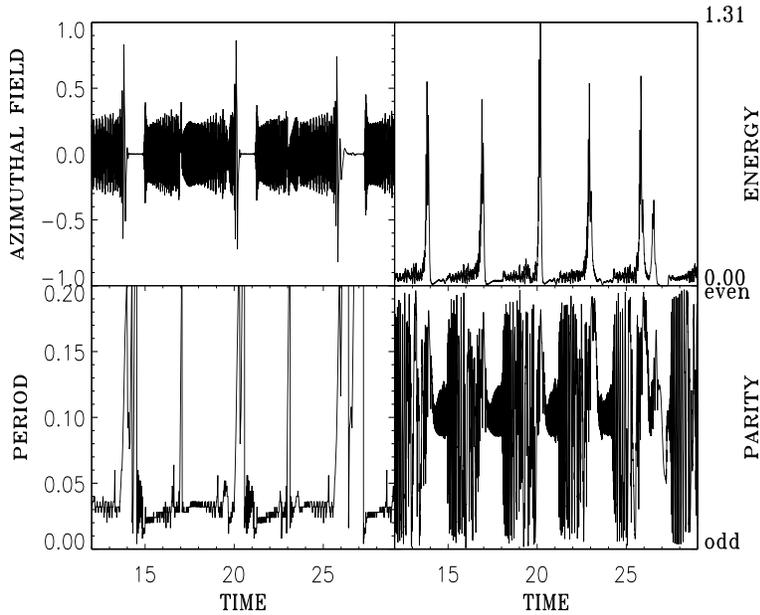,height=8truecm,width=12.5truecm,bbllx=-70pt,bblly=360pt,bburx=558pt,bbury=690pt}}
\caption{The same as in Fig.~\ref{f4} but for ${\rm Pm}=0.01$
showing much rarer occurrences of grand minima than for ${\rm Pm}=0.1$.}
\label{f8}
\end{figure}

The interplay of magnetic fields and differential rotation is
demonstrated in Fig.~\ref{f7} showing the variation of the 
radial rotational shear averaged over latitude,
$\langle(\partial\Omega/\partial r)^2\rangle$, versus time, 
compared with the toroidal field. A minimum in differential rotation is
accompanied by a decay of the magnetic field and followed by
a grand minimum. The differential rotation is being restored
during the grand minimum since the suppressing effect of
magnetic $\Lambda$-quenching is reduced. The behaviour of the
system can be evaluated in a phase diagram of the dipolar 
component $E_{\rm A}$ of the magnetic field energy, the quadrupolar 
component $E_{\rm S}$, and the mean angular velocity gradient, 
as also shown in Fig.~\ref{f7}. The dots in the interior of 
the diagram represent the actual time series; projections of 
the trajectory are given at the sides. 

The trajectory resides at strong differential rotation during
normal cyclic activity. The magnetic field oscillates in a 
wide range of energies. The trajectory moves down as the field
starts to suppress the differential rotation, with oscillation amplitudes
diminishing. The actual grand minimum is expressed by a loop
at very low dipolar energies; the quadrupolar component, however,
remains present throughout the minimum. The differential-rotation measure 
is already growing at that time.
At first glance, the phase graph may be associated with Type 1
modulation as classified in Knobloch \ea  (1998).

The effect of large-scale Lorentz forces on the differential
rotation as the only feedback of strong magnetic fields produces 
irregular grand minima with strong variations in cycle period. 
The complex time series turns into a single-period solution, if 
the suppression of dynamo action ($\alpha$-quenching) is included. 
If a strong feedback of small-scale flows on the generation of Reynolds 
stress (\L-quenching) is added, grand minima occur at a reasonable rate 
between 10 and 20 cycle times. The cycle period varies by a
factor of 3 or 4. 
The northern and southern hemispheres differ slightly in their
temporal behaviour. {\em This is a general characteristic of mixed-mode
dynamo explanations of grand minima.}  

The magnetic Prandtl number directs the intermittency of the activity 
cycle. Values smaller than unity are required for the existence of
grand minima, whereas the occurrence of grand minima again becomes
more and more exceptional for very small values of Pm (Fig.~\ref{f8}).

\section{Stellar Cycles}
The cycle period may be considered as a main test of stellar dynamo theory
because it reflects an essential property of the dynamo mechanism.
A realistic solar dynamo model should provide the correct 22-year cycle
period and, in the case of stellar cycles, the observed dependence
of the cycle period on the rotation period.

The following Sections will deal with a number of applications of
1D models illuminating some effects of more complicated
EMF. This will include the additional consideration of the
magnetic-diffusivity quenching, a dynamo-induced
$\alpha$-function, and the influence of temporal fluctuations of $\alpha$ and 
$\eta_{\rm T}$.

The relations between stellar parameters and the amplitude and
duration of magnetic activity cycles are fundamental for our
knowledge of stellar physics. Observational results cover a variety 
of findings which depend on how stars are grouped and how unknown 
parameters must be chosen. Generally, relations for the magnetic field 
and the cycle frequency are expressed by
\beg
\bar{B}_{\rm max} \propto \Omega^{*m}, \qq \omega_{\rm cyc}
\propto \Omega^{*n},
\label{0.1}
\ende
with different $m$ and $n$. The value 
discussed by Noyes \ea  (1984) was $n=1.28$.  Saar (1996)
derives a
very small $m$ (his Fig.~3), while Baliunas \ea (1996) find
$n\simeq 0.47$ for young stars and $n\simeq 1.97$ for old
stars. Brandenburg \ea (1998) derive $n_{\rm I}=1.46$ and
$n_{\rm A}=1.48$ for inactive (I) and active (A) stars. 
Saar and Brandenburg (1999) find 
$n=0.6$ for the young (``super-active'') stars and $n=1.5$ for the old ones. 
 Higher values are 
reported by Ossendrijver (1997) consistent with $n=2\dots2.5$.
All the reported exponents so far are positive. The relations
in (\ref{0.1}) can be reformulated with the dynamo number
${\cal D}$ as
\beg
B_{\rm max} = B_{\rm eq} {\cal D}^{m'} , \qq \omega_{\rm cyc} =
{\cal D}^{n'}/\tau_{\rm diff},
\label{0.3}
\ende
where $B_{\rm eq}$ means the `equipartition field' and
$\tau_{\rm diff}=H^2/\eta_0$ the diffusion time. 
Instead of the normalization used in (\ref{0.3})$_2$, Soon \ea
(1993) proposed the use of the basic rotation rate $\Omega_0$; a
procedure we do not follow here to retain consistency with 
conventional definitions of the dynamo number. 
							
The main issues of stellar activity and the relations to 
dynamo theory concerning cycle times were formulated 
by Noyes \ea (1984). Their argumentation concerns a zeroth-order 
\alf\Om-dynamo model for which the coefficient $n$ in the scaling
$
\omega_{\rm cyc} \propto {\cal D}^{{n/2}}
$
is determined. The linear case  yields $n=4/3$ for the most
unstable mode (Tuominen {\it et al.}, 1988). 
Different nonlinearities lead to different exponents.
For \alf-quenching, $n=0$ is found while for 
models with flux loss (rather than \alf-quenching) 
$n=1$ results or $n={2/ 3}$ -- if only the
toroidal field is quenched by magnetic buoyancy. Note
the absence of finite $n$ for simple
\alf-quenching. If the
theory is correct and the observations are following relations
such as (\ref{0.1}), then the nonlinear dynamo can never work with
\alf-quenching as the basic nonlinearity.

A number of more complex dynamo models confirming the
findings of Noyes \ea (1984) have been developed in the 
last decade. For large dynamo numbers Moss \ea (1990) also 
derived $n=1$ from a spherical dynamo saturated by magnetic 
buoyancy. Schmitt and Sch\"ussler (1989) showed with a 1D 
model (their Fig.~6) that there is
practically no dependence of the cycle frequency on the (large)
dynamo number for \alf-quenching, but they find a strong scaling
($n\simeq 2$) for their `flux-loss models'. Also for a 2D model
in spherical symmetry the dependence of the cycle frequency on
the dynamo number proved to be extremely weak, $n \lsim 0.1$
(R\"udiger {\it et al.}, 1994, dashed line in their Fig.~4).

The latter application, however, requires further consideration.
One finds a finite value for $n$ if the magnetic 
feedback is not only considered acting upon the \alf-effect, but
also for the eddy diffusivity tensor. What we assume here is
that the magnetic field always suppresses and deforms the
turbulence field, and this has consequences for both the
\alf-effect and the eddy diffusivity. The theory for a
second-order correlation-approximation is given in Kitchatinov
\ea (1994b), applications are summarized in R\"udiger
\ea (1994).

The turbulent-diffusivity quenching concept was also
described in Noyes \ea (1984). As the magnetic fields
become super-equipartitioned, however, a series expansion such as used 
cannot be adopted. Tobias (1998), with a 2D global model in
Cartesian coordinates, finds $n'$ varying between 0.38 and 0.67
depending on the nonlinearity. He also finds the
exponent growing with increasing effect of $\eta$-quenching.
The results for the Malkus-Proctor effect alone yield the
weakest dependence for the cycle period with ${\cal D}$.

Since the paper by Brandenburg \ea (1998),
`anti-quenching' expressions such as
\beg
\alpha \propto B^p, \qq \eta_{\rm T} \propto B^q
\label{36}
\ende
are also under consideration, with $p>0$ and $q>0$. The idea is to
find the consequences of EMF quantities being induced by the
magnetic field itself (see Section~7). Saturation of such
dynamos requires $q > p$. All their global models, however,
lead to exponents $n$ of order 0.5. It concerns both 1D and 2D
models, characteristic values for $q$ were numbers up to 6.

It is shown in the following that a special 1D slab dynamo,
as a generalization of Parker's zero-dimensional wave dynamo
without $\eta$-quenching, has a very 
low $n$  while $n$ exceeds unity if 
$\eta$-quenching is included. Buoyancy effects are therefore  not
the only possible addition to dynamo theory to explain
observations as given in (\ref{0.1}); the magnetic suppression of
the turbulent magnetic diffusivity also yields an explanation
of the observations.
\subsection{The Model}
Here we are not working with the simplest possibility, which would 
read $\alpha_{ij} = \alpha\delta_{ij}$ and  $\eta_{ijk} = \eta_{\rm
T}\varepsilon_{ijk}$. 
The full feedback of the induced magnetic field on
the turbulent EMF is included, i.e. the magnetic suppression
and deformation of both the tensors \alf\ and
$\eta$. In particular, the influence of the magnetic field on
the magnetic diffusivity is often ignored in dynamo computations, 
and we shall demonstrate the differences in the gross properties 
of the solutions with and without $\eta$-quenching.
The main consequence of the inclusion of $\eta$-quenching is
the appearance of a nonlinear `magnetic velocity' in $\vec{\cal{E}}$,
\beg
\vec{\cal{E}} = \cdots \ +{\bf U}^{\rm mag} \times \langle{\bf B}\rangle
\label{2.5}
\ende
with
\beg
{\bf U}^{\rm mag} = \hat\eta \nabla \log
\langle B\rangle + \eta_z
 {\rot \langle{\bf B}\rangle \times
\langle{\bf B}\rangle \over
\langle B\rangle^2} 
\label{2.6}
\ende
(Kitchatinov {\it et al.}, 1994b) and 
\beg
\eta_{\rm T} = \eta_0 \varphi , \q \eta_z = \eta_0 \varphi_z, \q \hat
\eta = \eta_0 \hat\varphi .
\label{2.7}
\ende
The coefficient functions $\varphi(\beta)$ are
\begin{eqnarray}
\varphi&=&\frac{3}{2\beta^2}\left(-\frac{1}{1+\beta^2}+\frac{1}{\beta}
  \arctan\beta\right),\\
\varphi_z&=&\frac{3}{8\beta^2}\left(1+\frac{2}{1+\beta^2}+
  \frac{\beta^2-3}{\beta}\arctan\beta\right),\\
\hat\varphi&=&\frac{3}{8\beta^2}\left(-\frac{5\beta^2+3}{(1+\beta^2)^2}+
  \frac{3}{\beta}\arctan\beta\right) 
\end{eqnarray}
with 
\beg
\beta = |\langle B\rangle|/B_{\rm eq}.
\label{44.1}
\ende
The components $\alpha_{\phi\phi}$ and $\alpha_{ss}$ of the 
\alf-tensor are taken from \R\ and Schultz (1997).
They are equal in our approximation and read
\beg
\alpha_{ss} = \alpha_{\phi\phi} = -{2\over 5} \Omega^* \ {d\log\rho \over
dz}\ u_{\rm T}^2 \ \tau_{\rm corr} \ \Psi 
\label{19.1}
\ende
with the $\alpha$-quenching function $\Psi$ given in (\ref{alfquench}).
These quenching functions are approximated by $
\Psi(\beta)\simeq 1-\frac{12}{7}\beta^2,
\varphi(\beta)\simeq 1-\frac{6}{5}\beta^2, 
\varphi_z(\beta) \simeq \frac{2}{5}\beta^2$ and $ 
\hat\varphi(\beta) \simeq \frac{3}{5}\beta^2$
for weak magnetic fields.

\begin{figure}
\psfig{figure=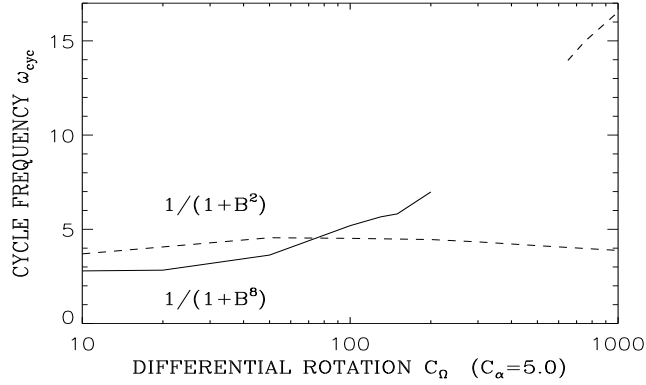,height=5.5truecm,width=11.8truecm,bbllx=-130pt,bblly=0pt,bburx=504pt,bbury=350pt}
\caption{\label{dynall3}The cycle frequency of oscillatory solutions of the
$\alpha^2\Omega$-dynamo versus $C_\Omega$ with different $\alpha$-quenching
functions. A second solution can be excited for the slower cut-off
function (dashed) for high $C_\Omega$. The steeper cut-off
function (solid) delivers
chaotic solutions above $C_\Omega\sim200$.}
\label{f10}
\end{figure}

The only coordinate about which our quantities are allowed to
vary is the direction of the rotation axis, i.e. $z$. 
The $z$-dependencies in (\ref{19.1}) are summarized in the form 
$
\alpha = - \hat \alpha \ \sin2z
$
where the lower and upper boundary are located at $z =
0$ and $z = \pi$. The plane has infinite extent in
the $x$ and $y$ directions
and is restricted by boundaries in the $z$-direction. The  normalized
dynamo  equations are  
\begin{eqnarray}
\frac{\partial A}{\partial t}&=&C_\alpha \hat\alpha(z)\,\Psi B
+\varphi\frac{\partial^2 A}{\partial z^2}-
w\frac{\partial A}{\partial z},\nonumber\\
\frac{\partial B}{\partial t}&=&-C_\alpha\frac{\partial}{\partial z}\left({
  \hat\alpha(z)\,\Psi\frac{\partial A}{\partial z}}\right)-
  C_\Omega\frac{\partial A}{\partial z}+
\frac{\partial}{\partial z}\left(\varphi\frac{\partial B}{\partial z}\right)-
\frac{\partial}{\partial z}\Bigl(w B\Bigr) 
\end{eqnarray}
with 
$w=(2\hat\varphi-\varphi_z)\partial \log B_{\rm tot}/\partial z$
and $B_{\rm tot}^2=B^2+(\partial A/\partial z)^2$.
The boundary conditions are $B=\partial A/\partial z= 0$ 
at $z=0$ and $z=\pi$ which limits the magnetic field 
to the disk. The dimensionless parameters
\beg
  C_\alpha=\frac{\alpha_0 H}{\eta_{\rm T}}, \q
  C_\Omega=\frac{\partial u_y}{\partial x}\frac{H^2}{\eta_{\rm
T}}
\label{CO}
\ende
represent the strength and sign of the $\alpha$-effect and the shear
flow. $C_\alpha$ is fixed to a value of 5 here. A positive 
$C_\alpha$ describes a positive \alf\ in the upper half of the layer 
and vice versa. We chose to vary $C_\Omega$ with positive values
which represent a positive shear.

\begin{figure}
\mbox{\psfig{figure=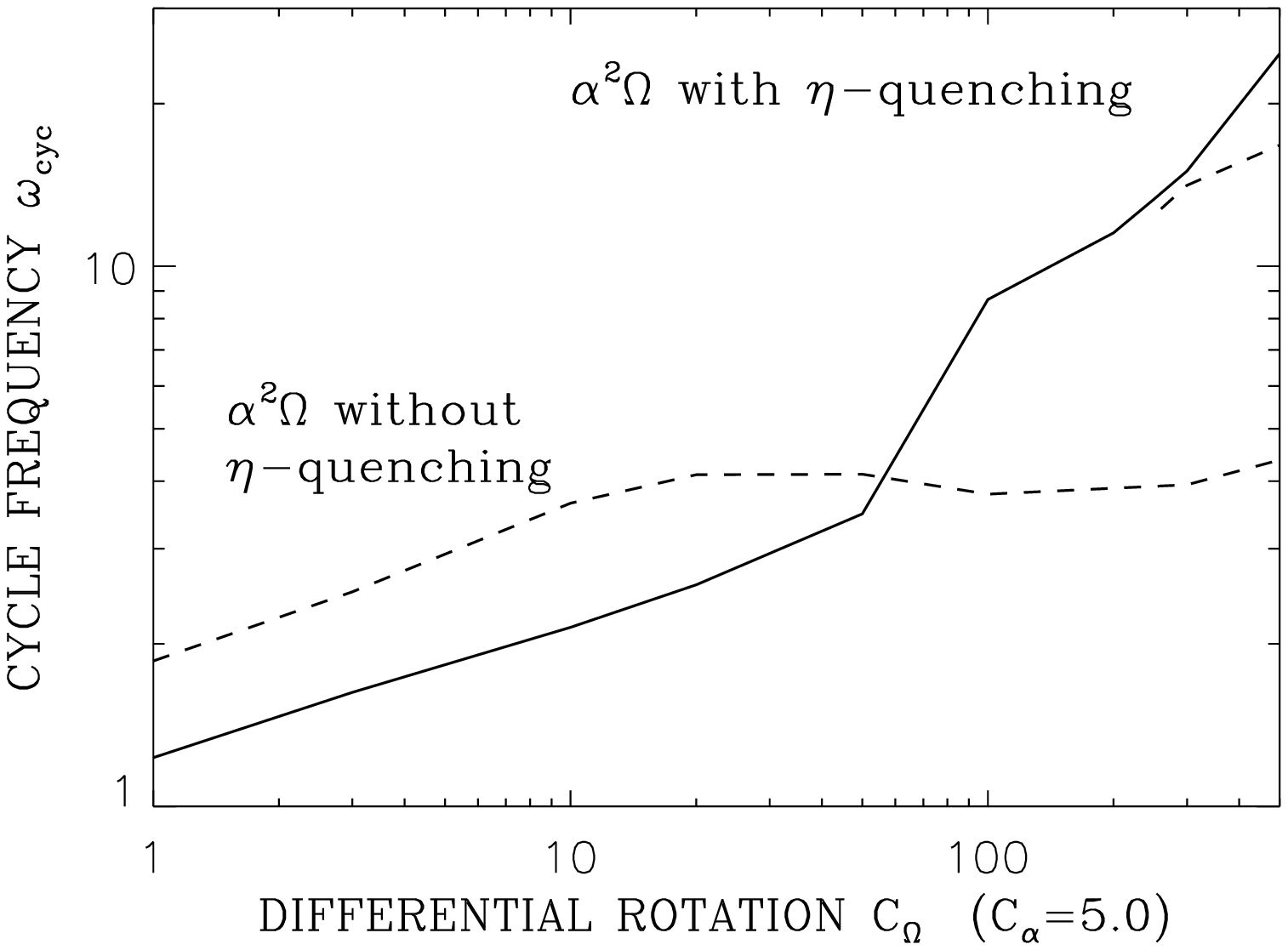,width=7cm,height=5.0cm} \hfill
     \psfig{figure=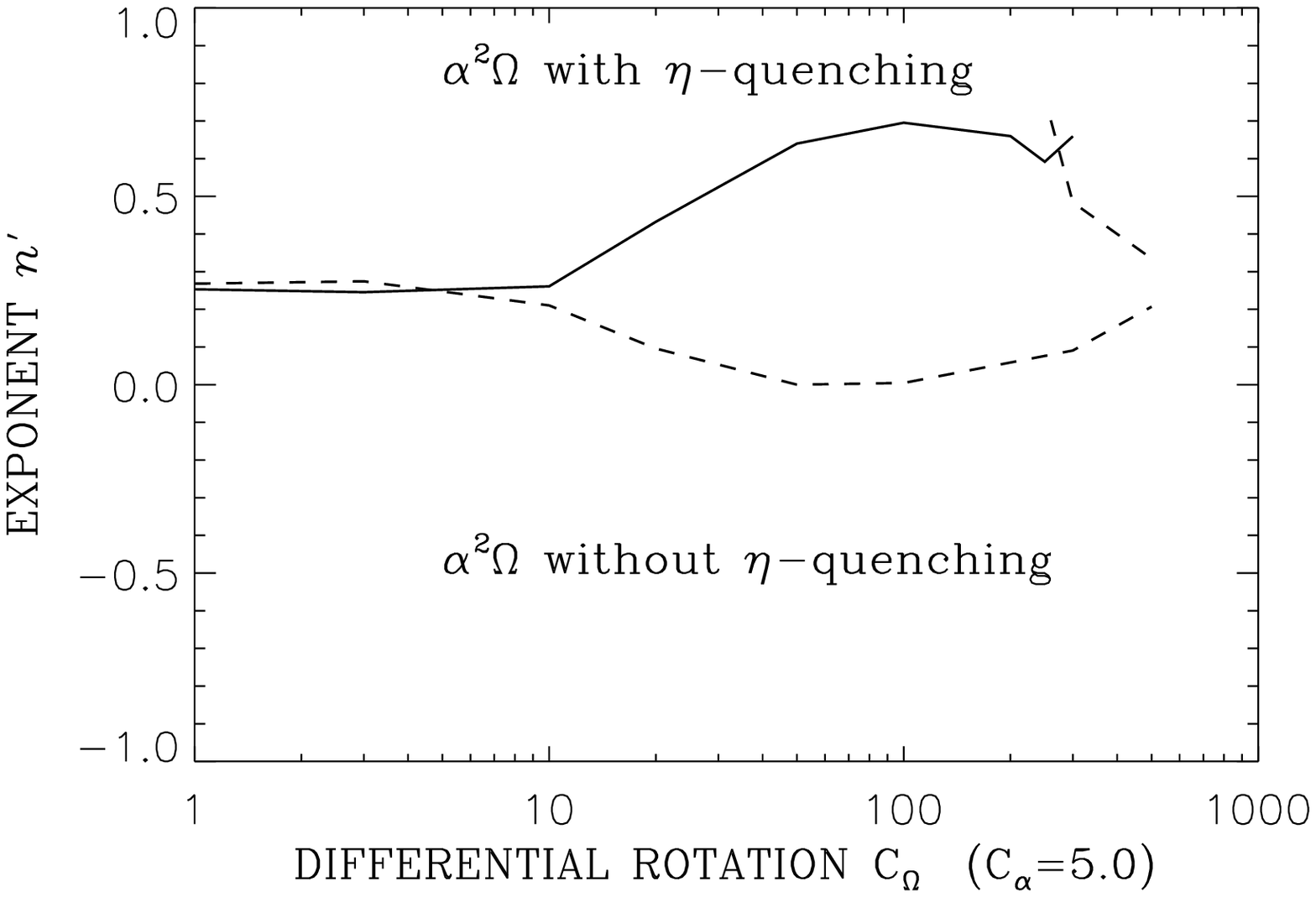,width=7cm,height=5cm}}
\caption{LEFT: The cycle frequency of oscillatory solutions of the
$\alpha^2\Omega$-dynamo vs. $C_\Omega$ with (solid) and without
(dashed) $\eta$-quenching. RIGHT: The exponent $n'$ for the scaling 
(36). Only diffusivity quenching yields exponents of order 0.5.
The model without $\eta$-quenching contains a second solution
which can be excited at high $C_\Omega$.}
\label{cycfreq}
\end{figure}

\subsection{The Results}
All the magnetic fields of the nonlinear 1D 
model are symmetric  with respect to the 
equatorial plane. In our domain of positive shear the solutions
of both the $\alpha^2$-dynamo  (small $C_\Omega$) and the 
$\alpha\Omega$-dynamo (large $C_\Omega$) are oscillatory. The 
transition between these regimes occurs at $C_\Omega \simeq 50$
(cf. R\"udiger and Arlt, 1996). The
cycle periods are always constant over time. Fig.~\ref{f10} shows the
resulting cycle frequencies for the conventional 
\alf-quenching expression. The cycle period is strikingly
independent of the dynamo number (see also Noyes {\it et al.}, 1984; Schmitt
and Sch\"ussler, 1989; Jennings and Weiss, 1991). On the other hand, 
Fig.~\ref{cycfreq} gives the resulting cycle frequencies of the 
oscillatory solutions
for the $\alpha^2\Omega$-dynamo with the $\eta$-quenching concept. The 
cycle periods are given for models with and without $\eta$-quenching. 
The  exponent $n'$ is variable and positive for the model with
$\eta$-quenching whereas it is much more complex for the model 
without $\eta$-quenching.

The resulting exponent $n'\simeq 0.5$ for large dynamo numbers 
in Fig.~\ref{cycfreq} is remarkably similar to the behaviour of
linear dynamos. In order to compare this value with the observed 
value of $n$ in (\ref{0.1}) we have to introduce the scaling of 
the physical quantities with $\Omega^*$. We adopt the scaling
$\eta_0\propto 1/\Omega^*$ from \K\ {\it et al.} (1994b) and use the 
parameterization
\beg
{\partial \log \Omega\over \partial \log r} \propto {\Omega^*}^\kappa.
\label{27}
\ende
Negative $\kappa$ describe a {\em decrease} of the unknown normalized
differential rotation in stellar convection zones with
increasing basic rotation.
Negative $\kappa$ are indeed  produced by theoretical models of the
differential rotation, though only if meridional flow is included.
The results in Fig.~\ref{fdr1} lead to $\kappa \simeq -1$. 
The dynamo number scales as
\beg
{\cal D} \propto {\Omega^*}^{\mu + \kappa +3},
\label{28}
\ende
if the \alf-effect is assumed as scaling with $\Omega^{*\mu}$. 
Then the relations (\ref{0.1}) and (\ref{0.3}) yield
\beg
n = (\mu +3+\kappa) n' - 1.
\label{29}
\ende
We apply the exponent $n'\simeq0.5$ from Fig.~\ref{cycfreq}
for large dynamo numbers and get $n= (\mu + \kappa)/2 + 0.5$, so that with 
$\kappa = - 1$ we find $\mu \simeq 2n$. Consequently, the observed 
exponents $n= 0.5 \dots 1.5$
would lead to 
\beg
\mu = 1 \dots 3
\label{mu}
\ende
for the \alf-effect running
with the Coriolis number which does not seem too unreasonable. 

\section{Dynamo-induced on-off Alpha-Effect}

A two-component model for the solar dynamo has been suggested
by Ferriz-Mas {\it et al.} (1994).
Toroidal magnetic fields are generated by differential
rotation in the overshoot layer whereas a dynamo acts in the
upper part of the convection zone. The thin overshoot layer is
assumed  to be stably stratified up to magnetic field strengths of
$10^5$~Gauss  (Ferriz-Mas and Sch\"ussler, 1995). The dynamo may 
operate in the turbulent convective zone and impose stochastic 
perturbations to the overshoot layer. 

Schmitt {\it et al.} (1996) discuss whether the
interaction of the weak-field convection zone dynamo and the
magnetic fields in the overshoot layer might explain the grand-minimum
behaviour of the solar cycle. They assumed a mean-field dynamo with an 
$\alpha$-effect working only when the magnetic field exceeds a
certain threshold. Grand minima are expected when the perturbations
from the convection zone do not keep the overshoot layer instabilities
above this threshold. Here we test whether a dynamo, once excited,
will really turn off in this sense.

The dynamo equations are studied in both 1D and 2D domains. In the 1D
approach the integration region extends only in the $z$-direction whence
the remaining radial and azimuthal components of the magnetic field depend
on $z$ only. Normalization of times with the diffusion time
$\tau_{\rm diff}=H^2/\eta_{\rm T}$
and  vertical distances $z$ with $H$ yields
\begin{eqnarray}
\label{dynamo1}
\frac{\partial A}{\partial t}&=&C_\alpha \hat\alpha(z)\,\Psi(B_{\rm tot
}) B
+\frac{\partial^2\!A}{\partial z^2},\\
\label{dynamo2}
\frac{\partial B}{\partial t}&=&-C_\alpha\frac{\partial}{\partial z}\left({
\hat\alpha(z)\,\Psi(B_{\rm tot})\frac{\partial A}{\partial z}}\right)-
C_\Omega\frac{\partial A}{\partial z}+\frac{\partial^2 B}{\partial z^2}.
\end{eqnarray}

\begin{figure}
\psfig{file=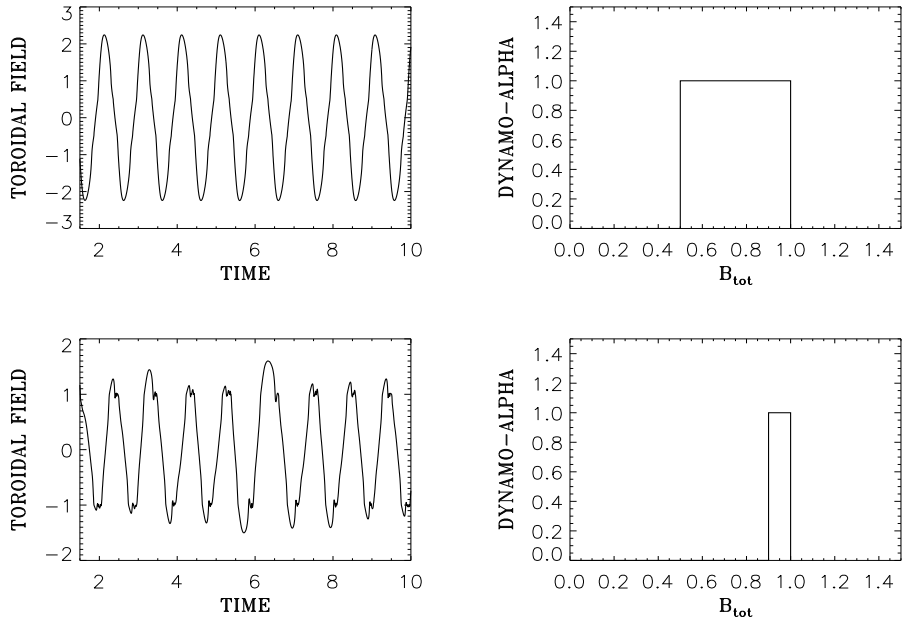,height=8.8truecm,width=13.6truecm,bbllx=30pt,bblly=360pt,bburx=558pt,bbury=720pt}
\caption{\label{onoff1d}The toroidal component of the
magnetic field (LEFT) at a fixed point of the domain for 
two different on-off $\alpha$-functions (RIGHT) in a 1D model. 
The initial conditions imply a field strong enough to
cover partly the `on'-range of $\alpha$.
}
\end{figure}

\begin{figure}
\psfig{file=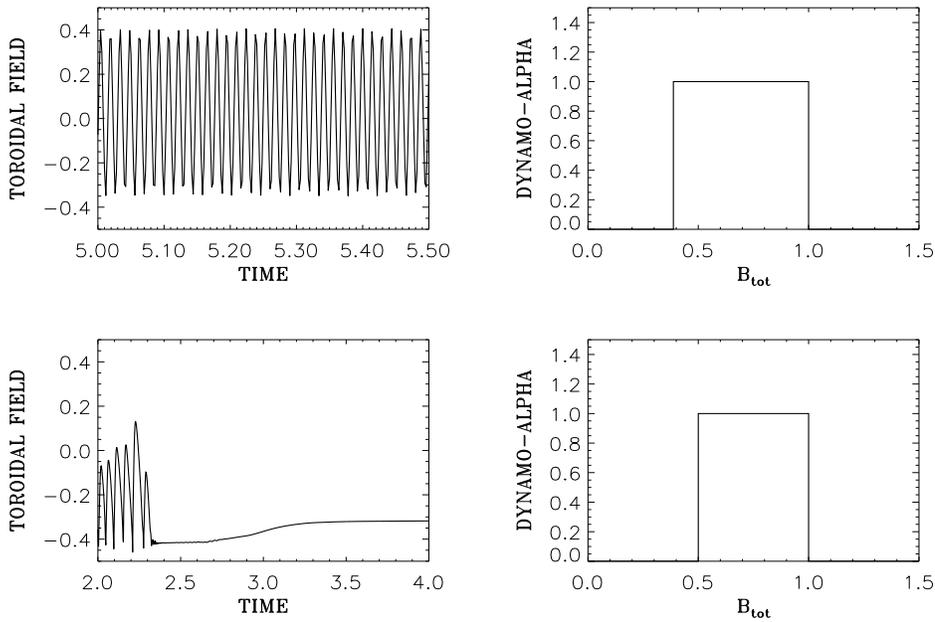,height=8.8truecm,width=13.6truecm,bbllx=30pt,bblly=360pt,bburx=558pt,bbury=720pt}
\caption{\label{onoff2d}The toroidal component of the
magnetic field (LEFT) for two different on-off $\alpha$-functions 
(RIGHT) in a 2D model. $B_{\rm max} = 1$ in both cases.}
\end{figure}

The $\alpha$ depends on the magnetic field as well as
on the location in the object. Hence we have two different dependencies
in $\alpha=\hat\alpha(z)\alpha_0\Psi(B_{\rm tot})$. Again, the first factor is
expressed by $\hat\alpha(z)=-\sin 2z$.

The $\alpha$-quenching by  magnetic fields
has usually been expressed by functions such as 
$1/[1+(B_{\rm tot}/B_{\rm max})^2]$,
with a cut-off field strength $B_{\rm max}$ related to the
energy of velocity fluctuations or the gas pressure. In the present
approach $\alpha$ acts only in a certain range of $B_{\rm tot}$
and it is zero otherwise, i.e.
\beg
\Psi(B_{\rm tot})=\left\{{\begin{array}{l}
1\quad\quad\mbox{for}~B_{\rm min}<B_{\rm tot}<B_{\rm max}\\
0\quad\quad\mbox{otherwise}.\end{array}}\right.
\ende
The lower threshold $B_{\rm min}$ represents the onset
of instability in the strong-field dynamo layer at the bottom
of the convection zone, and the upper threshold $B_{\rm max}$ is related
to the buoyancy-driven disappearance of flux tubes from the 
overshoot layer.
Again a vacuum surrounds the 1D model. 

The normalized induction equations of the 2D model read
\begin{eqnarray}
{\partial B\over\partial t}  &=&  {1\over r}{\partial\over\partial r}
\left(\eta_{\rm T}{\partial(B r)\over\partial r}\right) +
{\eta_{\rm T}\over r^2}{\partial\over\partial\theta}\bigg({1\over\sin\theta}
{\partial(B\sin\theta)\over\partial\theta}\bigg)
 +\nonumber\\
&  +&  {1\over r} {\partial\Omega\over\partial r}
{\partial A\over\partial\theta} -
{1\over r \sin\theta}\ {\partial \over \partial r} \left(\alpha
\ {\partial A \over  \partial r}\right) -
 {1\over r^3}\
{\partial \over \partial \theta} \left({\alpha \over
\sin\theta} \ {\partial A \over \partial \theta}\right) ,
\label{3}
\end{eqnarray}
\beg
{\partial A\over \partial t}  =  \eta_{\rm T}{\partial^2
A\over\partial r^2}+\eta_{\rm T}{\sin\theta\over
r^2}{\partial\over\partial\theta}
\left({1\over\sin\theta}{\partial A\over\partial\theta}\right) 
+ \alpha r\sin\theta B .
\label{4.1}
\ende
The computational domain extends from $r=0.5$ to $r=1.0$ measured
in solar radii, and from $\theta=0$ to $\theta=\pi$ in colatitude.
The dynamo effect is assumed to be placed between $r=0.7$ and
$r=0.8$ neighboured by a weakly dissipative layer with $\eta_{\rm T}=0.01$
below and a zone with strong turbulent diffusion ($\eta_{\rm T}=1$) above the
dynamo layer; $\eta_{\rm T}$ is also unity in the dynamo shell. 
Here we neglect the $\theta$-dependence of the angular velocity
and assume a constant $\partial\Omega/\partial r$ for simplicity.

An $\alpha^2\Omega$-dynamo provides oscillatory
and steady-state solutions. The first test in a 1D
model applies $C_\Omega=100$ and $C_\alpha=5$ and two
on-off functions with both $B_{\rm min} =
0.5$ and $B_{\rm min} = 0.9$. The dynamo was excited with an
initial magnetic field 
strong enough to reach the `on'-range of $\alpha$. Fig.~\ref{onoff1d}
shows the resulting oscillatory solutions. The dynamo never
dies since the $\alpha$-effect acts before
the magnetic fields completely `dives' through the on-off
function.

The same  is found in a 2D model as shown in Fig.~\ref{onoff2d}.
We used the parameters $C_\Omega=10^5$ and $C_\alpha=-10$ and 
two on-off functions with $B_{\rm min} = 0.38$ and $B_{\rm min}
= 0.5$, resp.
If the on-off function is too narrow the solutions change from
oscillatory to steady-state, although we never find the dynamo
stopping its operation. The parity of the oscillatory solution
is antisymmetric with respect to the equator but equator symmetry is
found for the steady-state solution. 

The results indicate that the dynamo, once initiated by
sufficiently strong magnetic fields, will not switch off on
its own and, hence, does not impose grand minima to its cyclic
behaviour. The dynamo will only cease operating if the flux tubes
migrate into the upper convection zone on a time scale much shorter 
than the magnetic diffusion time.

\section{Random Alpha}\label{acht}
Two different types of magnetic dynamos appear to be acting in the Sun
and the Earth. In contrast to the distinct oscillatory solar magnetic 
activity, the terrestrial magnetic field is
`permanent'. Indeed, two different sorts of dynamos in spherical
geometry have been constructed with such properties.
The $\alpha^2$-dynamo yields stationary solutions
while the $\alpha\Omega$-dynamo yields oscillatory solutions.
Observation and theory seem to be rather in agreement. There are,
however, differences in the time behaviour:  solar activity is not
strictly periodic and the Earth's dynamo is not strictly
permanent. The Earth's magnetic field also varies. There are irregular
changes of the field strength, the shortest period of
stability being 40\,000 years. The average length of the intervals of
constant field is almost 10 times larger than this minimum value (\cf
Krause and Schmidt, 1988). Such a reversal was simulated 
numerically with a 3D MHD code by Glatzmaier and Roberts (1995).
 
In Section~5 we explained grand minima such as the
Maunder minimum by various nonlinearities in the mean-field equations. 
The averages are taken over an `ensemble', \ie\
a great number of identical examples. The other possibility to explain the 
temporal irregularities
is to consider the characteristic turbulence values as a time
series. The idea is that the averaging procedure concerns only a
periodic spatial coordinate, \eg\ the azimuthal angle $\phi$. In
other words, when expanding in Fourier series such as
$e^{im\phi}$ the mode $m = 0$
is considered as the mean value. Again, if the time scale of
this mode does not vary significantly during the correlation time, local
formulations such as equation (\ref{1.7}) below are reasonable. Nevertheless, 
the turbulence intensity, the \alf-effect, and the eddy
diffusivity become time-dependent quantities (Hoyng, 1988;
Choudhuri, 1992; Hoyng, 1993; Moss {\it et al.}, 1992; Hoyng
{\it et al.}, 1994; Vishniac and Brandenburg, 1997;
Otmianowska-Mazur {\it et al.}, 1997).
 
However, questions concerning the amplitude, the time scales,
and the phase relation between (say) helicity and eddy diffusivity 
need to be studied. Are the fluctuations strong enough to 
influence the dynamo significantly? We construct here a random turbulence model 
and evaluate the complete turbulent electromotive force as a time series.
The consequences of this concept are then computed for a simple 
plane-layer dynamo with differential rotation. The main parameter
which we  vary is the number of cells in the unit length. 

One can define mean values of a field $F$ by integration over (say) 
space. Of particular interest here is an averaging procedure over 
longitude, i.e.
\beg
\langle
F\rangle = {1\over 2\pi}\ \int\limits_0^{2\pi} F\, d\phi
\label{1.31}
\ende
(Braginsky, 1964; Hoyng, 1993). The
turbulent EMF for a given position in the meridional plane
forms a time series with the correlation time $\tau_{\rm corr}$
as a characteristic scale. The peak-to-peak variations in the
time series should depend on the number of cells. They remain
finite if  the number of cells is restricted as it is in reality. 
For an infinite number of the turbulence cells the peak-to-peak
variation in the time series goes to zero but it will grow for a finite 
number of cells along the unit length.
 
The tensors constituting the local mean-field EMF in (\ref{2}), 
$\alpha$ and $\eta$, must be calculated from one and the
same turbulence field. We propose to define a helical
turbulence existing in a rectangular parcel 
and to compute simultaneously their related
components.
 
We restrict ourselves to the computation of the turbulent EMF 
in the high-conductivity limit. Then the second-order
correlation approximation yields 
\beg
\vec{\cal{E}} = \int\limits_0^\infty \left\langle{ {\bf u}'
({\bf x},t)
\times {\rm curl} \bigg({\bf u}'({\bf x},t - \tau) \times \langle
{\bf B} ({\bf x},t)\rangle\bigg)}\right\rangle \ d\tau ,
\label{1.6}
\ende
which for short correlation times can be written in the form (\ref{2}). 
A Cartesian frame is used where $y$ represents the
azimuthal direction over which the average is performed.
For the dynamo only the components ${\cal E}_x$ and
${\cal E}_y$ are relevant.
 
It would be tempting to apply (\ref{1.6}) as it stands. In components it reads
\begin{eqnarray}
{\cal E}_x &=& \alpha_{xx} \langle B_x\rangle + \eta_{\rm T} \
{\partial \langle B_y\rangle \over \partial z},\nonumber\\
\ \ \ \nonumber\\
{\cal E}_y &=& \alpha_{yy} \langle B_y\rangle - \eta_{\rm T} \
{\partial \langle B_x\rangle \over \partial z}.
\label{1.7}
\end{eqnarray}
$\eta_{\rm T}$ plays the role of a common eddy diffusivity. From
(\ref{1.6}) one can read
\begin{eqnarray}
\alpha_{xx} &=& \int\limits_0^\infty \langle u'_y(t) \ {\partial
u'_z (t-\tau) \over \partial x} - u'_z(t)\ {\partial u'_y (t-\tau)
\over \partial x} \rangle \ d\tau , \nonumber\\
\ \ \ \nonumber\\
\alpha_{yy} &=& \int\limits_0^\infty \langle u'_z(t) \ {\partial
u'_x (t-\tau) \over \partial y} - u'_x(t)\ {\partial u'_z (t-\tau)
\over \partial y} \rangle \ d\tau ,\nonumber\\
\ \ \ \nonumber\\
\eta_{\rm T} &=& \int\limits_0^\infty \langle u'_z(t) \
u'_z(t-\tau) \rangle \ d\tau 
\label{1.8}
\end{eqnarray}
(Krause and R\"adler, 1980).

\subsection{The Turbulence Model}
%
%
We study the time evolution of the dynamo coefficients (\ref{1.8}) 
generated by turbulent gas motions.
We analyse a parcel of the solar gas situated in the
convective zone. The rectangular coordinate system
has the $xy$-plane parallel to the solar equator, the $x$-axis
pointing parallel to the solar
radius, the $y$-axis tangential to the azimuthal direction and  the
$z$-axis directed to the south pole. The parcel is
permanently perturbed by vortices of the form of rotating columns of gas
randomly distributed at all angles in 3D space.
At every time step the resulting
velocity field is  used to calculate the EMF-coefficients after (\ref{1.8}).
\begin{figure}[h]
\psfig{figure=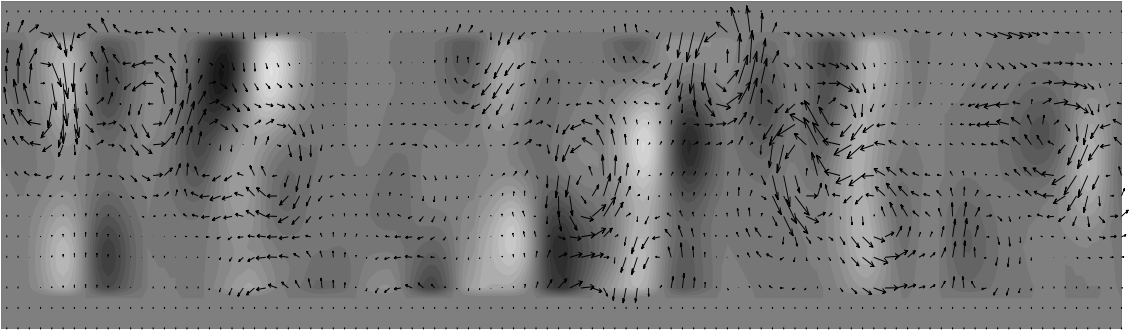,angle=90,height=4.3truecm,width=15truecm,bbllx=30pt,bblly=13pt,bburx=165pt,bbury=561pt}
\caption{Cross-section of the velocity field for the 100-eddy turbulence 
model ${\cal B}$ in the $xy$-plane at an arbitrary time of the 
simulation. White colour denotes positive $u_z$, black colour negative $u_z$.}
\label{f11}
\end{figure}
 
A single vortex rotates with the velocity $\omega$ around
its axis of rotation and moves with the velocity up and down 
along this axis. The eddy itself has also its own lifetime growing 
linearly with subsequent time steps. The vortex velocity $\omega$ 
as well as the vertical drift $w$ decrease with the distance $r$ 
from the axis and with the lifetime $t$ according to
\beg
(\omega,w)=(\omega_{0},w_0)\cdot
e^{-0.5\bigg(\big(r/l_{\rm corr}\big)^2
+ \big(z/z_{\rm corr}\big)^2\bigg)}\cdot e^{-t/\tau_{\rm corr}},
\label{1.9}
\ende
where
$l_{\rm corr}$ is the adopted vortex radius (Otmianowska-Mazur
{\it et al.}, 1992),
$z_{\rm corr}$
is its length  scale along the axis of rotation and $\tau_{\rm corr}$ is
the eddy decay time scale.
All  velocities  are
truncated at $3l_{\rm corr}$ and $3z_{\rm corr}$.
In the  model all vortices are orientated like right-handed screws. 
The resulting motion has maximum helicity. A uniform distribution of 
eddies in a 3D
space is approximated with 12 possible inclination angles of their
rotational axes to the $xy$-plane.
 
The initial state is a number $N$ of moving turbulent cells having
random inclinations and positions in the $xy$-plane. After an assumed 
period of time (one or more time steps), a fraction of them, 
$R_{\rm tur}$ (the ratio of the new-eddies number to $N$), is changed to
new ones with different randomly  given position, inclination
and  with  lifetime  starting  from  zero.  
The situation repeats itself continuously with time.

\begin{table}[b]
\caption[]{Input and output for the turbulence  models ${\cal A,B,C}$
and ${\cal D}$.   $N$ is the eddy
population of  the equator,  $R_{\rm tur}$ the eddy birth rate,
other  quantities and normalizations as
explained in the text.}
\begin{center}
\begin{tabular}{|c|llll|cccccc|}\hline
{\rule[-0mm]{0mm}{5mm}
}  &$l_{\rm corr}$ & $\tau_{\rm corr}$ & $N$ & $R_{\rm tur}$&
$\eta_{\rm T}$ & $\sigma(\eta_{\rm T})$
  & $\alpha_{xx}$ & $\sigma(\alpha_{xx})$ & $\alpha_{yy}$ &
$\sigma(\alpha_{yy})$ \\[0.5ex]
\hline
{\rule[-2mm]{0mm}{7mm}
${\cal A}$} & 1 & 10 & 200 & 1.0 & 0.191 & 0.086 &  $-$0.046 & 0.048&
$-$0.077 & 0.055 \\
\ ${\cal B}$ & 2 & 20 & 100  & 0.7 & 0.336 & 0.208 & $-$0.029 & 0.063&
$-$0.048 & 0.059 \\
\ ${\cal C}$ & 4 & 40 & \phantom{0}50 & 0.005& 1.045 & 0.860 & $-$0.029 & 0.119&
$-$0.039 & 0.110 \\
\ ${\cal D}$ & 8 & 80 & \phantom{0}25 & 0.0003& 1.793 & 2.278 & $-$0.015 & 0.136&
$-$0.017 & 0.110 \\[0.9ex]
\hline
\end{tabular}
\end{center}
\end{table}
The vortex radius $l_{\rm corr}$ as
well as the decay time $\tau_{\rm corr}$ and the fraction of the new
turbulent cells $R_{\rm tur}$ are varied for four cases given
in Tables~2 and 3 in normalized units
(cgs units for time, velocity 
and diffusivity  result after  multiplication  with $2.5 \cdot 10^4$~s,
$10^{5}$ cm/s and $2.5 \cdot 10^{14}$ cm$^2$/s). 

We made numerical experiments with the following values for the
vortex radius $l_{\rm corr}$ and the length scale $z_{\rm
corr}$ in the $z$-direction in normalized units: 1, 2, 4 and 8.
For successive length scales we choose the time scales 10, 20,
40 and 80.  The ratio of both scales is always $l_{\rm corr}/\tau_{\rm
corr}\cong 0.1$. In order to obtain the expected mean turbulent 
velocity value of 0.1, the fraction of new vortices $R_{\rm tur}$ is 
also varied (Table~2; cf. Otmianowska-Mazur {\it et al.}, 1997).

\subsection{Numerical Experiments}
The simulations deliver time series of the turbulence
intensity, the eddy diffusivity and the $\alpha$-coefficients,
and a standard deviation $\sigma$ from their temporal averages
\begin{equation}
\sigma=\sqrt{E[(X-E(X))^{2}]},
\label{2.1}
\end{equation}
\noindent where $E(X)$ is the time average of a random variable 
$X$. The standard deviation measures the amplitude of the fluctuations 
compared with the temporal mean of a given quantity. Table~2 lists the
input model parameters $l_{\rm corr}$, $\tau_{\rm corr}$, and $R_{\rm tur}$ 
and the coefficients obtained. The ratios of the standard deviations to
their mean values,
\beg
S_{\eta}=\sigma(\eta)/\mid\eta\mid,
\label{10.1}
\ende
are given in Table~3.

\begin{table}
\caption[]{The ratio $S$ of  fluctuations and  mean values.}
\begin{center}
\begin{tabular}{|c|lll|}\hline
{\rule[-0mm]{0mm}{5mm}
} model & $S_{\eta_{\rm T}}$ &  $S_{\alpha_{xx}}$ & $S_{\alpha_{yy}}$ \\[0.5ex]
\hline
{\rule[-2mm]{0mm}{7mm}
${\cal A}$} & 0.45 & 1.03 & 0.72  \\
\ ${\cal B}$ & 0.62 & 2.13 & 1.24  \\
\ ${\cal C}$ & 0.82 & 4.10 & 2.79  \\
\ ${\cal D}$ & 1.27 & 8.94 & 6.62 \\[0.9ex]
\hline
\end{tabular}
\end{center}
\end{table}
 
\begin{figure}
\psfig{figure=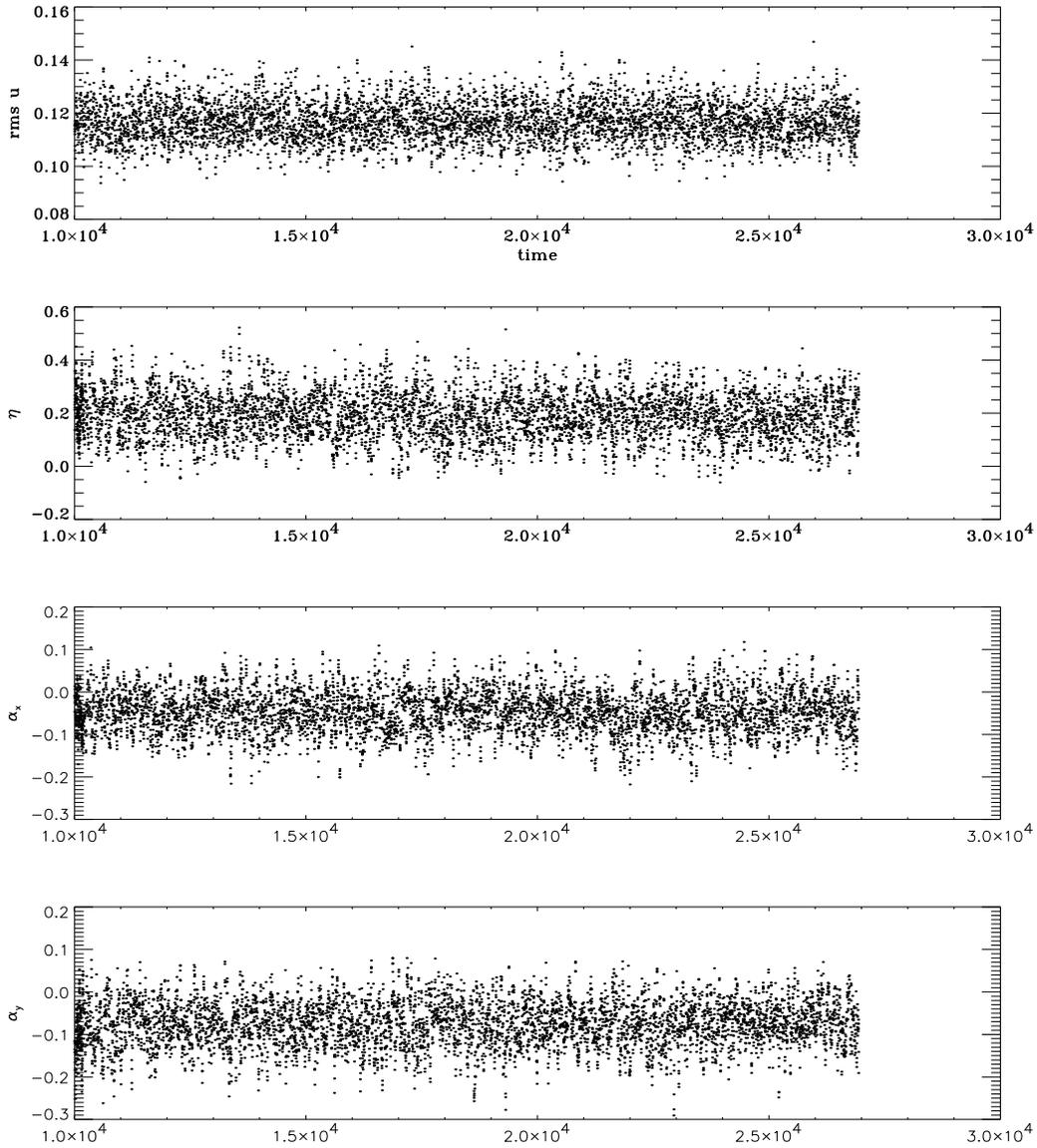,height=16truecm,width=14truecm}
\caption{
Time series for turbulence intensity,  eddy diffusivity  and  alpha-tensor components  $\alpha_{xx}$
and  $\alpha_{yy}$ for turbulence model $\cal A$.
For time, velocity ($u$ and $\alpha$)
 and diffusivity cgs units result after  multiplication  with
$2.5 \cdot 10^{4}$ s, 
$10^5 $ cm/s 
and 
$2.5 \cdot 10^{14}$ cm$^2$/s. 
}
\label{f12}
\end{figure}
\begin{figure}
\psfig{figure=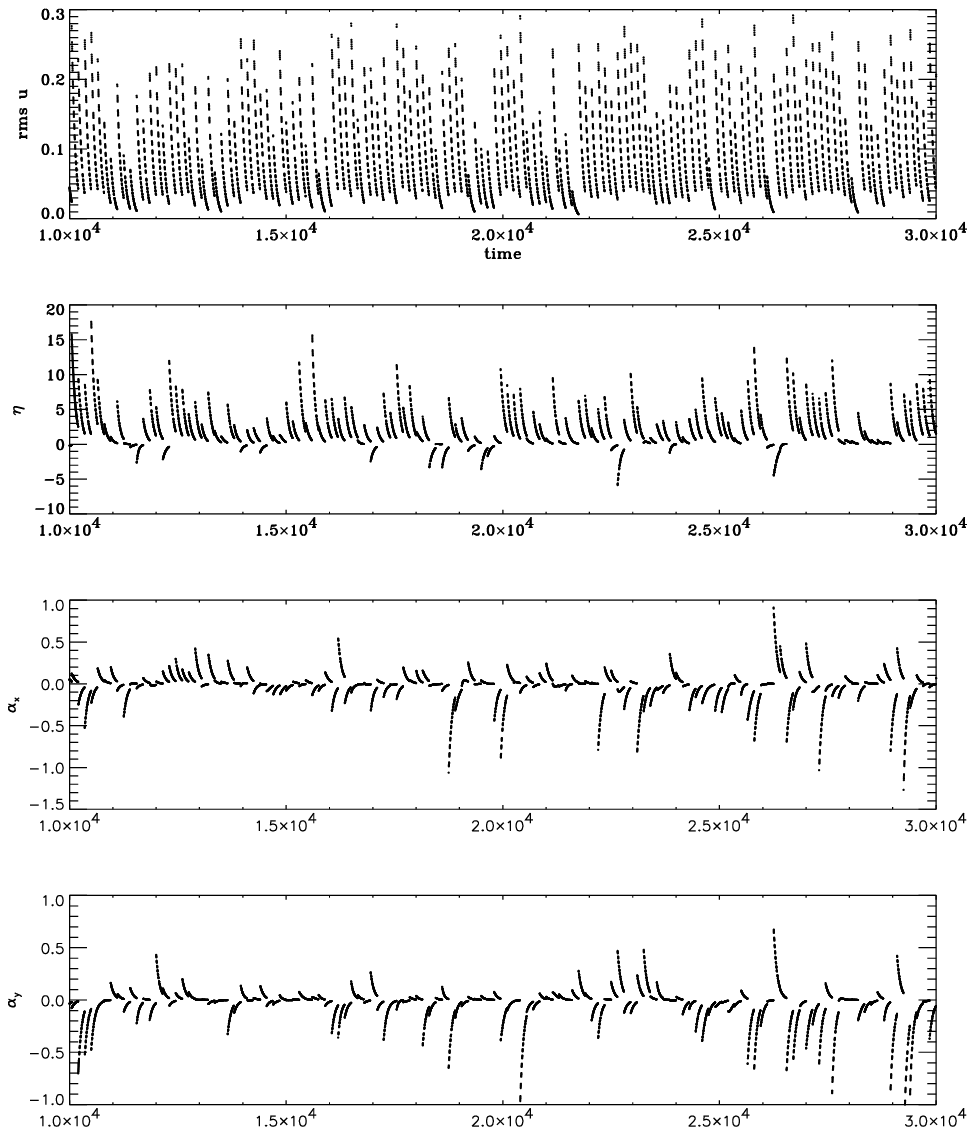,height=16truecm,width=14truecm}
\caption{
The same as in Fig.~\ref{f12} but for model ${\cal D}$.
}
\label{f14}
\end{figure}

The results for a model ${\cal A}$ with $l_{\rm corr}=1$ and $\tau_{\rm corr}=10$,
which is the sample of shortest lifetimes and smallest spatial dimensions 
of individual vortices in our simulation sequence is given
in Fig.~\ref{f12}. All computed quantities are drawn as a time
series, each in its own normalized units according to their definitions.
The three  EMF-coefficients as well as the  turbulence intensity
show fluctuations around their mean values in time.
The diffusion coefficient $\eta_{\rm T}$ possesses positive
values during most of the time. There are also short periods with a negative 
$\eta_{\rm T}$.
The negative values are not significant because the resulting mean
$\eta_{\rm T}$ is always positive yielding
0.191 (Table~2) in normalized units.
The ratio $S_{\eta_{\rm T}}$ is 0.45 (Table~3). Indeed, the value 0.191 corresponds to
$
\eta_{\rm T} \simeq \ \tau_{\rm corr}\  u_{\rm T}^2,
$
which gives 0.1 for this case. In contrast to $\eta_{\rm T}$,
the coefficients $\alpha_{xx}$ and $\alpha_{yy}$
possess negative values during the major part of the time;
for short periods, however, positive values are also present.
The negative $\alpha$ results from the assumed
non-zero, right-handed helicity of the vortices
and is in agreement with the expectations. Averaged in time
$\alpha_{xx}$  for model ${\cal A}$
is $-4.6 \cdot 10^{3} {\rm cm}/{\rm s}$. For $\alpha_{yy}$ we get
$-7.7 \cdot 10^{3} {\rm cm}/{\rm s}$ resulting in a dynamo
number of $C_\alpha = 10$.         
It can be seen that $\alpha_{xx}$ is always slightly
smaller than $\alpha_{yy}$. This fact is connected with the
assumed averaging only along the $y$-axis, which certainly
influences the value of $\alpha_{xx}$. 
Model ${\cal A}$ yields the ratios
$\sigma(\alpha_{xx})/\mid\alpha_{xx}\mid=1.03$ and
$\sigma(\alpha_{yy})/\mid\alpha_{yy}\mid=0.72$.

The second experiment, $\cal B$, uses doubled correlation lengths and 
times, $l_{\rm corr}=2$, $\tau_{\rm corr}=20$ (see also Fig.~\ref{f11}).
A fraction of 0.7 new vortices at every time step was
applied. The resulting fluctuations of $\eta_{\rm T}$ and
$\alpha$-coefficients are higher than in the case $\cal A$
(Table~2). 
The averaged value of $\eta_{\rm T}$ increases to 0.336,
yielding $8.5\cdot 10^{13}~ {\rm cm}^{2}/{\rm s}$.
The ratios $S_{\alpha_{xx}}$ and $S_{\alpha_{yy}}$
increase up to  $2.13$ and $1.24$, respectively (Table~3). It
means that the fluctuations of $\alpha$ are higher for larger
eddies with longer lifetimes -- as it should.

The experiment ${\cal D}$ works with largest correlation 
times and longest lifetimes, $l_{\rm corr}=8$ and $\tau_{\rm corr}=80$. 
The birth rate $R_{\rm tur}$ is 0.0003 per time step -- 
really low in comparison 
with the previous cases. Fig.~\ref{f14} presents the time
series of studied quantities. The resulting fluctuations 
of the alpha coefficients are extremely high.
The mean value of $\eta_{\rm T}$ is 1.793 ($4.5\cdot
10^{14}~ {\rm cm}^{2}/{\rm s}$). The ratio $S_{\eta_{\rm T}}=1.27$ indicates
the significant growth of the fluctuations
compared with the mean value.
The obtained ratios are $S_{\alpha_{xx}}=8.94$ and
$S_{\alpha_{yy}}=6.62$. The averaged 
in time values of $\alpha_{xx}$ and $\alpha_{yy}$ decrease again to
$-0.015$ ($-1.5 \cdot 10^{3} {\rm cm}^{2}/{\rm s}$) and
$-0.017$ ($-1.7 \cdot 10^{3} {\rm cm}^{2}/{\rm s}$).
 
Table~3 summarizes the main results of our simulations. The
fluctuations in the time series become more and more dominant
with the decreasing number of eddies together with the increasing
vortex size. The extreme case is plotted
in Fig.~\ref{f14}. For such models the fluctuations of both the
eddy diffusivity as well as the \alf-effect are {\em much higher than
the averages}. Even short-lived changes of the sign of the quantities 
were found.
 
The \alf-effect fluctuations exceed those of the eddy diffusivity
in all our models. The latter proves to be more stable than the 
\alf-effect against dilution of the turbulence. This is a 
confirmation, indeed, for those studies in which an \alf-effect time series 
is exclusively used in dynamo computations.
\subsection{A Plane-layer Dynamo}
Quite a few publications deal with the influence of
stochastic \alf-fluctuations upon dynamo-gener\-ated magnetic
fields for various models.
Choudhuri (1992) discussed a simple plane-wave dynamo basically
in the linear regime. The fluctuations adopted are as weak as
about 10\,\%. While in the $\alpha\Omega$-regime {\em the oscillations are
hardly influenced}, the opposite is true for the
$\alpha^2$-dynamo. In the latter regime the solution suffers
dramatic and chaotic changes even for rather weak disturbances.
In the region between the both regimes the remaining irregular
variations are suppressed in a model with nonlinear feedback
formulated as a traditional \alf-quenching.

A plane 1D $\alpha^2\Omega$-model similar to that of Section~6.1
is used to illustrate the influence of the fluctuating turbulence
coefficients on a mean-field dynamo. It is {\em not} an overshoot
dynamo (cf. Ossendrijver and Hoyng, 1996). The plane has infinite extent in
the radial ($x$) and the azimuthal ($y$) direction, the
boundaries are in the $z$-direction. Hence, we assume that the
magnetic field components in both azimuthal direction, ($B$), and
radial direction, ($\partial A/\partial z$),  depend  on $z$ only.
Note that $z$ points opposite to the colatitude $\theta$.
 
The $\alpha$-tensor has only one component (given in Section 2.3)
vanishing at the equator, and it is assumed here to vanish also 
at the poles. The magnetic feedback is considered to be conventional
$\alpha$-quenching: $\alpha=\hat\alpha \alpha_0 \Psi(B_{\rm
tot})$ with $\Psi(B_{\rm tot}) \propto B_{\rm tot}^{-2}$.
The diffusivity is spatially uniform. The normalized
dynamo equations are given in (\ref{dynamo1}) and (\ref{dynamo2}).

\begin{figure}
\psfig{figure=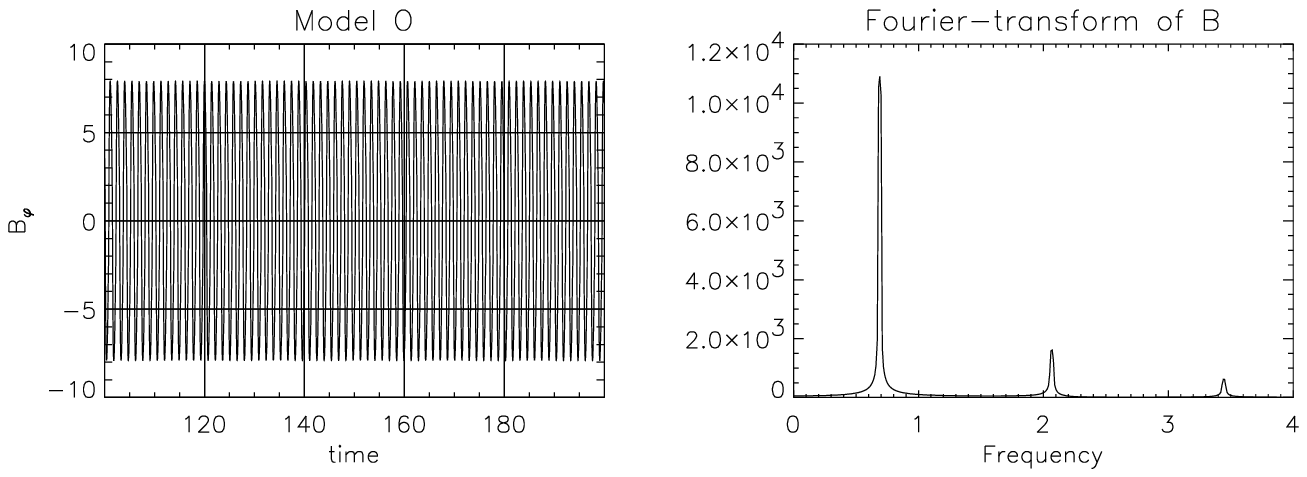}
\psfig{figure=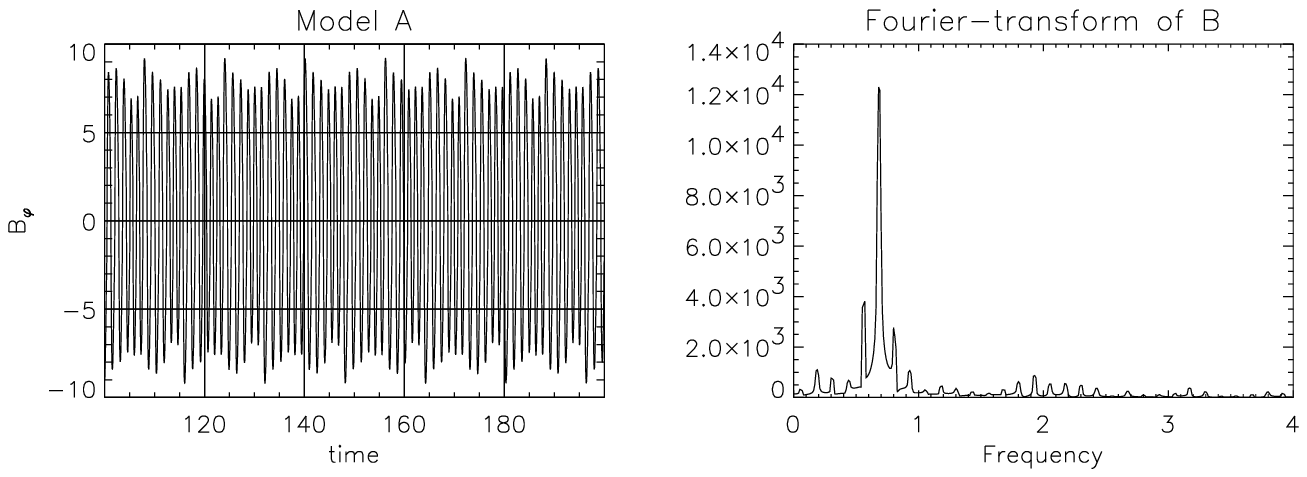}
\psfig{figure=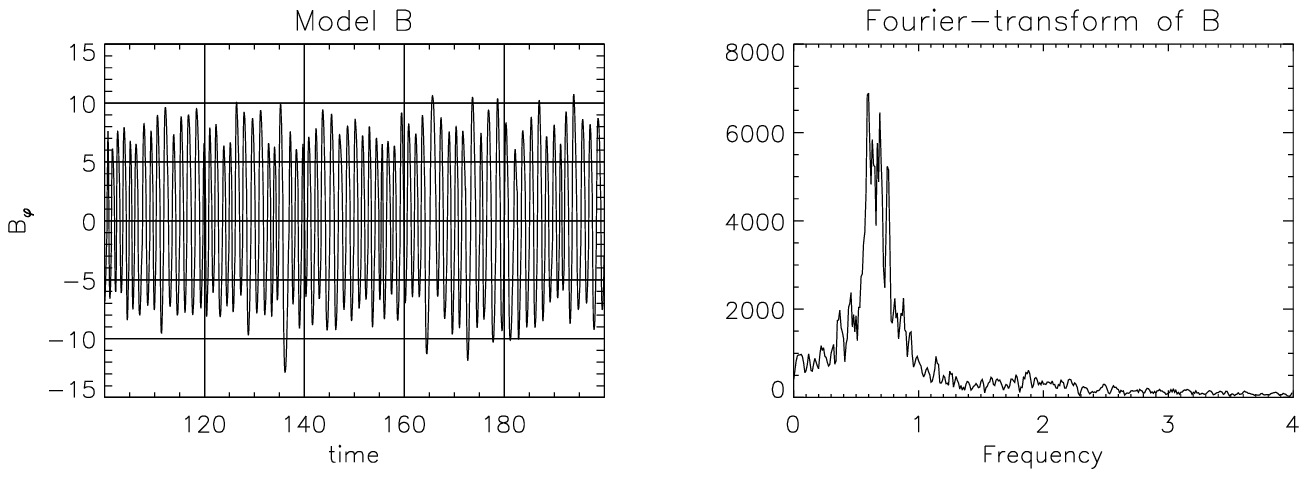}
\psfig{figure=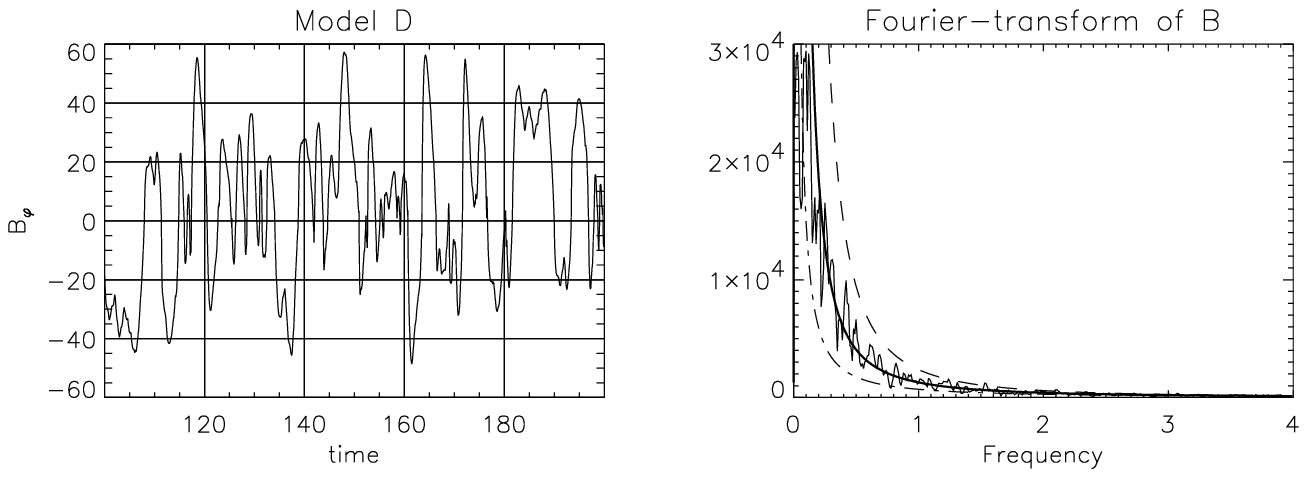}
\caption{Dynamo-induced magnetic toroidal fields  for the turbulence 
models $\cal O$ (27a), $\cal  A$ (27b), $\cal B$ (27c) and ${\cal D}$ (27d).
LEFT: time series, RIGHT: power spectra. The real time unit is 2.7~years.} 
\label{f15}
\end{figure}
 
Positive $C_\alpha$ describes a positive \alf\ in the northern 
hemisphere  and a negative one in the southern hemisphere.
$C_\Omega$ defines  the amplitude  of  the differential
rotation, positive
$C_\Omega$ represents positive shear $\partial \Omega/\partial
r$. The dynamo operates  with a positive \alf-effect in  the northern
hemisphere. Our  dynamo numbers are  $C_{\alpha}=5$ and $C_{\Omega}=200$. The
turbulence models ${\cal A, B}$ and ${\cal D}$ are applied, flow patterns
with  small (${\cal A}$), medium (${\cal B}$) and with very
large (${\cal D}$) eddies are used.
 
A magnetic dipole field  oscillating with a (normalized) period 
of 1.46 (Fig.~\ref{f15}a) is induced by a dynamo model without 
EMF-fluctuations ($N\to\infty$, called ${\cal O}$). The period corresponds 
to an activity cycle of 8 years in physical units. Also the turbulence 
model ${\cal B}$  produces an oscillating dipole but  with a 
much more complicated temporal behaviour (Fig.~\ref{f15}c). It is not 
a single oscillation, the power spectrum forms  a rather broad line with
substructures. The `quality',
\beg
Q = {\omega_{\rm cyc} \over \Delta \omega_{\rm cyc}} ,
\label{Q2}
\ende	
of this line (with $\Delta \omega_{\rm cyc}$ as its half-width) proves to be about 2.9.
This value  close to the observed quality  of the solar cycle
is produced here by a turbulence model with about 100 eddies along
the equator (see also Ossendrijver {\it et al.}, 1996).  There are
also remarkable variations of the magnetic cycle amplitude.

The turbulence field $\cal D$ lead, not suprisingly, to a highly
irregular temporal behaviour in the magnetic quantities. Its power 
spectrum peaks at several periods (Fig.~\ref{f15}d). The overall 
shape of the spectrum, however,  does not longer suggest any 
oscillations. The power of the lower frequencies is strongly 
increased, the high-frequency power decreases as 
$\omega_{\rm cyc}^{-5/3}$ like a Kolmogoroff spectrum, indicating the
existence of chaos.
 
Our experiments basically show how the dilution of
turbulence is able to transform a single-mode oscillation (for
very many cells) to an oscillation with low quality (for moderate
eddy population) and finally to a temporal behaviour close
to chaos (for very few cells). It is possible that the
observed low quality (\ref{dynb}) of the solar cycle  indicates the  finite
number of the giant cells driving the large-scale solar dynamo. 
Implications for the temporal evolution of the solar rotation 
law should be another output of such a cell number statistics. 
The consequences for rotation will be an independent
test of the presented theory  leading to the same
number of eddies contributing to the turbulent angular momentum
transport. For only `occasional' turbulence a
nontrivial time series for both the turbulence EMF-coefficients
and the magnetic field is an unavoidable consequence.

\section{References}
\begin{description}
\parskip0pt
\parsep0pt
\itemsep0pt
\item Abramenko, V.I., Wang, T. and V.B. Yurchishin, Analysis
of electric current helicity in active regions on the basis of
vector magnetograms. {\it Solar Phys.}, {\bf 168}, 75 (1996).
\item Baliunas, S.L. and A.H. Vaughan, Stellar activity
cycles. {\it Annual Review of Astronomy and Astrophysics}, {\bf
23}, 379 (1985).
\item Baliunas, S.L., E. Nesme-Ribes, D. Sokoloff and W.H. Soon,
A dynamo interpretation of stellar activity cycles. {\it
Astrophys. J.}, {\bf 460}, 848 (1996).
\item Bao, S. and H. Zhang, Patterns of current helicity for
solar cycle 22. {\it Astrophys. J.}, {\bf 496}, L43 (1998).
\item Barnes, J.R., A.C. Cameron, D.J. James, C.A. Watson and
J.-F. Donati, Starspot coverage and differential rotation on PZ
Tel, in {\it Proc. Stellar clusters and associations: Convection,
rotation, and dynamos}, Palermo (1999).
\item Beer, J., S.M. Tobias and N.O. Weiss, An active Sun
throughout the Maunder minimum. {\it Solar Phys.}, {\bf 181},
237 (1998).
\item Belvedere, G., G. Lanzafame and M.R.E. Proctor, The
latitude belts of solar activity as a consequence of a
boundary-layer dynamo. {\it Nature}, {\bf 350}, 481 (1991).
\item Braginsky, S.I., Self excitation of a magnetic field
during the motion of a highly conducting fluid. {\it Sov. Phys.
JETP}, {\bf 20}, 726 (1964). 
\item Brandenburg, A., I. Tuominen and D. Moss, On the
nonlinear stability of dynamo models. {\it Geophys. Astrophys.
Fluid Dynam.}, {\bf 49}, 129 (1989).
\item Brandenburg, A., \AA. Nordlund, P. Pulkkinen, R.F. Stein
 and I. Tuominen, 3-D simulation of turbulent cyclonic
magneto-convection. {\it Astron. Astrophys.}, {\bf 232}, 277
(1990).
\item Brandenburg, A., Simulating the solar dynamo, in {\it The cosmic
dynamo} (eds. F. Krause, K.-H. R\"adler and G. R\"udiger),
Kluwer, Dordrecht, 111 (1993). 
\item Brandenburg, A., Solar dynamos: Computational background,
in {\it Lectures on solar and 
planetary dynamos} (eds. M.R.E. Proctor and A.D. Gilbert),
Cambridge University Press, 117 (1994a).
\item Brandenburg, A., Hydrodynamical simulations of the solar
dynamo, in {\it Advances in solar physics} (eds. G. Belvedere and
M. Rodono), Springer, Heidelberg, 73 (1994b).
\item Brandenburg, A., Large scale turbulent dynamos, in {\it
Stellar and planetary magnetoconvection} (eds. J.~Brestensk\'{y}, 
S.~\v{S}ev\v{c}ik), {\it Acta Astr.~et
Geophys.~Univ.~Comenianae}, {\bf XIX}, 235 (1997). 
\item Brandenburg, A. and K.J. Donner, The dependence of the
dynamo alpha on vorticity. {\it Mon. Not. R. Astr. Soc.}, {\bf
288}, L29 (1997). 
\item Brandenburg, A., S.H. Saar and C.R. Turpin, Time
evolution of the magnetic activity cycle period. {\it
Astrophys. J.}, {\bf 498}, 51 (1998).
\item Brandenburg, A. and D. Schmitt, Simulations of an
alpha-effect due to magnetic buoyancy. {\it Astron.
Astrophys.}, {\bf 338}, L55 (1998). 
\item Brandenburg, A., Dynamo-generated turbulence and outflows
from accretion discs. {\it Phil. Trans. R. Soc. Lond. A}, {\bf
358}, 759 (2000).
\item Caligari, P., F. Moreno-Insertis and M. Sch\"ussler,
Emerging flux tubes in the solar convection zone. I. Asymmetry,
tilt and emergence latitude. {\it Astrophys. J.}, {\bf 441},
886 (1995).
\item Choudhuri, A.R., The evolution of loop structures in flux
rings within the solar convection zone. {\it Solar Phys.}, {\bf
123}, 217 (1989).
\item Choudhuri, A.R., On the possibility of an
$\alpha^2\omega$-type dynamo in a thin layer inside the Sun.
{\it Astrophys. J.}, {\bf 355}, 733 (1990).
\item Choudhuri, A.R., Stochastic fluctuations of the solar
dynamo. {\it Astron. Astrophys.}, {\bf 253}, 277 (1992).
\item Choudhuri, A.R., M. Sch\"ussler and M. Dikpati, The solar
dynamo with meridional circulation. {\it Astron. 
Astrophys.}, {\bf 303}, L29 (1995).
\item Christensen-Dalsgaard, J. and J. Schou, Differential
rotation in the solar interior, in {\it Seismology of the Sun
and Sun-like stars} (eds. V. Domingo and E.J. Rolfe), ESA
SP-286, 149 (1988).
\item DeLuca, E.E. and P.A. Gilman, The solar dynamo, in {\it Solar
interior and atmosphere} (eds. A.N. Cox, W.C. Livingston and
M.S. Matthews), Arizona University Press, Tucson, 275 (1991).
\item Dikpati, M. and P. Charbonneau, A Babcock-Leighton flux
transport dynamo with solar-like differential rotation. {\it
Astrophys. J.}, {\bf 518}, 508 (1999).
\item Donahue, R.A. and S.L. Baliunas, Evidence of
differential surface rotation in the solar-type star HD 114710.
{\it Astrophys. J.}, {\bf 393}, L63 (1992).
\item Donati, J.-F. and A.C. Cameron, Differential rotation and
magnetic polarity patterns on AB Doradus. {\it Mon. Not.
R. Astr. Soc.}, {\bf
291}, 1 (1997).
\item Durney, B.R. and R.D. Robinson, On an estimate of the
dynamo-generated magnetic fields in late-type stars. {\it
Astrophys. J.}, {\bf 253}, 290 (1982).
\item Durney, B.R., On the behavior of the angular velocity in
the lower part of the solar convection zone. {\it Astrophys.
J.}, {\bf 338}, 509 (1989).
\item Durney, B.R., On a Babcock-Leighton dynamo model with a
deep-seated generating layer for the toroidal magnetic field.
{\it Solar Phys.}, {\bf 160}, 213 (1995).
\item Durney, B.R., On a Babcock-Leighton dynamo model with a
deep-seated generating layer for the toroidal magnetic field, II.
{\it Solar Phys.}, {\bf 166}, 231 (1996).
\item Dziembowski, W.A. and P.R. Goode, Seismology for the fine
structure in the Sun's oscillations varying with its activity
cycle. {\it Astrophys. J.}, {\bf 376}, 782 (1991). 
\item Fan, Y., G.H. Fisher and E.E. DeLuca, The origin of
morphological asymmetries in bipolar active regions. {\it
Astrophys. J.}, {\bf 405}, 390 (1993).
\item Ferri\`{e}re, K., The full alpha-tensor due to supernova explosions
and superbubbles in the galactic disk. {\it Astrophys. J.},
{\bf 404}, 162 (1993).
\item Ferriz-Mas, A. and M. Sch\"ussler, Instabilities of
magnetic flux tubes in a stellar convection zone. I. Equatorial
flux rings in differentially rotating stars. {\it Geophys.\
Astrophys.\ Fluid Dynam.}, {\bf 72}, 209 (1993).
\item Ferriz-Mas, A., D. Schmitt and M. Sch\"ussler, A dynamo
effect due to instability of magnetic flux tubes. {\it
Astron. Astrophys.}, {\bf 289}, 949 (1994).
\item Ferriz-Mas,\,A. and M.\,Sch\"ussler, Instabilities of
magnetic flux tubes in a stellar convection zone.\ 
II.\,Flux\,rings\,outside\,the\,equatorial\,plane.\ 
{\it Geophys.\,Astrophys.\,Fluid\,Dynam.},\,{\bf 81},\,233\,(1995).
\item Frick, P., D. Galyagin, D.V. Hoyt, E. Nesme-Ribes, 
K.H. Schatten, D. Sokoloff and V. Zakharov, Wavelet analysis
of solar activity recorded by sunspot groups. {\it Astron. 
Astrophys.}, {\bf 328}, 670 (1997a).
\item Frick, P., E. Nesme-Ribes and D. Sokoloff, Wavelet
analysis of solar activity recorded by sunspot groups and solar
diameter data, in {\it Stellar and planetary magnetoconvection}
(eds. J. Bresten\-sk\'{y} and S. \v{S}ev\v{c}ik), {\it Acta
Astron. et Geophys. Univ. Comenianae}, {\bf XIX}, 113 (1997b).
\item Gilman, P.A., What can we learn about solar cycle
mechanisms from observed velocity fields, in {\it The solar
cycle} (ed. K.L. Harvey), {\it ASP Conf. Ser.}, {\bf 27}, 241 (1992).
\item Glatzmaier, G.A. and R. Roberts, A three-dimensional
convective dynamo solution with rotating and finitely
conducting inner core and mantle. {\it Phys. Earth Planet.
Inter.}, {\bf 91}, 63 (1995).
\item Hale, G.E., The fields of force in the atmosphere of the
Sun. {\it Nature}, {\bf 119}, 708 (1927). 
\item Hempelmann, A., J. Schmitt and K. St\c{e}pie\`{n}, Coronal X-ray
emission of cool stars in relation to chromospheric activity
and magnetic cycles. {\it Astron. Astrophys.}, {\bf 305},
284 (1996).
\item Hood, L.L. and J.L. Jirikowic, A probable approx. 2400
year solar quasi-cycle in atmospheric delta C-14, in {\it
Climate impact of solar variability}, NASA, 98 (1990).
\item Howard, R.F. and B.J. LaBonte, The sun is observed to be a
torsional oscillator with a period of 11 years. {\it Astrophys. J.}, {\bf
239}, L33 (1980). 
\item Hoyng, P., Turbulent transport of magnetic fields. III.
Stochastic excitation of global magnetic modes. {\it Astrophys.
J.}, {\bf 332}, 857 (1988).
\item Hoyng, P., Helicity fluctuations in mean field theory: an
explanation for the variability of the solar cycle. {\it
Astron. Astrophys.}, {\bf 272}, 321 (1993).
\item Hoyng, P., D. Schmitt and L.J.W. Teuben, The effect of
random alpha-fluctuations and the global properties of the
solar magnetic field. {\it Astron. Astrophys.}, {\bf 289},
265 (1994).
\item Jennings, R.L. and N.O. Weiss, Symmetry breaking in stellar
dynamos. {\it Mon. Not. R. Astr. Soc.}, {\bf 252}, 249 (1991).
\item Kaisig, M., G. R\"udiger and H.W. Yorke, The alpha-effect
due to supernova explosions. {\it Astron. Astrophys.}, {\bf
274}, 757 (1993).
\item Keinigs, R.K., A new interpretation of the alpha effect.
{\it Phys. Fluids}, {\bf 76}, 2558 (1983). 
\item Kitchatinov, L.L., Turbulent transport of angular
momentum and differential rotation. {\it Geophys. Astrophys.
Fluid Dynam.}, {\bf 35}, 93 (1986).
\item Kitchatinov, L.L. and G. R\"udiger, Magnetic-field
advection in inhomogeneous turbulence. {\it Astron. 
Astrophys.}, {\bf 260}, 494 (1992).
\item Kitchatinov, L.L., Turbulent transport of magnetic fields
and the solar dynamo, in {\it The cosmic
dynamo} (eds. F. Krause, K.-H. R\"adler and G. R\"udiger),
Kluwer, Dordrecht, 13 (1993).
\item Kitchatinov, L.L. and V.V. Pipin, Mean-field buoyancy.
{\it Astron. Astrophys.}, {\bf 274}, 647 (1993).
\item Kitchatinov, L.L. and G. R\"udiger, Lambda-effect and
differential rotation in stellar convection zones. {\it Astron.
 Astrophys.}, {\bf 276}, 96 (1993).
\item Kitchatinov, L.L., G. R\"udiger and M. K\"uker,
$\Lambda$-quenching as the nonlinearity in stellar-turbulence
dynamos. {\it Astron. Astrophys.}, {\bf 292}, 125 (1994a).
\item Kitchatinov, L.L., V.V. Pipin and G. R\"udiger, Turbulent
viscosity, magnetic diffusivity, and heat conductivity under
the influence of rotation and magnetic field. {\it Astron.
Nachr.}, {\bf 315}, 157 (1994b).
\item Kitchatinov, L.L. and G. R\"udiger, Differential
rotation in solar-type stars: revisiting the Taylor number
puzzle. {\it Astron. Astrophys.}, {\bf 299}, 446 (1995).
\item Kitchatinov, L.L. and G. R\"udiger, Differential
rotation models for late-type dwarfs and giants. {\it Astron.
 Astrophys.}, {\bf 344}, 911 (1999).
\item Knobloch, E. and A.S. Landsberg, A new model of the solar
cycle. {\it Mon. Not. R. Astr. Soc.}, {\bf 278}, 294 (1996).
\item Knobloch, E., S.M. Tobias and N.O. Weiss, Modulation
and symmetry changes in stellar dynamos. {\it Mon. Not.
R. Astr. Soc.}, {\bf 297}, 1123 (1998).
\item K\"ohler, H., The solar dynamo and estimates of the
magnetic diffusivity and the $\alpha$-effect. {\it Astron. Astrophys.}, {\bf
25}, 467 (1973).
\item Kosovichev, A.G., J. Schou, P.H. Scherrer, {\it et al.},
Structure and rotation of the solar interior: Initial results
from the MDI medium-l program. {\it Solar Phys.}, {\bf 170}, 43
(1997).
\item Krause, F., Eine L\"osung des Dynamoproblems auf der
Grundlage einer linearen Theorie der magnetohydrodynamischen
Turbulenz. {\it Thesis}, Universit\"at Jena (1967). 
\item Krause, F. and K.-H. R\"adler, {\it Mean-field
magnetohydrodynamics and dynamo theory}, Pergamon Press, Oxford (1980).
\item Krause, F. and R. Meinel, Stability of simple nonlinear
$\alpha^2$-dynamos. {\it Geophys. Astrophys. Fluid Dynam.}, {\bf
43}, 95 (1988).
\item Krause, F. and H.-J. Schmidt, A low-dimensional attractor
for modeling the reversals of the Earth's magnetic field. {\it Phys.
Earth Planet. Inter.}, {\bf 52}, 23 (1988).
\item Krivodubskij, V.N. and M. Schultz, Complete alpha-tensor
for solar dynamo, in {\it The cosmic
dynamo} (eds. F. Krause, K.-H. R\"adler and G. R\"udiger),
Kluwer, Dordrecht, 25 (1993). 
\item K\"uker, M., G. R\"udiger and L.L. Kitchatinov, An
alpha-omega model of the solar differential rotation. {\it
Astron. Astrophys.}, {\bf 279}, 1 (1993).
\item K\"uker, M., G. R\"udiger and V.V. Pipin, Solar torsional
oscillations as due to the magnetic quenching of the Reynolds
stress. {\it Astron. Astrophys.}, {\bf 312}, 615 (1996).
\item K\"uker, M., R. Arlt and G. R\"udiger, The Maunder
minimum as due to magnetic $\Lambda$-quenching. {\it Astron.
Astrophys.}, {\bf 343}, 977 (1999).
\item Kurths, J., A. Brandenburg, U. Feudel and W. Jansen,
Chaos in nonlinear dynamo models, in {\it The cosmic dynamo} (eds.
F. Krause, K.-H. R\"adler and G. R\"udiger), Kluwer, Dordrecht,
83 (1993).
\item Kurths, J., U. Feudel, W. Jansen, U. Schwarz and H. Vos, Solar
variability: simple models and proxy data, in {\it Proc. Int.
School Physics Enrico Fermi, Course CXXXIII} (eds. G.
Castagnoli and A. Provenzale), IOS Press, Amsterdam, 247 (1997).
\item Levy, E.H., Physical assessment of stellar dynamo theory,
in {\it Cool stars, stellar systems, and the Sun} (eds. M.S.
Giampapa and J.A. Bookbinder), {\it ASP Conf. Ser.}, {\bf
26}, 223 (1992).
\item Low, B.C., Solar activity and the corona. {\it Solar Phys.}, {\bf 167}, 217 (1996).
\item Malkus, W.V.R. and  M.R.E. Proctor, The macrodynamics of
$\alpha$-effect dynamos in rotating fluids. {\it J. Fluid
Mech.}, {\bf 67}, 417 (1975).
\item Markiel, J.A., Thomas, J.H., Solar interface dynamo models
with a realistic rotation profile. {\it Astrophys. J.}, {\bf 523},
827 (1999). 
\item Moffatt, K.H., {\it Magnetic field generation in electrically
conducting fluids}, Cambridge University Press (1978).
\item Moreno-Insertis, F., Rise times of horizontal magnetic
flux tubes in the convection zone of the sun. {\it Astron. Astrophys.},
{\bf 122}, 241 (1983).
\item Moss, D., I. Tuominen and A. Brandenburg, Nonlinear
dynamos with magnetic buoyancy in spherical geometry. {\it
Astron. Astrophys.}, {\bf 228}, 284 (1990).
\item Moss, D., A. Brandenburg, R. Tavakol and I. Tuominen,
Stochastic effects in mean field dynamos. {\it Astron.
Astrophys.}, {\bf 265}, 843 (1992). 
\item Nesme-Ribes, E., D. Sokoloff, J.C. Ribes and
M. Kremliovsky, The maunder minimum and the solar dynamo, in
{\it The solar engine and its influence on terrestrial
atmosphere and climate} (ed. E. Nesme-Ribes), Springer-Verlag,
Berlin, 71 (1994). 
\item Noyes, R.W., N.O. Weiss and A.H. Vaughan, The relation
between stellar rotation rate and activity cycle periods. {\it
Astrophys. J.}, {\bf 287}, 769 (1984).
\item Ossendrijver, A.J.H. and P. Hoyng, Stochastic and
nonlinear fluctuations in a mean field dynamo. {\it Astron.
Astrophys.}, {\bf 313}, 959 (1996). 
\item Ossendrijver, A.J.H., P. Hoyng and D. Schmitt, Stochastic
excitation and memory of the solar dynamo. {\it Astron.
Astrophys.}, {\bf 313}, 938 (1996). 
\item Ossendrijver, A.J.H., On the cycle periods of stellar
dynamos. {\it Astron. Astrophys.}, {\bf 323}, 151 (1997).
\item Otmianowska-Mazur, K., M. Urbanik and A. Terech, Magnetic
field in a turbulent galactic disk. {\it Geophys. Astrophys.
Fluid Dynam.}, {\bf 66}, 209 (1992).
\item Otmianowska-Mazur, K., G. R\"udiger, D. Elstner and
R. Arlt, The turbulent EMF as a time series and the `quality'
of dynamo cycles. {\it Geophys. Astrophys. Fluid Dynam.}, {\bf
86}, 229 (1997).
\item Parker, E.N., The generation of magnetic fields in
astrophysical bodies. X - Magnetic buoyancy and the solar
dynamo. {\it Astrophys. J.}, {\bf 198}, 205 (1975).
\item Parker, E.N., The dynamo dilemma. {\it Solar Phys.}, {\bf
110}, 11 (1987).
\item Pevtsov, A.A., Canfield, R.C. and T.R. Metcalf,
Latitudinal variation of helicity of photospheric magnetic
fields. {\it Astrophys. J.}, {\bf 440}, L109 (1995).
\item Pipin, V.V., The Gleissberg cycle by a nonlinear $\alpha\Lambda$ dynamo.
{\it Astron.\ Astrophys.},\,{\bf 346}, 295\,(1999).
\item Prautzsch, T., The dynamo mechanism in the deep
convection zone of the Sun, in {\it Theory of solar and
planetary dynamos} (eds. P.C. Matthews and A.M. Rucklidge),
Cambridge University Press, 249 (1993).
\item R\"adler, K.-H., Zur Elektrodynamik turbulent bewegter
leitender Medien. Teil II. {\it Zeitschr. f. Naturforschg.},
{\bf 23a}, 1851 (1968).
\item R\"adler, K.-H. and N. Seehafer, Relations between
helicities in mean-field dynamo models, in {\it Topological
Fluid Mechanics} (eds. H.K. Moffatt and A. Tsinober) Cambridge
University Press, 157 (1990).
\item Ribes, J.C. and E. Nesme-Ribes, The solar sunspot cycle
in the Maunder minimum AD1645. {\it Astron. Astrophys.},
{\bf 276}, 549 (1993).
\item Roberts, P.H., Kinematic dynamo models. {\it Phil. Trans.
R. Soc. Lond. A.}, {\bf 272}, 663 (1972).
\item Roberts, P.H. and M. Stix, $\alpha$-Effect dynamos by the
Bullard-Gellman formalism. {\it Astron. Astrophys.}, {\bf 18}, 453 (1972).
\item Rozelot, J.P., On the chaotic behaviour of the solar
activity. {\it Astron. Astrophys.}, {\bf 297}, L45 (1995).
\item R\"udiger, G., I. Tuominen, F. Krause and H. Virtanen,
Dynamo-generated flows in the Sun. {\it Astron. Astrophys.},
{\bf 166}, 306 (1986).
\item R\"udiger, G., {\it Differential rotation and stellar
convection: Sun and solar-type stars}, Gordon and Breach Science
Publishers, New York (1989).
\item R\"udiger, G. and L.L. Kitchatinov, $\alpha$-effect and
$\alpha$-quenching. {\it Astron. Astrophys.}, {\bf 269}, 581
(1993).
\item R\"udiger, G., L.L. Kitchatinov, M. K\"uker and
M. Schultz, Dynamo models with magnetic diffusiv\-ity-quenching.
{\it Geophys. Astrophys. Fluid Dynam.}, {\bf 78}, 247 (1994).
\item R\"udiger, G. and A. Brandenburg, A solar dynamo in the
overshoot layer: cycle period and butterfly diagram. {\it
Astron. Astrophys.}, {\bf 296}, 557 (1995).
\item R\"udiger, G. and R. Arlt, Cycle times and magnetic
amplitudes in nonlinear 1D $\alpha^2\Omega$-dynamos. {\it
Astron. Astrophys.}, {\bf 316}, L17 (1996).
\item R\"udiger, G. and L.L. Kitchatinov, The slender solar
tachocline: a magnetic model. {\it Astron. Nachr.}, {\bf 318},
273 (1997).
\item R\"udiger, G. and M. Schultz, Nonlinear galactic dynamo
models with magnetic-supported interstellar gas-density
stratification. {\it Astron. Astrophys.}, {\bf 319}, 781 (1997).
\item R\"udiger, G. and V.V. Pipin, {\it Astron. Astrophys.} in
prep. (2000).
\item Saar, S.H. and S.L. Baliunas, Recent advances in stellar
cycle research, in {\it The solar cycle} (ed. K.L.
Harvey), {\it ASP Conf. Ser.}, {\bf 27}, 150 (1992a). 
\item Saar, S.H. and S.L. Baliunas, The magnetic cycle of
$\kappa$ Ceti (G5V), in {\it The solar cycle} (ed.
K.L. Harvey), {\it ASP Conf. Ser.}, {\bf 27}, 197 (1992b).
\item Saar, S.H., Recent measurements of stellar magnetic
fields, in {\it Stellar surface structure} (eds. K.G.
Strassmeier and J.L. Linsky), Kluwer, Dordrecht, 237 (1996).
\item Saar, S.H. and A. Brandenburg, Time evolution of the
magnetic activity cycle period. II. Results for an expanded stellar
sample. {\it Astrophys. J.}, {\bf 524}, 295 (1999).
\item Schlichenmaier, R. and M. Stix, The phase of the radial
mean field in the solar dynamo. {\it Astron. Astrophys.}, {\bf
302}, 264 (1995).
\item Schmitt, D., {\it Thesis}, Universit\"at G\"ottingen (1985).
\item Schmitt, D., An $\alpha\omega$-dynamo with an
$\alpha$-effect due to magnetostrophic waves. {\it Astron.
Astrophys.}, {\bf 174}, 281 (1987). 
\item Schmitt, D. and M. Sch\"ussler, Non-linear dynamos. I.
One-dimensional model of a thin layer dynamo. {\it Astron.
Astrophys.}, {\bf 223}, 343 (1989). 
\item Schmitt, D., The solar dynamo, in {\it The cosmic
dynamo} (eds. F. Krause, K.-H. R\"adler and G. R\"udiger),
Kluwer, Dordrecht, 1 (1993). 
\item Schmitt, D., M. Sch\"ussler and A. Ferriz-Mas, 
Intermittent solar activity by an on-off dynamo. {\it Astron.
Astrophys.}, {\bf 311}, L1 (1996). 
\item Sch\"ussler, M., The solar torsional oscillation and
dynamo models of the solar cycle. {\it Astron. Astrophys.},
{\bf 94}, 17 (1981).
\item Sch\"ussler, M., Magnetic fields and the rotation of the
solar convection zone, in {\it The internal solar angular
velocity} (ed. R. Durney), Reidel, Dordrecht, 303 (1987).
\item Schwarz, U., Zeitreihenanalyse astrophysikalischer
Aktivit\"atsph\"anomene. {\it Thesis}, Universit\"at 
Potsdam (1994).
\item Seehafer, N., Electric current helicity in the solar
atmosphere. {\it Solar Phys.}, {\bf 125}, 219 (1990). 
\item Soon, W.H., S.L. Baliunas and Q. Zhang, An interpretation
of cycle periods of stellar chromospheric activity. {\it
Astrophys. J.}, {\bf 414}, L33 (1993).
\item Spiegel, E.A. and N.O. Weiss, Magnetic activity and
variations in solar luminosity. {\it Nature}, {\bf
287}, 616 (1980).
\item Sp\"orer, G., \"Uber die Periodizit\"at der Sonnenflecken
seit dem Jahre 1618. {\it Vierteljahresschrift
der Astron. Ges.}, {\bf 22}, 323 (1887).
\item Steenbeck, M. and F. Krause, Zur Dynamotheorie stellarer
und planetarer Magnetfelder. I. Berechnung sonnen\"ahnlicher
Wechselfeldgeneratoren. {\it Astron. Nachr.}, {\bf 291}, 49 (1969).
\item Stix, M., Differential rotation and the solar dynamo.
{\it Astron. Astrophys.}, {\bf 47}, 
243 (1976).
\item Stix, M. and D. Skaley, The equation of state and the
frequencies of solar p modes. {\it Astron. Astrophys.}, {\bf
232}, 234 (1990).
\item Stix, M., The solar dynamo. {\it Geophys. Astrophys.
Fluid Dynam.}, {\bf 62}, 211 (1991). 
\item Tobias, S.M., Grand minima in nonlinear dynamos. {\it
Astron. Astrophys.}, {\bf 307}, L21 (1996).
\item Tobias, S.M., The solar cycle: parity interactions and
amplitude modulation. {\it Astron. Astrophys.}, {\bf 322},
1007 (1997).
\item Tobias, S.M., Relating stellar cycle periods to dynamo
calculations. {\it Mon. Not. R. Astr. Soc.}, {\bf 296}, 653 (1998).
\item Tuominen, I., G. R\"udiger and A. Brandenburg,
Observational constraints for solar-type dynamos, in {\it
Activity in cool star envelopes} (eds. Havnes et al.), Kluwer,
Dordrecht, 13 (1988).
\item Vainshtein, S.I. and F. Cattaneo, Nonlinear restriction
on dynamo action. {\it Astrophys. J.}, {\bf 393}, 165 (1992).
\item van Ballegooijen, A.A., The overshoot layer at the base
of the solar convective zone and the problem of magnetic flux
storage. {\it Astron. Astrophys.}, {\bf 113}, 99 (1982).
\item van Ballegooijen, A.A., Understanding the solar cycle, in
{\it Synoptic solar physics} (eds. K.S. Balasubramaniam, J.
Harvey and D. Rabin), {\it ASP Conf. Ser.}, {\bf 140}, 17 (1998).
\item Verma, V.K., On the north-south asymmetry of solar
activity cycles. {\it Astrophys. J.}, {\bf 403}, 797 (1993).
\item Vishniac, E.T. and A. Brandenburg, An incoherent
$\alpha\omega$-dynamo in accretion disks. {\it Astrophys. J.},
{\bf 475}, 263 (1997).
\item Vos, H., J. Kurths and U. Schwarz, Reconstruction of
grand minima of solar activity from radiocarbon data -- linear and
nonlinear signal analysis. {\it J.
Geophys. Res. A}, {\bf 101}, 15637 (1996). 
\item Vos, H., A. Sanchez, B. Zolitschka, A. Brauer and
J.F.W. Negendank, Solar activity variations recorded in varved
sediments from the crater lake of Holzmaar -- A maar lake in
the Westeifel volcanic field, Germany. {\it Surv. Geophys.},
{\bf 18}, 163 (1997).
\item W\"alder, M., W. Deinzer and M. Stix, Dynamo action
associated with random waves in a rotating stratified fluid. {\it J.
Fluid Mech.}, {\bf 96}, 207 (1980).
\item Weiss, N.O., F. Cattaneo and C.A. Jones, Periodic and
aperiodic dynamo waves. {\it Geophys. Astrophys. Fluid Dynam.},
{\bf 30}, 305 (1984).
\item Weiss, N.O., Solar and stellar dynamos, in {\it Lectures on solar and
planetary dynamos} (eds. M.R.E. Proctor and A.D. Gilbert),
Cambridge University Press, 59 (1994).  
\item Wittmann, A., The sunspot cycle before the maunder
minimum. {\it Astron.\ Astrophys.},\,{\bf 
66},\,93\,(1978).
\item Yoshimura, H., Phase relation between the poloidal and
toroidal solar-cycle general magnetic fields and location of
the origin of the surface magnetic fields. {\it Solar Phys.},
{\bf 50}, 3 (1976). 
\item Yoshimura, H., Solar cycle Lorentz force waves and the
torsional oscillations of the sun. {\it Astrophys. J.}, {\bf
247}, 1102 (1981).
\item Ziegler, U., H.W. Yorke and M. Kaisig, The role of
supernovae for the galactic dynamo: The full alpha-tensor. {\it
Astron. Astrophys.}, {\bf 305}, 114 (1996).

\end{description}

\end{document}